\documentclass[10pt]{article}
\pdfoutput=1
\usepackage{jpub}
\usepackage[utf8]{inputenc}
\usepackage{amsmath,amssymb,amsbsy,amstext,amsthm,simplewick,amsfonts}
\usepackage{graphicx, placeins}
\usepackage{subcaption}
\usepackage{lscape}
\usepackage{cancel}
\usepackage{wrapfig}
\usepackage{upgreek}
\usepackage{bm} 
\usepackage{framed}
\usepackage{bbm}
\usepackage{textcomp}
\usepackage{mathtools}
\usepackage{stackengine}
\setstackEOL{\\}

\usepackage{tikz}
\tikzset{every picture/.style={line width=0.75pt}} 

\tikzset{graviton/.style={decorate, decoration={snake, amplitude=.4mm, segment length=1.5mm, pre length=.5mm, post length=.5mm}, double}}

\usepackage{pifont}
\usetikzlibrary{matrix,shapes,fit,tikzmark,calc}
\usepackage{tikz-cd, tikz-feynman} 
\usepackage{adjustbox}
\usepackage{makecell}
\usepackage{tcolorbox}
\usepackage{physics}
\usepackage{empheq}
\usepackage[normalem]{ulem}
\usepackage{enumitem}
\usepackage{braket} 
\usepackage{array}
\usepackage{dsfont}
\usepackage{ulem}
\usepackage{titlesec}
\titleformat{\section}{\normalfont\fontsize{12}{16}\bfseries}{\thesection}{1em}{}

\usepackage[framemethod=TikZ]{mdframed}
\mdfdefinestyle{boxed}{%
linecolor=black,
roundcorner=5pt,
innerleftmargin=0,%
innerrightmargin=0,
backgroundcolor=gray!10,%
}

\definecolor{blue3}{RGB}{31,119,180}
\definecolor{red3}{RGB}{214,39,40}
\definecolor{orange3}{RGB}{255,127,14}
\definecolor{green3}{RGB}{44,160,44}

\usepackage{hyperref}
\hypersetup{colorlinks=true,linkcolor=teal,citecolor=orange3,urlcolor=green3,pdfencoding=auto}

\def\Disc{\text{Disc}}
\def\E{{\mathcal{E}}}
\def\th{{\vartheta}}
\def\bfb{{\bm b}}
\def\bfx{{\bm x}}

\def\bfr{{\bm r}}
\def\bfk{{\bm k}}
\def\bfp{{\bm p}}
\def\bfq{{\bm q}}
\def\bfl{{\bm \ell}}
\def\bfP{{\bm P}}
\def\bfX{{\bm X}}

\def\bfL{{\bm L}}

\def\d{{\rm d}}

\def\E{{\cal E}}

\def\bfp{\textbf{p}}

\def\beq{\begin{equation}}
\def\eeq{\end{equation}}
\def\bfK{\mathbf{K}}

\allowdisplaybreaks[1]
\begin{document}


\title{Orbital Precession and Hidden Symmetries in Scalar-Tensor Theories}

\author[a]{Anne-Christine Davis}
\author[b,a]{ and Scott Melville}

\affiliation[a]{DAMTP, University of Cambridge, Wilberforce Road, Cambridge, CB3 0WA, U.K.}
\affiliation[b]{Astronomy Unit, Queen Mary University of London, Mile End Road, London, E1 4NS, U.K.}


\date{today}
\abstract{
We revisit the connection between relativistic orbital precession, the Laplace-Runge-Lenz symmetry, and the $t$-channel discontinuity of scattering amplitudes. 
Applying this to scalar-tensor theories of gravity, we compute the conservative potential and orbital precession induced by both conformal/disformal-type couplings at second Post-Minkowskian order ($\mathcal{O} \left( G_N^2 \right)$), complementing the known third/first order Post-Newtonian results.
There is a particular tuning of the conformal coupling for which the precession vanishes at leading PN order, and we show that this coincides with the emergence of a Laplace-Runge-Lenz symmetry and a corresponding soft behaviour of the amplitude.
While a single scalar field inevitably breaks this symmetry at higher PN orders, certain supersymmetric extensions have recently been shown to have an have an exact Laplace-Runge-Lenz symmetry and therefore classical orbits do not precess at any PN order.
This symmetry can be used to relate scattering amplitudes at different loop orders, and we show how this may be used to bootstrap the (classically relevant part of the) three-loop $2 \to 2$ scattering of charged black holes in $\mathcal{N}=8$ supergravity from existing two-loop calculations.
}


\setcounter{tocdepth}{3}
\maketitle


\section{Introduction}

Following the advent of gravitational wave astronomy \cite{LIGOScientific:2014pky, VIRGO:2014yos, KAGRA:2013rdx, LIGOScientific:2021djp},
 there has been a greatly renewed interest in the two-body problem. 
Accurate analytical calculations of how a binary system evolves during the inspiral phase (in which the two objects are widely separated and slowly moving) are essential for the efficient waveform modelling required to extract transient gravitational wave signals from data \cite{LIGOScientific:2019hgc, Ossokine:2020kjp, Pratten:2020ceb, Gamba:2021ydi}. This is particularly the case for next generation observatories such as LISA, which will observe a much longer portion of this inspiral phase \cite{LISA:2017pwj, LISA:2022yao, LISA:2022kgy}, as well as future terrestrial detectors \cite{Reitze:2019iox, Maggiore:2019uih}  \\

Gravitational two-body systems are a particularly interesting laboratory in which to probe our best theories of gravity and look for signs of any deviation from General Relativity (GR). 
For one thing, the strong gravitational fields near compact objects represent an extreme environment in which we have yet to accurately test the predictions of GR. 
Furthermore, gravitationally bound states can be particularly sensitive to new light degrees of freedom, since they generically affect both the binding energy and the radiative losses. 
The binary system of two astrophysical compact objects can therefore play an analogous role to the Hydrogen atom in the development of quantum mechanics, since theoretical calculations in this idealised system are now being confronted by increasingly precise spectra from gravitational wave observatories \cite{LIGOScientific:2021sio, LIGOScientific:2016lio, LIGOScientific:2018dkp, LIGOScientific:2019fpa, LIGOScientific:2020tif}.  \\

In this work, we study the evolution of binary systems in which the two compact objects couple to an effective metric $\tilde{g}_{\mu\nu}$ built from the $g_{\mu \nu}$ of General Relativity and one additional (massless) scalar field $\phi$.
Adding a single degree of freedom in this way is arguably the simplest modification one can make to GR. 
Such scalar fields (non-minimally coupled to matter) arise naturally in a host of different gravity theories, including string theory and supergravity.
In cosmology, these scalar-tensor theories are routinely used to model the effects of dark matter or dark energy on large scales; or deviations from Einstein's theory in the strong-field regime (see, e.g., \cite{Clifton:2011jh, Bamba:2012cp, Joyce:2014kja, Hui:2016ltb, Urena-Lopez:2019kud, Berti:2015itd} for reviews of this extensive literature). \\

The most general effective metric compatible with causality is the Bekenstein metric \cite{Bekenstein:1992pj}, 
\begin{align}
 \tilde{g}_{\mu\nu} = e^{ C \left( \frac{\phi}{M_P} \right) } \eta_{\mu\nu} + D \left( \frac{\phi}{M_P} \right) \,  \frac{ \nabla_\mu \phi \nabla_\nu \phi }{ M_P^2 M_{\partial}^2 } \; .
 \label{eqn:geff_def}
\end{align}
The dimensionless functions $C$ and $D$ describe so-called \emph{conformal} and \emph{disformal} couplings between $\phi$ and matter. $M_P$ and $M_{\partial}$ are fixed energy scales which control the Effective Field Theory expansion at low energies. 
The conformal-type coupling $C (\phi)$ is tightly constrained by local tests of gravity in both the laboratory and the Solar System~\cite{Bertotti:2003rm, Adelberger:2009zz, Burrage:2017qrf, Hofmann2018, Berge:2017ovy}. 
The disformal-type coupling $D (\phi)$ has also been constrained experimentally\cite{Koivisto:2008ak, Zumalacarregui:2010wj, Koivisto:2012za, vandeBruck:2013yxa, Neveu:2014vua, Sakstein:2014isa, Sakstein:2014aca, Ip:2015qsa, Sakstein:2015jca, vandeBruck:2015ida, vandeBruck:2016cnh, Kaloper:2003yf, Brax:2014vva, Brax:2015hma, Brax:2012ie, vandeBruck:2012vq, Brax:2013nsa}, although due to the additional derivatives its effects are suppressed at large distances and hence the viable parameter space for $D/M_{\partial}^2$ remains much larger than that of its conformal counterpart.
Recent work~\cite{Brax:2018bow, Brax:2019tcy, Kuntz:2019zef, Melville:2019wyy, Brax:2020vgg, Brax:2021qqo} has studied the effects of such a disformal coupling in binary systems, where it leads to non-trivial corrections to the two-body potential that, although higher-order in a post-Newtonian expansion in powers of velocity, can nonetheless compete with the conformal coupling for sufficiently low $M_{\partial}$.  
\\

These previous studies have focussed on the leading post-Newtonian (PN) effects of the disformal interaction by expanding in small relative velocity. Here we find the complementary post-Minkowskian (PM) expansion of the potential and orbital motion by expanding instead in the weak coupling of the fields (powers of $1/M_P$).
This is made possible thanks to recent advances in matching field-theory scattering amplitudes to the classical potential and observables
\cite{Cheung:2018wkq,Bern:2019nnu,Bern:2019crd,Bjerrum-Bohr:2018xdl,Ciafaloni:2018uwe,Bjerrum-Bohr:2019kec,Cachazo:2017jef,Cristofoli:2019neg,Damgaard:2019lfh,Cristofoli:2020uzm,Kosower:2018adc,Maybee:2019jus,KoemansCollado:2019ggb,Mougiakakos:2020laz,Parra-Martinez:2020dzs,Bern:2020buy,Bern:2021dqo,Herrmann:2021tct,DiVecchia:2020ymx,Kalin:2019rwq,Kalin:2019inp,Bjerrum-Bohr:2021vuf,Cheung:2020gyp,Damour:2020tta,Kalin:2020lmz,DiVecchia:2021bdo,Liu:2021zxr,DiVecchia:2021ndb,Cho:2021mqw,Bjerrum-Bohr:2021din,Dlapa:2021npj,Cristofoli:2021vyo,Bautista:2021wfy,Kosmopoulos:2021zoq,delaCruz:2020bbn,delaCruz:2021gjp}.
This machinery has been applied to great effect in both General Relativity and in various extensions, such as including higher-curvature corrections \cite{Brandhuber:2019qpg,AccettulliHuber:2019jqo,Emond:2019crr,AccettulliHuber:2020oou,AccettulliHuber:2020dal} or further degrees of freedom like the helicity-0 mode \cite{Carrillo-Gonzalez:2021mqj}, which behaves as a Galileon scalar-tensor theory.
We find that this amplitude approach is particularly well-suited to studying the disformal interaction, since while the time derivatives can be laborious to treat in more traditional approaches, they are treated covariantly in the amplitude and amount to a simple $p_1^\mu p_{2}^{\nu}$ factor compared to the conformal-type coupling.
\\

Interestingly, there is a particular tuning of $C(\phi)$ for which the leading orbital precession in this scalar-tensor theory \emph{vanishes} and the classical orbits remain closed ellipses at 1PN. 
The fact that orbits close in Newtonian mechanics can be explained by the presence of a ``hidden symmetry'' called the Laplace-Runge-Lenz (LRL) symmetry \cite{goldstein2002classical}.
This symmetry is usually broken by relativistic corrections, and the anomalous change in its Noether charge is closely related to the precession of bound orbits~\cite{Nabet:2014kva}.
It is therefore natural to ask: does the tuning of $C(\phi)$ which removes the orbital precession introduce some additional symmetry for this scalar-tensor theory?
We are able to answer this question, at least partially.
The Newtonian symmetry and its conserved LRL vector do remain unbroken/conserved at 1PN, and this is tied to a particular soft behaviour of the amplitude. 
Our 2PM results also show how this symmetry is inevitably broken beyond leading PN order: classical orbits in these scalar-tensors always precess at a sufficiently high order in perturbation theory.  \\

Concretely, our main results in this work are to: 
\begin{itemize}

\item[(i)] write down explicit maps between the (Post-Minkowskian expansions of the) conservative two-body potential, the $2 \to 2$ scattering amplitude, the precession of bound orbits and the breaking of the Laplace-Runge-Lenz symmetry. 
This network is summarised in Figure~\ref{fig:overview}. 
Many of these connections are either well-known or have appeared recently in the literature, though here we emphasise both the connection to the LRL vector and the role played by derivative couplings. 

\item[(ii)] compute for the first time the 2PM potential and orbital precession for a scalar-tensor theory in which the scalar has the general conformal/disformal coupling to matter~\eqref{eqn:geff_def}. These agree with earlier 1PN results \cite{Brax:2018bow, Brax:2019tcy}, (and recent 3PN results for the conformal coupling only \cite{Julie:2022qux}) where the expansions overlap at low relative velocity.

\item[(iii)] explore the consequences of restoring the LRL symmetry in a relativistic theory. For the above scalar-tensor theory, this is achieved at 1PN for the special value $C (\phi ) = \sqrt{6} \phi$, at which the leading relativistic precession \emph{vanishes} as a result of an emergent LRL symmetry and corresponding soft limit for the scattering amplitude. We show that to restore the LRL symmetry beyond 1PN requires additional fields, and give $\mathcal{N} = 8$ supergravity as an example which achieves this (at least at 2PM). 
	
\end{itemize}

\noindent There is a fairly intuitive explanation for why the leading relativistic precession can vanish in scalar-tensor theories. 
Generally, due to the relativistic nature of the fields which mediate the interaction, classical orbits are no longer closed ellipses. 
From periastron to periastron, the orbit sweeps out an angle $\Theta = 2 \pi + \Delta \Theta$. 
It turns out that the \emph{sign} of $\Delta \Theta$ is tied to the spin of the underlying fields. 
In metric theories of gravity like GR, $\Delta \Theta$ is positive, but if gravity were instead mediated solely by a scalar field, then $\Delta \Theta$ would be negative. 
For a general scalar-tensor theory of gravity, the sign of $\Delta \Theta$ therefore depends on the relative strength of the coupling to the metric and to the scalar. 
In particular, there is a special tuning for which these effects exactly cancel, at least at leading post-Newtonian order. 
However, this cancellation cannot be achieved for arbitrary velocities because the precession introduced by a scalar or metric mediator will scale differently with velocity.
Consequently, beyond 1PN a scalar-tensor theory of the form \eqref{eqn:geff_def} will always lead to precessing orbits.   \\

The restoration of the LRL symmetry (no precession) at 1PN corresponds to a particular soft behaviour for the classical part of the one-loop amplitude\footnote{
In this context, the ``classical part'' corresponds to the terms that $\sim 1/\sqrt{-t}$. 
}: it must vanish in the limit $p^2 \to 0$, where $p$ is the spatial momentum in the centre-of-mass frame. 
%
In order to restore the LRL symmetry beyond 1PN, the classical part of the one-loop amplitude would have to vanish for all values of $p^2$. 
While this cannot happen for any conformal/disformal coupling to a single scalar field, one could ask whether adding additional fields might allow for such a cancellation at any order in $p^2$. 
One solution was found in \cite{Caron-Huot:2018ape}: $\mathcal{N} = 8$ supergravity adds to \eqref{eqn:geff_def} a coupling to a further vector field (the graviphoton), which provides a velocity-dependence precession that can be tuned to exactly cancel that of the metric and scalar(s) in the one-loop amplitude, and hence restore the LRL symmetry at 2PM. 
In order for the LRL symmetry to remain unbroken up to 3PM, we show here that the classical parts of the two- and three-loop amplitudes must be related in a non-trivial way.
Since the two-loop amplitude for black hole scattering in supergravity has recently been computed in \cite{Bern:2020gjj}, this immediately gives a prediction for the (classical part of the) three-loop scattering for two extremal black holes in $\mathcal{N} = 8$ supergravity. \\

We end this introduction with our conventions and a brief technical summary of our results. 
In Section~\ref{sec:overview}  we provide an overview of the general formalism which relates potentials/amplitudes/precession and the LRL symmetry. 
Then in Section~\ref{sec:ST} we use this formalism to compute our main results for the scalar-tensor theory \eqref{eqn:geff_def}, and finally in Section~\ref{sec:integrable} we compare these with some known results for (exactly integrable) supersymmetries theories. 
Finally, we conclude in Section~\ref{sec:disc} with a discussion of future directions.

\paragraph{Conventions.}
For each spatial vector $\bfp$ we denote its magnitude by $p \equiv |\bfp|$ and the corresponding unit vector $\hat{\bfp} = \bfp/p$. 
The masses of the two compact objects, $m_A$ and $m_B$, will often be written in terms of the total and reduced mass,
\begin{align}
	M &= m_A + m_B  \; , \;\; &\frac{1}{\mu} &= \frac{1}{m_A} + \frac{1}{m_B}  \; , \;\; &\nu &= \frac{m_A m_B}{M^2} = \frac{\mu}{M}
\end{align}
For relativistic motion, it will also be useful to define the shorthand,
\begin{align}
	\sigma = \frac{s - m_A^2 - m_B^2}{2 m_A m_B} = \frac{ \eta_{\mu \nu} p_1^\mu  p_2^\nu }{ m_A m_B} = 1 + \frac{p^2}{ 2 \mu^2} + (4 \nu -1 ) \frac{ p^4 }{\mu^4} + ( 8 \nu^2 - 6 \nu + 1 ) \frac{p^6}{16 \mu^6} + ... 
\end{align}
We use the mostly minus convention for the metric tensor.
Finally, $G_N = 1/(8 \pi M_P^2)$ relates Newton's constant to the Planck mass, $\kappa = G_N m_A m_B$ is the dimensionless coupling that controls our PM expansion, and we work in natural units where the speed of light $c=1$ throughout. 

%
%
%
%

\subsection{Summary of main results}
\label{sec:summary}

After briefly reviewing how unitarity and the optical theorem may be used to compute the classical two-body potential, via the $t$-channel discontinuity of the $2 \to 2$ scattering amplitude, we turn to the question of how to extract the orbital precession. 

\paragraph{Orbital invariants from the LRL vector.}
We take a somewhat novel approach in which the Newtonian Laplace-Runge-Lenz vector $\bfK[r,p]$, together with the Hamiltonian $H[r,p]$ and angular momentum $\bfL [r, p]$, is used to construct the on-shell solutions for $r (t)$ and $p (t)$ by iteratively solving the equations,
\begin{align}
 H [ r, p ] &= E \; , \;\; &\bfL [r, p] &= \bfL  \;, \;\; &\hat{\bfr} \cdot \hat{\bfK} [r, p] &=  \cos \th (t)
\end{align}
where $E$ and $\bfL$ are fixed constants and $\th (t)$ is the one dynamical degree of freedom left in the problem. 
The magnitude of the LRL vector,
\begin{align}
 | \bfK [r, p] | = K ( E, L , \th )
\end{align}
then acts as a generating function for the periastron-to-periastron precession and period of the orbit,
\begin{align}
\Delta \Theta &=  \oint d \th \, \frac{ \cos \th }{K} \frac{\partial K}{\partial \cos \th} \; ,   &T &= L \oint \frac{d\th}{K} \left[ \frac{\kappa^2}{r^2} \frac{\partial K}{\partial E} \right]^{-1} 
\end{align}
where $\kappa = G_N m_A m_B$ is the usual gravitational coupling constant. 
This makes precise the connection between closed orbits (i.e. $\Delta \Theta = 0$) and the LRL symmetry, of which $\bfK$ is the conserved charge (i.e. $\partial_\th K = 0$ if the symmetry is unbroken).
Using this formalism, we re-derive the connection between $\Delta \Theta$ and the scattering amplitude up to 2PM, and extend these to 3PM.
We also pay particular attention to \emph{derivative} couplings between matter and the force-mediating fields, and perform a consistent power counting which allows for these couplings to be suppressed by a scale $M_{\partial} \ll M_P$.

\begin{figure}[htpb!]
\centering
	\begin{tikzpicture}[scale=0.6,baseline=0.6cm]
	\begin{feynman}
		
		\vertex (mLU) at (0.3,0);
		\vertex (mLD) at (-0.3,0);
		\vertex (mRU) at (0.3,2);
		\vertex (mRD) at (-0.3,2);
		
		\vertex (mU) at (0,0);
		\vertex (mD) at (0,2);
		
		\node at (0.9,1) {$ g_{\mu\nu}$};
		\node at (-0.8,1) {};
		
		\vertex (a) at (-2,-1) {$B$};
		\vertex (b) at ( -2, 3) {$A$};
		\vertex (c) at (2,-1) {$B$};
		\vertex (d) at ( 2, 3) {$A$};
		\diagram* {
			(a) -- [fermion] (mLD),
			(mU) -- [graviton] (mD),
			(mLU) -- [fermion] (c), 
			(b) -- [fermion] (mRD), 
			(mRU)-- [fermion] (d),
		};
		
		\node [thick, draw=black, fill=black!10, ellipse, minimum width=0.75cm,minimum height=0.25cm,align=center] at (0,0);
		\node [thick, draw=black, fill=black!10, ellipse, minimum width=0.75cm,minimum height=0.25cm,align=center] at (0,2);
				
	\end{feynman}
\end{tikzpicture} $+$ 
	\begin{tikzpicture}[scale=0.6,baseline=0.6cm]
	\begin{feynman}
		
		\vertex (mLU) at (0.3,0);
		\vertex (mLD) at (-0.3,0);
		\vertex (mRU) at (0.3,2);
		\vertex (mRD) at (-0.3,2);
		
		\vertex (mU) at (0,0);
		\vertex (mD) at (0,2);
		
		\node at (0.8,1) {$\phi$};
		\node at (-0.8,1) {};
		
		\vertex (a) at (-2,-1) {$B$};
		\vertex (b) at ( -2, 3) {$A$};
		\vertex (c) at (2,-1) {$B$};
		\vertex (d) at ( 2, 3) {$A$};
		\diagram* {
			(a) -- [fermion] (mLD),
			(mU) -- [scalar] (mD),
			(mLU) -- [fermion] (c), 
			(b) -- [fermion] (mRD), 
			(mRU)-- [fermion] (d),
		};
		
		\node [thick, draw=black, fill=black!10, ellipse, minimum width=0.75cm,minimum height=0.25cm,align=center] at (0,0);
		\node [thick, draw=black, fill=black!10, ellipse, minimum width=0.75cm,minimum height=0.25cm,align=center] at (0,2);
				
	\end{feynman}
\end{tikzpicture}
$ = \frac{  4 G_N m_A^2 m_B^2 }{ r } \left[  2 \sigma^2 - 1 +  \alpha^2  \right]  $  
\\[15pt]
	\begin{tikzpicture}[scale=0.6,baseline=0.6cm]
	\begin{feynman}
		
		\vertex (mLU) at (0.3,0);
		\vertex (mLD) at (-0.3,0);
		\vertex (mRU) at (0.3,2);
		\vertex (mRD) at (-0.3,2);
		
		\vertex (mU) at (0.3,1);
		\vertex (mD) at (-0.3,1);
		
		\node at (0.8,1) {};
		\node at (-0.8,1) {};
		
		\vertex (a) at (-2,-1) {$B$};
		\vertex (b) at ( -2, 3) {$A$};
		\vertex (c) at (2,-1) {$B$};
		\vertex (d) at ( 2, 3) {$A$};
		\diagram* {
			(a) -- [fermion] (mLD) -- [graviton] (mRD),
			(mRU) -- [graviton] (mLU) -- [fermion] (c), 
			(b) -- [fermion] (mRD), 
			(mRU)-- [fermion] (d),
		};
		
		\node [thick, draw=black, fill=black!10, ellipse, minimum width=0.75cm,minimum height=0.25cm,align=center] at (0,0);
		\node [thick, draw=black, fill=black!10, ellipse, minimum width=0.75cm,minimum height=0.25cm,align=center] at (0,2);
				
	\end{feynman}
\end{tikzpicture} $+$ 
	\begin{tikzpicture}[scale=0.6,baseline=0.6cm]
	\begin{feynman}
		
		\vertex (mLU) at (0.3,0);
		\vertex (mLD) at (-0.3,0);
		\vertex (mRU) at (0.3,2);
		\vertex (mRD) at (-0.3,2);
		
		\vertex (mU) at (0.3,1);
		\vertex (mD) at (-0.3,1);
		
		\node at (0.8,1) {};
		\node at (-0.8,1) {};
		
		\vertex (a) at (-2,-1) {$B$};
		\vertex (b) at ( -2, 3) {$A$};
		\vertex (c) at (2,-1) {$B$};
		\vertex (d) at ( 2, 3) {$A$};
		\diagram* {
			(a) -- [fermion] (mLD) -- [graviton] (mRD),
			(mRU) -- [scalar] (mLU) -- [fermion] (c), 
			(b) -- [fermion] (mRD), 
			(mRU)-- [fermion] (d),
		};
		
		\node [thick, draw=black, fill=black!10, ellipse, minimum width=0.75cm,minimum height=0.25cm,align=center] at (0,0);
		\node [thick, draw=black, fill=black!10, ellipse, minimum width=0.75cm,minimum height=0.25cm,align=center] at (0,2);
				
	\end{feynman}
\end{tikzpicture}
$+$ 
	\begin{tikzpicture}[scale=0.6,baseline=+0.6cm]
	\begin{feynman}
		
		\vertex (mLU) at (0.3,0);
		\vertex (mLD) at (-0.3,0);
		\vertex (mRU) at (0.3,2);
		\vertex (mRD) at (-0.3,2);
		
		\vertex (mU) at (0.3,1);
		\vertex (mD) at (-0.3,1);
		
		\node at (0.8,1) {};
		\node at (-0.8,1) {};
		
		\vertex (a) at (-2,-1) {$B$};
		\vertex (b) at ( -2, 3) {$A$};
		\vertex (c) at (2,-1) {$B$};
		\vertex (d) at ( 2, 3) {$A$};
		\diagram* {
			(a) -- [fermion] (mLD) -- [scalar] (mRD),
			(mRU) -- [scalar] (mLU) -- [fermion] (c), 
			(b) -- [fermion] (mRD), 
			(mRU)-- [fermion] (d),
		};
		
		\node [thick, draw=black, fill=black!10, ellipse, minimum width=0.75cm,minimum height=0.25cm,align=center] at (0,0);
		\node [thick, draw=black, fill=black!10, ellipse, minimum width=0.75cm,minimum height=0.25cm,align=center] at (0,2);
				
	\end{feynman}
\end{tikzpicture}
$ = \frac{ G_N^2 m_A^3 m_B^3 }{ r^2 } \left[  3 ( 5 \sigma^2 - 1 ) + 8 \alpha^2 - 4 \alpha^4  + \lambda  \alpha^2  \frac{  \sigma^2 - 1 }{r^2 M_{\partial}^2}  \right]  $ 
\caption{The tree- and one-loop $2\to 2$ scattering amplitudes for a pair of spin-less compact objects coupled to the effective metric~\eqref{eqn:geff_def}, where $8 \pi G_N = 1/M_P^2$ and the grey blob represents all tree-level subdiagrams.}
\label{fig:amps}
\end{figure}
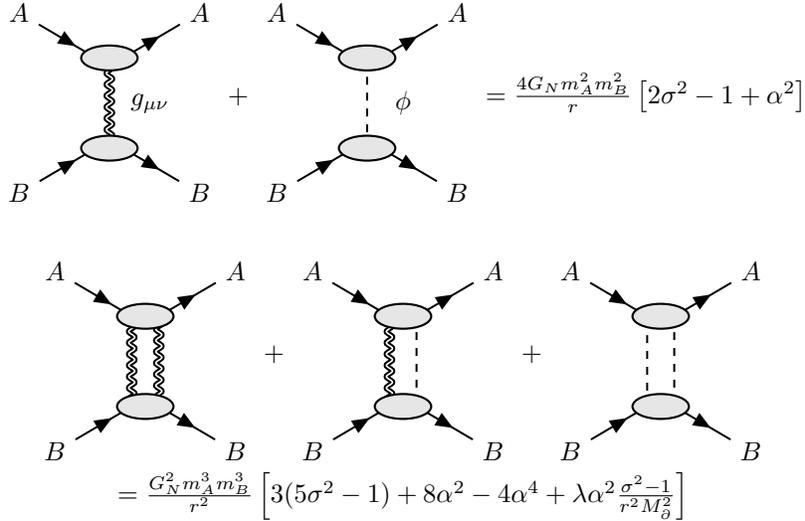

\paragraph{Scalar-tensor precession.}
As an illustrative example, we consider the scalar-tensor theory in which matter couples to the effective metric~\eqref{eqn:geff_def}.
We focus in the main text on a linear $C (\phi) =  \sqrt{2}  \alpha \phi $ and a constant $D ( \phi ) = \lambda$, and study more general couplings in Appendix~\ref{app:ST}.
The conservative part of the classical potential at large distances takes the form, 
\begin{align}
 V ( p^2 , r ) &= \frac{\kappa}{r} \left( V^{(1)}_{\rm GR} (p^2) + \alpha^2 V^{(1)}_{\alpha^2} (p^2)   \right)   \nonumber \\
 & +  \frac{\kappa^2}{r^2} \left(  V^{(2)}_{\rm GR} (p^2) + \alpha^2 V^{(2)}_{\alpha^2} (p^2)   + \alpha^4 V^{(2)}_{\alpha^4} (p^2)    \right) + \frac{\kappa^2}{ r^4 M_{\partial}^2} \, \lambda \alpha^2 \, V_{\lambda \alpha^2}^{(2)} (p^2)  
 \label{eqn:intro_V}
\end{align}
plus terms which are higher order in $\kappa/r$.
We determine this potential by computing the scattering amplitudes shown in Figure~\ref{fig:amps}. The resulting potential coefficients are given in \eqref{eqn:v2_GR} for GR, (\ref{eqn:v1_con}, \ref{eqn:v2_con}) for the conformal coupling, and \eqref{eqn:v2_dis} for the disformal coupling, and all agree with the currently known PN expansions when expanded at small $p^2$.

In the Newtonian limit, orbits in this theory are Keplerian ellipses determined by an effective $\tilde{\kappa}  = \kappa ( 1 + \alpha^2 )$, i.e. with a semi-latus rectum $\mu \tilde{r}_c = L^2/ \tilde{\kappa}$.
Beyond the Newtonian limit, the corrections to $V( p^2, r)$ lead to a precession of this orbit of the form, 
\begin{align}
\frac{  \Delta \Theta }{2\pi} = \frac{\kappa^2}{L^2} \left[ \theta_{\rm GR} (\E)  
 + \theta_{\rm con} (\E)  + \frac{\mu^2}{ M_{\partial}^2}  \frac{1}{ L^2 } \theta_{\rm dis} (\E)
 \right] + \mathcal{O} \left(  \frac{\kappa^4}{L^4} \right)
 \label{eqn:intro_pre}
\end{align}
where the usual GR precession is quoted in~\eqref{eqn:pre_GR}, and we determine the conformal and disformal additions to be,
\begin{align}
\theta_{\rm con} (\E) &= \frac{2 \alpha^2 - \alpha^4}{1 + \nu \E}  \; , 
& \theta_{\rm dis} ( \E ) &=  - \lambda \alpha^2 \;  \frac{ \E^2  (2 + \nu \E )^2 (4 + 2 \E  + \nu \E^2  )^2 }{ 128 (1 + \nu \E )^3 } \;  
 \label{eqn:pre_dis_PM}
\end{align}
where $E = M + \mu \E$ separates the energy into a total rest mass and a relative part. 
This result captures all orders in $\mathcal{E}$ (i.e. $v^2/c^2$), at a fixed order in $\kappa^2/L^2$ (i.e. $G_N$). 
Due to the derivative nature of the disformal coupling, the PN expansion of this PM result must be done carefully. In particular, since $\theta_{\lambda \alpha} (\E) \sim \E^2$ at small velocities, it is the same order as higher PM corrections which $\sim \frac{\kappa^4}{L^4} \E$ and $\sim \frac{\kappa^6}{L^6}$. 
Carefully extracting the leading PN disformal precession from the potential $V_{\lambda \alpha^2}^{(2)}$, we find,
\begin{align}
\frac{  \Delta \Theta }{2 \pi}  = \frac{\kappa^2}{L^2} ( 1 + \alpha^2 ) (3 - \alpha^2 ) +  \frac{\kappa^2}{\mu^2 \tilde{r}_c^2 }  \;  \frac{ 5 \, \lambda \alpha^2  }{   \tilde{r}_c^2 \,  M_{\partial}^2 }   + ... 
\end{align}
in complete agreement with the earlier 1PN result of \cite{Brax:2018bow}.

\paragraph{Hidden symmetries.}
We point out the curious feature that the leading PN precession vanishes for the special value $\alpha^2 = 3$. 
That $\Delta \Theta$ can be set to zero by tuning $\alpha$ is by no means trivial, since (as demonstrated by the potential $V^{(1)} \propto (1 + \alpha^2)$) the dependence on coupling constants can be monotonic: often stronger couplings simply lead to more precession. 
But since the scalar and the metric contributions pull the orbit in opposite directions, there is a value of $\alpha^2$ for which they precisely cancel. 
Thanks to our new relations between the precession, the amplitude and the LRL vector, we interpret this cancellation as the restoration of the LRL symmetry at this first PN order, which corresponds to the one-loop scattering amplitude having the vanishing soft limit,
\begin{align}
  \lim_{ \substack{  p^2 \to 0 } }  \text{Disc}_t \,\mathcal{A}^{\text{1-loop}}_{A B \to A B}  = 0 \; 
  \label{eqn:soft}
\end{align}  
at small positive momentum transfer $t$, the region which dominates the classical potential. 
However, at higher PN orders (finite values of $p^2$), our amplitude result shows that the LRL symmetry is inevitably broken in any scalar-tensor theory of this kind.

\paragraph{LRL bootstrap.}
Finally, we find that \eqref{eqn:soft} is the first in an infinite tower of relations which the scattering amplitude must satisfy as the result of an unbroken LRL symmetry.
We therefore propose an \emph{LRL bootstrap} for integrable theories in which classical orbits close: for such theories, the classical part\footnote{
Below we define the ``classical part'' of an amplitude to be the part which scales with the appropriate power of $t$ to enter the classical potential. 
} of every odd-loop $2 \to 2$ amplitude is uniquely fixed in terms of the even-loop $2 \to 2$ amplitudes. 
At present, the only known theories that have this symmetry beyond 1PN are $\mathcal{N} =4$ super-Yang-Mills \cite{Caron-Huot:2014gia, Alvarez-Jimenez:2018lff} and $\mathcal{N} = 8$ supergravity \cite{Caron-Huot:2018ape, deNeeling:2023egt}. 
Since the two-loop supergravity amplitude for black hole scattering was recently computed \cite{Bern:2020gjj}, our condition that for vanishing orbital precession (i.e. the Ward identity for the LRL symmetry) immediately fixes the relevant part of the three-loop amplitude for describing classical orbits to be~\eqref{eqn:LRL_boot_a4}.

\section{Overview of general formalism}
\label{sec:overview}

We begin by outlining general relations which can be applied to any two-body system, regardless of the underlying field content. 
These results therefore apply not only to gravitationally bound binary systems, but also to electromagnetic systems or to modified gravity theories. 

We focus on effective field theories of the general form\footnote{
For electromagnetic interactions, one would include a coupling between the compact objects and an effective $\tilde{A}_{A, B}^{\mu}$  with the same power counting, i.e. force-mediating fields come suppressed by $M_P$, derivatives come suppressed by $M_{\partial}$. 
},
\begin{align}
 S = \int d^4 x \; \left\{
  M_P^2 M_{\partial}^2 \; \mathcal{L} 
 + \mathcal{L}_A \left[ \tilde{g}_A^{\mu\nu} , \chi_A \right] + \mathcal{L} \left[ \tilde{g}_B^{\mu\nu} , \chi_B \right]
 \right\}   
 \label{eqn:EFT_form}
\end{align}
where the field Lagrangian $\mathcal{L}$ and the effective metrics $\tilde{g}^{\mu\nu}_{A,B}$ depend on the massless fields that mediate long-range forces (e.g. the metric fluctuations $h_{\mu\nu}$, a dark scalar $\phi$, ...) and their derivatives through the dimensionless ratios $\{  \frac{ h_{\mu\nu}}{M_P} ,  \frac{\phi}{M_P} , ... , \frac{\partial_\mu}{M_{\partial}}  \}$,  i.e. we adopt a power-counting in which field insertions are suppressed by the scale $M_P$ and derivatives are suppressed by $M_{\partial}$.
The degrees of freedom of the two compact objects are represented by two auxilliary fields $\chi_A$ and $\chi_B$. 
For instance, for a spin-less compact object, 
\begin{align}
\mathcal{L}_A \left[ \tilde{g}_A^{\mu\nu} , \chi_A \right] = \frac{1}{2} \sqrt{- \tilde{g}_A } \left[  \tilde{g}_A^{\mu\nu} \partial_\mu \chi_A \partial_\nu \chi_A   -  m_A^2 \chi_A^2 \right] \; ,
\label{eqn:LA_form}
\end{align}
where $\tilde{g}^{\mu\nu}_A = \eta^{\mu\nu} + \mathcal{O} \left( \frac{h_{\mu\nu}}{M_P} , \frac{\phi}{M_P}, ... , \frac{\partial_\mu}{M_{\partial}} \right)$ so that the inertial motion of special relativity is obtained in the decoupling limit $M_P \to \infty$.

By now there are many ways of efficiently computing, from a field theory action, the resulting orbits in binary systems. 
We adopt the approach of \cite{Iwasaki:1971vb, Feinberg:1988yw}, in which the classical potential between two point particles can be extracted from a quantum mechanical scattering amplitude. 
In particular, a loop expansion of the amplitude corresponds roughly to a $1/r$ expansion of the potential.

\begin{figure}[t]
	\begin{tikzcd}[column sep=3.5em, row sep=2em]
		& 		\fbox{\Centerstack[c]{Amplitude discontinuity \\ $\Disc \, \mathcal{A} \sim \sum_n a_n (s) ( \sqrt{t} )^{n-d}$ }}  \arrow{ddr}[sloped, anchor=south]{ \eqref{eqn:pre_from_amp} } \arrow{ddl}[sloped, anchor = south]{\eqref{eqn:LRL_from_amp}}  \arrow{d}[anchor=west]{\eqref{eqn:amp_to_pot}} \\[1.5em]
   & \fbox{\Centerstack[c]{Two-body potential \\ $V \propto  \sum_n v_n (p^2) r^{-n} $ }} \arrow{u} \arrow{dl} \arrow{dr} &  \\[-10pt] 
		\fbox{\Centerstack[c]{LRL symmetry breaking \\ $\frac{d}{dt} \left| \bfK \right| =  \sum_n k_n (E) r^{-n} $ \\ $ | \frac{d}{dt} \hat{\bfK} | =  \sum_n \hat{k}_n (E) r^{-n} $  }} 
	\arrow{uur}	\arrow{ur}[sloped, anchor=south]{\eqref{eqn:LRL_from_pot}} 
  \arrow[rr,"\eqref{eqn:pre_from_LRL}"] 
  && 
				 \fbox{\Centerstack[c]{Orbital precession  \\ $\frac{\Theta}{2\pi} = 1 + \sum_n \theta_n (E) L^{-2n}$ \\ $\frac{T}{T_0} = 1 + \sum_n t_n (E) L^{-2n} $ }} \arrow[ll] \arrow{ul}[sloped, anchor=south]{\eqref{eqn:pre_from_pot}} \arrow{uul}
	\end{tikzcd}
	\caption{There is a one-to-one correspondence between: the conservative part of the two-body potential, the classical part of the $2 \to 2$ scattering amplitude's $t$-channel discontinuity, the relativistic precession and dilation of bound orbit, and the breaking (/anomalous time dependence) of the Laplace-Runge-Lenz symmetry (/vector).  Explicit maps between the Post-Minkowski coefficients of these various quantities are given in the equations indicated. 
 }
	\label{fig:overview}
\end{figure}
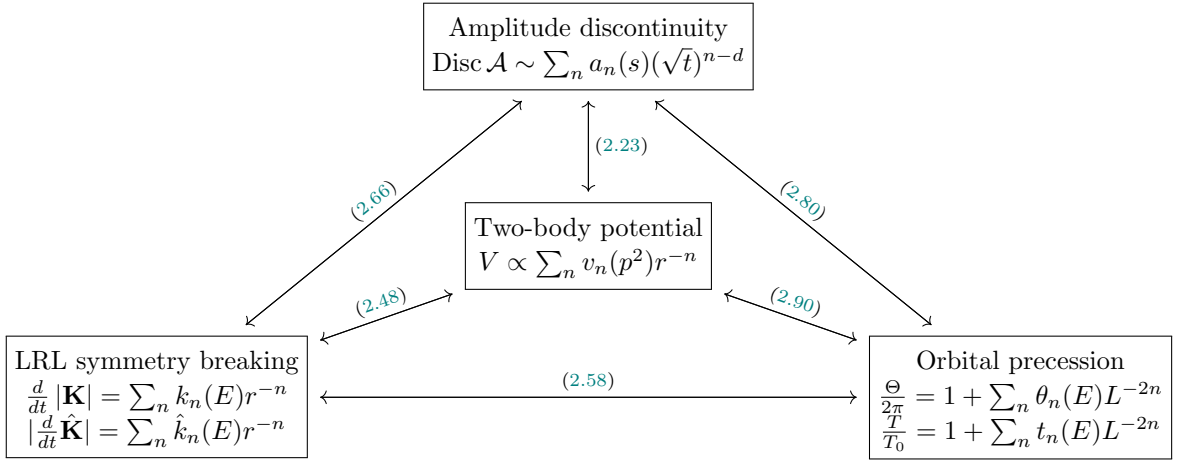

Concretely, our goal in this section will be to outline the network of relations shown in Figure~\ref{fig:overview}, establishing simple formulae which extract observables like the orbital precession directly from the Hamiltonian or scattering amplitude and which will be useful in next section when we specialise to scalar-tensor theories.
Most of these relations are not novel---they have appeared in various forms throughout many recent works~\cite{Cheung:2018wkq, Bern:2019crd, Cristofoli:2019neg, Kalin:2019rwq, Kalin:2019inp}. 
However, we emphasis the role played by the Laplace-Runge-Lenz vector and this perspective connects this hidden symmetry of the theory to the orbital precession / amplitude in a useful way. 
We also allow for a scale hierarchy $M_{\partial} \ll M_P$ for the derivative interactions, which breaks the usual $1/r^n$ scaling at $n^{\rm th}$ PM order.

Finally, note that since we are going to work perturbatively in both $M_P$ and $M_{\partial}$, we do not consider screening mechanisms that exploit a resummation of either derivative or field insertions. See \cite{Carrillo-Gonzalez:2021mqj} for a discussion of how to incorporate such screening effects into the amplitude/Post-Minkowskian amplitude framework.

\subsection{Two-body Hamiltonian and the interaction potential}

The general two-body problem consists of solving for the dynamical evolution of a position and momentum ($\bfx_A (t)$ and $\bfp_A (t)$) for each particle.
However, Poincar\'{e} invariance implies the conservation of the total energy $H$, the total spatial momentum $\bfP$, the total angular momentum $\bfL$ and the centre-of-mass co-ordinate $\bfX - \bfP t$.
We can therefore work in the centre-of-mass frame, setting $\bfP = 0$ and $\bfX = 0$, which leads to a \emph{reduced} two-body problem in which the only degrees of freedom are the relative separation $\bfr = \bfx_1 - \bfx_2$ and its conjugate momentum $\bfp$. 
Our first task is to describe the form of the reduced two-body Hamiltonian which arises from \eqref{eqn:EFT_form}.

\paragraph{Non-derivative couplings.}
Let us begin by considering a \emph{non-derivative} coupling between the compact objects and the force-mediating fields (i.e. take $M_{\partial} \to \infty$ to decouple any derivative interactions). 
The reduced two-body Hamiltonian can then be written as a perturbative expansion in powers of $1/M_P$, the weak coupling that suppresses the fields. 
It is convenient to define the dimensionless coupling,
\begin{align}
 \kappa \equiv G_N m_A m_B \equiv \frac{m_A m_B}{8 \pi M_P^2}
 \label{eqn:kappa_def}
\end{align}
and write the Hamiltonian as,
\begin{align}
   H [ r , p ] = \mu \sum_{n=0} H_n \left( \frac{p^2}{2 \mu^2} ,  \frac{p_r^2}{2 \mu^2} \right) \left( \frac{\kappa}{\mu r} \right)^n  
    =  \mu \sum_{ \substack{ n = 0 \\ a,b = 0 } } H_{n}^{(a,b)} \left(  \frac{p^2}{2 \mu^2} \right)^a \left( \frac{p_r^2}{2 \mu^2} \right)^b \left( \frac{\kappa}{\mu r} \right)^n
    \label{eqn:H_PM}
\end{align}
where each $H_n$ is a function of the momentum $p$, the radial momentum $p_r = (\bfr \cdot \bfp)/r$ and the mass ratio $\nu$, while each $H_n^{(a,b)}$ is a function of the mass ratio only.
$n$ is the Post-Minkowskian (PM) order of each term, and $a+b+n$ is its Post-Newtonian (PN) order.

Canonical transformations---redefinitions of $r$ and $p$ that preserve their canonical Poisson bracket---can be used to simplify this Hamiltonian \cite{Hiida:1972xs}. This is the analogue of performing field redefinitions and removing total derivatives in the Lagrangian picture. 
We review this explicitly in Appendix~\ref{app:canon}, and will often refer to performing a canonical transform as ``changing gauge''.

\paragraph{Isotropic gauges.}
One particularly convenient family of gauges are the so-called ``isotropic gauges'', in which a canonical transformation is used to remove the $p_r$ dependence from the Hamiltonian, leaving an expansion of the form,
\begin{align}
 H [ r , p ]  = \mu \sum_{n=0} h_n \left(  \frac{p^2}{ 2 \mu^2 } \right)  \left( \frac{\kappa}{\mu r} \right)^n \; .
 \label{eqn:isotropic_gauge}
\end{align}
One particular advantage of these gauges is that it is straightforward to re-organise the expansion using the Newtonian Hamiltonian $H_N = \frac{p^2}{2\mu^2} - \frac{\kappa}{\mu r}$, 
\begin{align}
 H [ r , p ]  = \mu \sum_{n=0} \tilde{h}_n \left( H_N \right)  \left(  \frac{\kappa}{\mu r} \right)^n \; ,
\end{align}
where the expansion coefficients are related by, 
\begin{align}
 \tilde{h}_0 &= h_0 \; , \;\; &\tilde{h}_1 &= h_1 + h_0'  \; , \;\; &\tilde{h}_2 &= h_2 + h_1' + \frac{1}{2} h_0'' 
\end{align}
and so on. 
This re-organisation is particularly useful because the LRL vector (defined in \eqref{eqn:K_def} below and reviewed in Appendix~\ref{app:LRL}) commutes with $H_N$ and therefore commutes with each $\tilde{h}_n$ coefficient. 

Removing the $p_r$ dependence does not completely specify a gauge: there is a residual freedom to fix one of the $h_n$ or $\tilde{h}_n$ coefficients. 
There are two particular isotropic gauges we will make use of in this work, which we refer to as the \emph{amplitude} and the \emph{LRL} gauges.

\paragraph{Amplitude gauge.}
The amplitude gauge is specified by fixing $h_0$ so that it coincides with the relativistic motion of two free particles. 
As a result, the Hamiltonian in this gauge reads,
\begin{align}
	H [ r , p ] = \sqrt{p^2 + m_A^2} + \sqrt{ p^2 + m_B^2} + \mu V ( p^2, r )
	\label{eqn:amp_gauge}
\end{align}
where the \emph{two-body potential} can be expanded,
\begin{align}
	V ( p^2, r ) = \sum_{n=1} v_n \left( \frac{p^2}{2 \mu^2} \right) \left( \frac{\kappa}{\mu r} \right)^n  
	=  \sum_{\substack{n=1 \\ a = 0}} v_{n}^{(a)} \left(  \frac{p^2}{2 \mu^2} \right)^a  \left( \frac{\kappa}{\mu r} \right)^n
	\label{eqn:V_PM}
\end{align}
The PM functions $v_n$ and their PN coefficients $v_n^{(a)}$ are given in terms of the $H_n$ in Appendix~\ref{app:canon}. 
Note that fixing $h_0$ in this way has completely specified the gauge and as a result these particular combinations of the $C_n$ are invariant under canonical transformations\footnote{
Though note that \eqref{eqn:amp_gauge} is not invariant under a gauge transform. When we say $v_n$ are gauge-invariant, this is analogous to saying that the centre-of-mass energy $s$ is Lorentz-invariant: in any Lorentz frame, one can always compute the total energy the particles would have in their centre-of-mass frame, and this is an invariant notion which all observers agree upon. Similarly, given a general $H$, one can always compute the potential which would be felt in the amplitude gauge: this is likewise a gauge-invariant notion.  
}.

This gauge is most convenient for matching to the underlying field theory, since there is a close connection between the Post-Minkowskian expansion of $V$ and a loop expansion of the $2 \to 2$ scattering amplitude \cite{Cheung:2018wkq}.

\paragraph{LRL gauge.}
Another way to fix the residual freedom of the isotropic gauge is to impose that $\tilde{h}_1 = 0$. 
This produces a Hamiltonian,
\begin{align}
\frac{ H [ r , p ] }{\mu}  = \tilde{b}_0 \left( H_N  \right) + \sum_{n=2} \tilde{b}_n \left( H_N \right)  \left(  \frac{\kappa}{\mu r} \right)^n 
\label{eqn:LRL_gauge}
\end{align}
where the first few $\tilde{b}_n$ coefficients are related to the original $H_n$ in Appendix~\ref{app:canon}. These particular combinations are again invariant under canonical transformations. 

This gauge is useful for analysing the precession of orbits, which is sensitive only to $\tilde{b}_{n \geq 2}$.
In particular, closed orbits (no precession) occur if and only if all $\tilde{b}_{n \geq 2} = 0$, and in that case the LRL vector of the Newtonian limit becomes a conserved quantity in the full theory.

\paragraph{Derivative coupling.}
When we include a coupling that involves derivatives, we introduce a new scale $M_{\partial}$ which may be $\ll M_P$. 
We can therefore refine the above expansions so that $1/M_P$ and $1/M_{\partial}$ are expanded separately.
For instance, we can express the two-body potential in the amplitude gauge as,
\begin{align}
 V ( p^2 , r ) = \sum_{\substack{n=1 \\ n'=0}} v_{n,n'}  \left( \frac{p^2}{2 \mu^2} \right) \left( \frac{\kappa}{\mu r} \right)^n \left( \frac{1}{ r M_{\partial} } \right)^{n'}  \; .
 \label{eqn:V_PM_der}
\end{align}
In this case, we will still refer to the $\mathcal{O} ( \kappa^n )$ part of the potential as the $n^{\rm th}$ PM coefficient, but note that this need no longer coincide with a $1/r^n$ scaling at large distances. 
Some general features of this expansion are:
\begin{itemize}

\item[(i)] $v_{n, n'} = 0$ whenever $n'$ is odd, at least for the bosonic interactions considered here, since derivatives always come contracted in pairs (so we can never have an odd power of $M_{\partial}$),

\item[(ii)] for interactions with $D$ derivatives per field, $v_{n, n'} = 0$ whenever $n' > D n$ (since then the interactions contain only the combinations $M_P$ and $M_P M_{\partial}^D$),

\item[(iii)] $v_{1,n'} = 0$ for all $n' \neq 0$, since derivative interactions cannot modify a $1/r$ potential on dimensional grounds (this is also transparent in the amplitude construction given below, in which the $t$-channel pole is responsible for the $1/r$ part of the potential and this is simply removed whenever $\partial^2 \sim t$ is inserted).

\end{itemize}

\subsection{Scattering amplitude and $t$-channel discontinuity}

We will now describe how to determine the $v_n$ coefficients appearing in the two-body potential from the underlying field theory that mediates the interaction between the two bodies. 
There are many approaches to this problem.
We choose to extract the classical $H$ from the quantum mechanical scattering amplitude, since the amplitude has the advantage that:
\begin{itemize}
    \item[(i)] it is insensitive to redundancies like canonical/gauge transformations,
    \item[(ii)] it is straightforward to include additional fields or effective field theory interactions,
    \item[(iii)] it provides an explicit connection with symmetries (Ward identities) and soft theorems. 
\end{itemize}
The drawback is that the quantum amplitude contains far more information than we need for the classical problem, and extracting the relevant classical part can be non-trivial. 

Various maps between the amplitude and the classical Hamiltonian have been developed recently: for instance based on EFT matching \cite{Cheung:2018wkq} or the Born approximation in old-fashioned perturbation theory \cite{Iwasaki:1971vb, Cristofoli:2019neg}. 
Here we follow the traditional route of old-fashioned perturbation theory, which makes use of the Lippmann-Schwinger equation.
This is described in classic textbooks \cite{Sakurai:2011zz, Weinberg:1995mt}, and we provide a pedagogical review in Appendix~\ref{app:amp}.
In the main text we will simply quote relevant formulae.

\paragraph{Scattering amplitude.}
By Lorentz-invariance, the relativistic $\bfp_1 \bfp_2 \to \bfp_3 \bfp_4$ scattering amplitude $\mathcal{A}$ is a function of the two Mandelstam variables $s =  (p_1^\mu + p_2^\mu )^2$ and $t= (p_1^\mu - p_3^\mu )^2$ only,
\begin{align}
 \mathcal{A} ( \bfp_1, \bfp_2 ; \bfp_3 ,\bfp_4 ) = \mathcal{A} (s, t) \; . 
\end{align} 
In the centre-of-mass frame, the momenta are given by,
\begin{align}
	p_1^\mu &= ( E_A (p^2) , \bfp ) \; , \;\;
	&p_2^\mu &= ( E_B ( p^2 ) , -\bfp ) \; ,  
	&p_3^\mu &= ( E_A ( p'^2) , \bfp' ) \; , \;\; &p_4^\mu &= ( E_B (p'^2) , -\bfp' ) \; .
	\label{eqn:p_CoM}
\end{align}
where $E_A (p^2) = \sqrt{p^2 + m_A^2}$ and energy conservation therefore imposes $p^2 = p'^2$.
For comparison with the classical potential, it is convenient to normalise the scattering states with factors of $1/\sqrt{2 E_{A,B}}$, which explicitly break Lorentz-invariance.
In the centre-of-mass frame, this gives an amplitude,
\begin{align}
 A ( p^2 , q^2 ) = \frac{ \mathcal{A} ( s (p^2) ,  t(q^2) ) }{4 E_A (p^2) E_B (p^2) }
\end{align}
where $\bfq = \bfp - \bfp'$ is the momentum transfer, and the Mandelstam variables are now $s (p^2 ) =\left( E_A (p^2) + E_B (p^2) \right)^2$ and $t (q^2 ) = -q^2$. 
So once the usual perturbative expansion in Feynman diagrams has been used to compute $\mathcal{A} (s,t)$ from a given Lagrangian, the function $A(p^2, q^2)$ corresponds to evaluating it at the centre-of-mass kinematics \eqref{eqn:p_CoM} and dividing by an overall factor of $4 E_A E_B$.

\paragraph{Lippmann-Schwinger equation.}
The two-body potential may also be written in terms of the momentum transfer $q$ via a Fourier transform, 
\begin{align}
 V (   p^2 , q^2 ) = \int \d^3 \bfr \, e^{i \bfq \cdot \bfr} V ( p^2, r ) \; .
\end{align}
The functions $A (p^2, q^2)$ and $V (p^2, q^2)$ are related by the celebrated Lippmann-Schwinger equation,
\begin{align}
 A (p^2 , q^2 ) =  - V (p^2 , q^2) + \int \frac{d^d \bfk}{(2\pi)^d} \frac{ A (p^2, | \bfp - \bfk |^2 ) V ( k^2 , | \bfk - \bfp' |^2 ) }{ E_A (p^2) + E_B (p^2 ) - E_A (k^2) - E_B (k^2) + i \epsilon } + ... 
 \label{eqn:main_LS}
\end{align}
where the $+...$ are terms which capture operator ordering ambiguities and particle production (both of which can be neglected when considering the classical conservative potential).
We provide a careful derivation of \eqref{eqn:main_LS}, together with a definition of $A$ and $V$ in terms of quantum-mechanical matrix elements, in Appendix~\ref{app:amp}.

\paragraph{Discontinuity and the PM expansion.}
To perform the Post-Minkowskian expansion of \eqref{eqn:main_LS}, we require the Fourier transform of the $(\kappa/r)^n$ series \eqref{eqn:V_PM}.
This could be evaluated by brute force (e.g. using the integral identities given in Appendix~\ref{app:int_id} produces \eqref{eqn:V_PM_2}), but
a particularly efficient method proposed by Feinberg and Sucher \cite{Feinberg:1988yw} is to instead consider the \emph{$t$-channel discontinuity} of the potential/amplitude, defined by,
\begin{align}
 \text{Disc}_t \, \mathcal{A} (s,t) \equiv \lim_{\epsilon \to 0 }  \left( \mathcal{A} (s , t + i \epsilon ) - \mathcal{A} (s, t - i \epsilon )   \right) \; .
 \label{eqn:Disc_def}
\end{align}
Exploiting analyticity of the amplitude in the complex $t$ plane, \cite{Feinberg:1988yw} showed that,
\begin{align}
  \int \frac{d^d \bfq}{(2\pi)^3} e^{-i \bfq \cdot \bfr } \mathcal{A} (s, -q^2) = \int_0^{\infty} \frac{dt}{2 \pi  i} \frac{ e^{- \sqrt{t} r} }{4 \pi r} \, \text{Disc}_t \mathcal{A} (s,t) + ... 
  \label{eqn:FS_disc_trick}
\end{align}
where again the $+...$ are small-distance quantum corrections that do not affect the classical potential.
We briefly review this argument in Appendix~\ref{app:amp}. 
The utility of this relation is that applying the same argument to $V ( p^2, q^2 )$ allows us to immediately PM expand the Fourier transform\footnote{
Using the identity $\int_0^{\infty} \frac{dt}{2 \sqrt{t}} e^{-\sqrt{t} r} ( \sqrt{t} )^{n} = n! / r^{n+1}$, the integral \eqref{eqn:FS_disc_trick} of \eqref{eqn:Disc_V_PM} reproduces \eqref{eqn:V_PM}. 
},
\begin{align}
 \text{Disc}_t \, V ( p^2 , -t ) = 4 \pi^2 i \left[ 2 \kappa \,  v_1 \; \delta \left( \frac{t}{\mu^2} \right) +  \sum_{n=0}^{\infty} \frac{ \kappa^{n+2} v_{n+2} }{n!}  \left( \frac{ \sqrt{t} }{\mu} \right)^{n-1} 
 \right]
 \label{eqn:Disc_V_PM}
\end{align}
where $t \geq 0$. 
For non-derivative couplings, the $v_n$ in \eqref{eqn:Disc_V_PM} are functions of $p^2$ only and coincide with the coefficients of \eqref{eqn:V_PM}. 
For derivative couplings, the $v_n$ coefficients appearing in \eqref{eqn:Disc_V_PM} become the following analytic functions of $t$,
\begin{align}
 \frac{v_{2+n} \left( \frac{p^2}{2\mu^2}  , t \right) }{n!} = \sum_{n' = 0} \frac{ v_{2+n, 2n'}  \left( \frac{p^2}{2\mu^2} \right) }{ (n+2 n')! }  \left( \frac{t}{M_{\partial}^2} \right)^{n'} \; 
\end{align}
where the $v_{n,n'}$ coefficients are those of \eqref{eqn:V_PM_der}.

\paragraph{Classicality.}
\eqref{eqn:Disc_V_PM} represents the classical potential that we wish to extract from the quantum mechanical scattering amplitude.
We therefore \emph{define} $\mathcal{A}^{\rm cl}$, the ``classical part'' of the scattering amplitude, as the part of the $t$-channel discontinuity which scales with the appropriate power of $t$ at each order in perturbation theory to match \eqref{eqn:Disc_V_PM}\footnote{
See \cite{Cachazo:2017jef} for further discussion of this definition of classicality.
}. Explicitly, we define,
\begin{align}
 \text{Disc}_t \mathcal{A}^{\rm cl} (s, t) = \frac{ 4 \pi^2 i }{\nu} \left[ 2 \kappa \, a_1 (s) \, \delta \left( \frac{t}{\mu^2} \right) + \sum_{n=0}^{\infty} \frac{ \kappa^{n+2} a_{n+2} (s, t) }{ n!}  \left( \frac{  \sqrt{t} }{ \mu } \right)^{n-1}  
 \right] \; 
 \label{eqn:Disc_A_PM}
\end{align}
where each $a_{2+n} (s,t)$ is an analytic function of $t$ that can be written as, 
\begin{align}
\frac{ a_{2+n} (s, t) }{n!} = \sum_{n' = 0} \frac{ a_{2+n, 2n'} (s) }{ (n+2 n')! }  \left( \frac{t}{M_{\partial}^2} \right)^{n'}
\label{eqn:ann'_def}
\end{align}
and we have factored out an overall $1/\nu$ so that each $a_n$ is finite in the probe limit $\nu \to 0$. 
Here we see explicitly that the $a_1$ coefficient cannot receive derivative corrections since any powers of $t$ would vanish on the support of the $\delta(t)$ function.

Note that our use of ``classical'' should not be confused with ``tree-level''. 
In particular, at tree-level $\mathcal{A} (s, t)$ can have only simple poles in $t$, and therefore can only contain $a_1 (s)$ (i.e. the $v_1 \kappa/r$ part of the potential).
The $1/\sqrt{t}$ part of the discontinuity comes from Feynman diagrams that contain one loop, and similarly each $a_{2+n}$ coefficient corresponds to an $(n+1)$-loop Feynman diagram. 
The ``classical part'' of the amplitude, in this context, is the part whose Fourier transform back to position space yields the long-range potential which would be inferred from solving the classical equations of motion, and which takes the form \eqref{eqn:V_PM_der} for the class of EFTs we are considering.

\paragraph{From amplitude to potential.}
Altogether, we can now insert the PM expansions \eqref{eqn:Disc_A_PM} and \eqref{eqn:Disc_V_PM} into the Lippmann-Schwinger equation \eqref{eqn:main_LS} and relate the coefficients $v_n$ and $a_n$ at any given order in perturbation theory.
For instance, for non-derivative couplings the first two orders are,
\begin{align}
    v_1 \left( \frac{p^2}{2 \mu^2} \right) &= - \frac{ m_A m_B}{4 E_A E_B } a_1 ( s )   \nonumber \\ 
    v_{2,0} \left( \frac{p^2}{2 \mu^2} \right) &= - \frac{ m_A m_B}{4 E_A E_B } \left[ a_{2,0} ( s )  +   \frac{ \mu a_1 (s) }{ 2 \sqrt{s} }  \hat{D}_{p^2} \; v_{1} \left( \frac{p^2}{2\mu^2} \right)  \right] \label{eqn:amp_to_pot}  
\end{align}
where we have introduced the differential operator $\hat{D}_{p^2} =  \tfrac{ s - 3 E_A E_B}{4 E_A E_B }  + E_A E_B \partial_{p^2} $.
These agree with subtraction procedures already performed in the literature, e.g. \eqref{eqn:amp_to_pot} is equivalent to \cite[(23)]{Cheung:2018wkq}. 
For derivative couplings, there are no Born subtractions at 2PM, 
\begin{align}
 v_{2,2n} \left( \frac{p^2}{2 \mu^2} \right) &= - \frac{ m_A m_B}{4 E_A E_B } a_{2,2n} ( s )  
\end{align}
as a straightforward consequence of $v_{1,n} = 0$ for non-zero $n$. Though subtractions do appear at 3PM, for instance
\begin{align}
 v_{3,2} \left( \frac{p^2}{2 \mu^2} \right) &= - \frac{ m_A m_B}{4 E_A E_B } \left[ a_{3,2} ( s )  +  \frac{ \mu a_{2,2} (s) }{ 2 \sqrt{s} }  \hat{D}_{p^2} \; v_{1} \left( \frac{p^2}{2\mu^2} \right)  + \frac{ \mu a_{1} (s) }{ 2 \sqrt{s} }  \hat{D}_{p^2} \; v_{2,2} \left( \frac{p^2}{2\mu^2} \right)   \right] \; . 
 \label{eqn:amp_to_pot_der}
\end{align}
While the Born subtractions~\eqref{eqn:amp_to_pot} are well-known, the analogous Born subtractions in \eqref{eqn:amp_to_pot_der} for derivatively coupled interactions are somewhat new.

\begin{figure}

	\begin{tikzpicture}
	\begin{feynman}
		
		\vertex (mLU) at (0.4,0);
		\vertex (mLD) at (-0.4,0);
		\vertex (mRU) at (0.4,2);
		\vertex (mRD) at (-0.4,2);
		
		\vertex (mU1) at (0.4,0.9);
		\vertex (mU2) at (0.4,1.1);
		\vertex (mD1) at (-0.4,0.9);
		\vertex (mD2) at (-0.4,1.1);
		
		\node at (0.8,1) {$\ell^\mu_2$};
		\node at (-0.8,1) {$\ell^\mu_1$};
		
		\vertex (a) at (-2,-1) {$p_2^\mu = ( E_B , - \bfp )$};
		\vertex (b) at ( -2, 3) {$p_1^\mu = ( E_A , \bfp )$};
		\vertex (c) at (2,-1) {$p_4^\mu = ( E_B, -\bfp' )$};
		\vertex (d) at ( 2, 3) {$p_3^\mu = ( E_A,  \bfp' )$};
		\diagram* {
			(a) -- [fermion] (mLD) -- [photon] (mD1), (mD2) -- [photon] (mRD),
			(mU1) -- [photon] (mLU) -- [fermion] (c), 
			(mU2) -- [photon] (mRU),
			(b) -- [fermion] (mRD), 
			(mRU)-- [fermion] (d),
		};
		
		\node [thick, draw=black, fill=black!10, ellipse, minimum width=1.5cm,minimum height=0.5cm,align=center] at (0,0);
		\node [thick, draw=black, fill=black!10, ellipse, minimum width=1.5cm,minimum height=0.5cm,align=center] at (0,2);
		
		\draw [thin, dashed] (0.55,1) -- (-0.55,1);
		
	\end{feynman}
\end{tikzpicture}
\hfill
	\begin{tikzpicture}
		\begin{feynman}

			\vertex (mLU) at (0,0.4);
			\vertex (mLD) at (0,-0.4);
			\vertex (mRU) at (2.4,0.4);
			\vertex (mRD) at (2.4,-0.4);

			\vertex (mU1) at (1.1, 0.4);
			\vertex (mU2) at (1.3, 0.4);
			\vertex (mD1) at (1.1, -0.4);
			\vertex (mD2) at (1.3, -0.4);

			\node at (1.2,0.8) {$\tilde{\ell}^\mu_1 = (\tilde{E} , \bfl )$};
			\node at (1.2,-0.8) {$\tilde{\ell}^\mu_2 = (\tilde{E} , -\bfl )$};

			\vertex (a) at (-1,-2) {$\tilde{p}_2 = ( \tilde{E}, -\tilde{\bfp}_A)$};
			\vertex (b) at ( 3.4,-2) {$\tilde{p}_4 = ( \tilde{E}, - \tilde{\bfp}_B)$};
			\vertex (c) at (-1, 2) {$\tilde{p}_1^\mu = (\tilde{E}, \tilde{\bfp}_A )$};
			\vertex (d) at ( 3.4, 2) {$\tilde{p}_3 = ( \tilde{E},  \tilde{\bfp}_B )$};
			\diagram* {
				(a) -- [fermion] (mLD) -- [photon, with arrow = 0.8] (mD1), (mD2) -- [photon, with arrow = 0.2] (mRD),
				(c) -- [fermion] (mLU) -- [photon, with arrow = 0.8] (mU1), 
				(mU2) -- [photon, with arrow=0.2] (mRU),
				(mRD) -- [fermion] (b), 
				(mRU)-- [fermion] (d),
			};
		
		\node [thick, draw=black, fill=black!10, ellipse, minimum width=0.5cm,minimum height=1.5cm,align=center] at (0,0);
		\node [thick, draw=black, fill=black!10, ellipse, minimum width=0.5cm,minimum height=1.5cm,align=center] at (2.4,0);
		
			\draw [thin, dashed] (1.2,0.55) -- (1.2,-0.55);
		
		\end{feynman}
	\end{tikzpicture}
	\caption{The analytic continuation of \cite{Feinberg:1988yw} maps any $t$-channel diagram to an $s$-channel diagram. For concreteness, we show above the case of two internal lines (with unitarity cut shown) and given the explicit kinematics that we use. Time flows from left to right.}
	\label{fig:FS}
\end{figure}
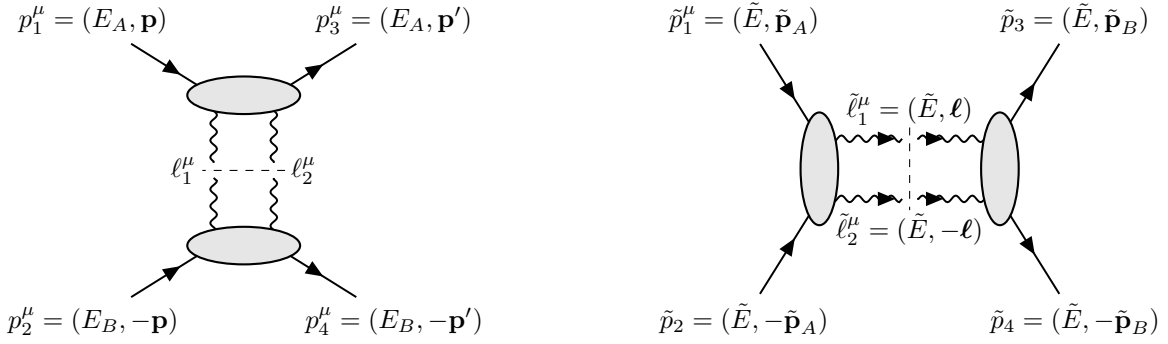

\paragraph{Crossing.}
To compute the $\text{Disc}_t \mathcal{A} (s,t)$, Feinberg and Sucher made use of ``crossing'': an analytic continuation (complex boost) of the momenta, $p^\mu \to \tilde{p}^\mu$, such that,
\begin{align}
A_{\chi_A \chi_B \to \chi_A \chi_B } (s,t) \equiv 
A_{\chi_A \chi_B \to \chi_A \chi_B } ( \bfp_1, \bfp_2 ; \bfp_3, \bfp_4 ) = 
A_{\chi_A \chi_A \to \chi_B \chi_B } ( \tilde{\bfp}_1 , \tilde{\bfp}_2 ; \tilde{\bfp}_3 , \tilde{\bfp}_4 ) \equiv A_{\chi_A \chi_A \to \chi_B \chi_B } (t, s) \; ,
\end{align}
when $\chi_{A,B}$ are real scalars.
These new $\tilde{p}$ are fixed by the requirements that they remain on-shell, conserved and are related to the original $p$ by:
\begin{align}
	s &= (p_1^\mu + p_2^\mu )^2 = (\tilde{p}_1^\mu - \tilde{p}_3^\mu )^2  \; , 
	&t &= (p_1^\mu - p_3^\mu )^2 = (\tilde{p}_1^\mu + \tilde{p}_2^\mu )^2  \; .
\end{align}
The explicit kinematics for these $\tilde{p}$ in their new centre-of-mass frame is shown in Figure~\ref{fig:FS}, where the energy and momentum of each particle is given by,
\begin{align}
	\tilde{E} = \frac{ \sqrt{t} }{2} \; , \;\; \tilde{p}_A = \frac{\sqrt{t - 4m_A^2}}{2} \; , \;\; \tilde{p}_B = \frac{\sqrt{t-4m_B^2}}{2} \; .
\end{align}
and the angle between their spatial momenta is,
\begin{align}
	\hat{\bfp}_A \cdot \hat{\bfp}_B =  \frac{2 s + t - 2 m_A^2 - 2 m_B^2}{4 \tilde{p}_A \tilde{p}_B}  =  -\sigma + \mathcal{O} (t)  \; . 
	\label{eqn:hatpApB_def}
\end{align}
This analytic continuation is essentially a Wick rotation in both time and space (which is why a $t$-channel diagram ``flips'' into an $s$-channel diagram), and as a result both the energy and the magnitude of the spatial momenta would be complex\footnote{
The branch cuts are chosen so that, $2 \tilde{E} = i \sqrt{-t}$, $2\tilde{\bfp}_A = i \hat{\bfp}_A \sqrt{4m_A^2 - t} $ and $2 \tilde{\bfp}_B = i \hat{\bfp}_B \sqrt{4m_B^2 - t} $ when $t < 0$, where $\hat{\bfp}_A$ and $\hat{\bfp}_B$ are real unit-vectors satisfying \eqref{eqn:hatpApB_def}.
} for the physical region $t < 0$. 
But crucially, for the $t$-channel discontinuity at $t > 0$ which determines the Fourier transform~\eqref{eqn:FS_disc_trick}, the energy of these vectors become real.

\paragraph{Unitarity.}
The virtue of this crossed kinematics is that now $\chi_A \chi_A \to \chi_B \chi_B$ scattering proceeds through intermediate states with real on-shell momenta. 
As a result, the optical theorem can be used to express the $\text{Disc}_t$ as a sum over a complete set of $N$-particle states,
\begin{align}
	\text{Disc}_t \, \mathcal{A}_{\chi_A \chi_B \to \chi_A \chi_B} ( s , t ) &= \text{Disc}_t \, \mathcal{A}_{\chi_A \chi_A \to \chi_B \chi_B} ( t , s ) \nonumber \\ 
	&= \sum_N \int d \Pi_N \mathcal{A}_{\chi_A \chi_A \to N} ( \tilde{\bfp}_1 , \tilde{\bfp}_2 ;  \{ \tilde{\bfl} \}_N ) \mathcal{A}^*_{\chi_B \chi_B \to N} ( \tilde{\bfp}_3, \tilde{\bfp}_4 ;  \{ \tilde{\bfl} \}_N  )
	\label{eqn:Disc_unit}
\end{align}
where each term in the sum includes a Lorentz-invariant integral over all continuous quantum numbers of the $N$-particle state, subject to energy- and momentum-conservation. 
Explicitly, for a state with $N$ massless fields that mediate the long-range potential with spatial momenta $\{ \tilde{\bfl}_1, ... , \tilde{\bfl}_N \}$, we have,
\begin{align}
 \int d \Pi_N =  \frac{1}{N!} \left[ \prod_{a=1}^N \int \frac{d^3 \tilde{\bfl}_a}{ (2\pi)^3 2 \tilde{\ell}_a } \right] ( 2 \pi )^4 \delta^4 \left(  \tilde{p}_1^\mu + \tilde{p}_2^\mu  - \sum_{a=1}^N \tilde{\ell}_a^\mu  \right) \; .
\end{align}
Since every field insertion introduces a factor of $1/M_P$, this power counting ensures that each term in \eqref{eqn:Disc_unit} corresponds to a fixed PM order. 
We therefore define the classical part of each integral as,
\begin{align}
 \left[ \int d \Pi_N \; \mathcal{A}_{\chi_A \chi_A \to N} ( \tilde{\bfp}_1 , \tilde{\bfp}_2 ;  \{ \tilde{\bfl} \}_N ) \mathcal{A}^*_{\chi_B \chi_B \to N} ( \tilde{\bfp}_3, \tilde{\bfp}_4 ; \{ \tilde{\bfl} \}_N ) \right]^{\rm cl} 
= 4 \pi^2 i \kappa^{N}  \frac{a_{N} (s,t) }{ (N-2)!} \left( \frac{\kappa \sqrt{t} }{\mu} \right)^{N-3} \; . 
\end{align}
While the original $\mathcal{A}_{\chi_A \chi_B \to \chi_A \chi_B}$ required computing loop-level Feynman diagrams, unitarity has reduced this to a computation of tree-level diagrams for the $2 \to N$ process. 
While generalised unitarity cuts have been used to great effect in computing loop corrections in General Relativity~(see e.g. \cite{Bjerrum-Bohr:2013bxa, Neill:2013wsa, Cachazo:2017jef, Guevara:2017csg, Bjerrum-Bohr:2018xdl} and many further works), the above approach of Feinberg and Sucher seems to have fallen out of fashion (though see \cite{Holstein:2016cfx, Holstein:2016fxh, Caron-Huot:2018ape}).
One advantage is that this implementation of the optical theorem makes a clear connection with the on-shell intermediate states mediating the interaction, and therefore provides a systematic way to include additional degrees of freedom beyond the metric tensor.
In particular, note that the $t$-channel discontinuity was used in \cite{Brax:2017xho, Banks:2020gpu} to study the potential arising from new dark sector fields in a similar spirit to the present work.

\paragraph{Computing 1PM amplitudes.}
At leading order in field insertions (i.e. tree-level Feynman diagrams for $\mathcal{A}_{\chi_A \chi_B \to \chi_A \chi_B}$), the unitarity integral becomes trivial since,
\begin{align}
 \int d \Pi_1 =  2 \pi i \delta (t) \; . 
\end{align}
The $a_1$ coefficient can therefore be written as\footnote{
Momentum conservation fixes $\tilde{\bfl} = - \tilde{\bfp}_1 - \tilde{\bfp}_2 = -\tilde{\bfp}_3 - \tilde{\bfp}_4$ in \eqref{eqn:a1_unit}.
},
\begin{align}
\frac{m_A^2 m_B^2 }{2 M_P^2} \, a_1 (s ) = \sum_X  \mathcal{A}_{\chi_A \chi_A \to X} ( \tilde{\bfp}_1, \tilde{\bfp}_2 ; \tilde{\bfl} ) \mathcal{A}_{\chi_B \chi_B \to X}^* ( \tilde{\bfp}_3, \tilde{\bfp}_4 ; \tilde{\bfl} )  |_{t=0}
\label{eqn:a1_unit}
\end{align}
where the sum is over single-particle states, i.e. the sum is over each massless field $X$ that mediates the long-range interaction between the compact objects. This is consistent with the expectation that at tree-level, $\mathcal{A}_{\chi_A \chi_B \to \chi_A \chi_B} \supset Z^2/(-t - i \epsilon)$ and locality guarantees the factorisation of $Z^2$ into two three-particle amplitudes.

\paragraph{Computing 2PM amplitudes.}
At next-to-leading order (i.e. one-loop Feynman diagrams for $\mathcal{A}_{\chi_A \chi_B \to \chi_A \chi_B}$), the unitary integral is no longer trivial but can be written as an angular average,
\begin{align}
 \int d \Pi_2 \; f ( \tilde{\bfl}_1, \tilde{\bfl}_2 )  = \frac{1}{16 \pi} \int \frac{d^2 \hat{\bfl} }{ 4 \pi } \; f (\tilde{\bfl} , -\tilde{\bfl} )  \equiv \frac{1}{8 \pi} \left\langle  f ( \tilde{\bfl}, - \tilde{\bfl} ) \right\rangle \; . 
\end{align}
We therefore define the angular averages,
\begin{align}
\rho_{X_1 X_2} (s, t)  = \left\langle  \mathcal{A}_{\chi_A \chi_A \to X_1 X_2} ( \tilde{\bfp}_1, \tilde{\bfp}_2 ; \tilde{\bfl}_1 , \tilde{\bfl}_2 )  \mathcal{A}^*_{\chi_B \chi_B \to X_1 X_2}  ( \tilde{\bfp}_3 , \tilde{\bfp}_4 ; \tilde{\bfl}_1 , \tilde{\bfl}_2 ) \right\rangle
\label{eqn:rho_def}
\end{align}
with the kinematics shown in Figure~\ref{fig:FS}, since the amplitude coefficient $a_2$ then corresponds to the sum over possible massless intermediate states\footnote{
As before, the superscript ${\rm cl}$ indicates that any term in $\rho$ which does not appear on the left-hand-side (i.e. does not fit the power counting~\eqref{eqn:ann'_def}) should be discarded. 
},
\begin{align}
\frac{\kappa^2 M}{\sqrt{t} } a_{2} (s , t)  = \frac{1}{64 \pi^3} \sum_{X_1, X_2} \rho_{X_1 X_2}^{\rm cl} (s, t) \; . 
\label{eqn:a2_unit}
\end{align}
In the centre-of-mass frame, the kinematics of each cut diagram is specified by $t$ and three independent momenta: $\hat{\bfp}_A$, $\hat{\bfp}_B$ and $\hat{\bfl}$. 
We therefore define three angular variables,
\begin{align}
 x_A &= \hat{\bfp}_A \cdot \hat{\bfl} \; , \;\;  &x_B &= \hat{\bfp}_B \cdot \hat{\bfl} \; , \;\;  &y &= \hat{\bfp}_A \cdot \hat{\bfp}_B \; .
\end{align}
where $y$ is given by~\eqref{eqn:hatpApB_def}. 
In practice, the angular averages~\eqref{eqn:rho_def} are then straightforward to evaluate since the only $x_A, x_B$ dependence which can contribute to the classical limit are those of the form,
\begin{align}
	\left\langle \frac{ x_B^{2n} }{ x_A^2 + \frac{t}{4m_A^2}  } \right\rangle 
	=
	\frac{m_A}{\sqrt{t}} \frac{  \sqrt{\pi} \Gamma \left( \frac{1}{2} + n \right) }{ \Gamma ( 1 + n ) } ( 1 - y^2 )^n + \mathcal{O} (t^0 ) 
	\label{eqn:ang_int_id}
\end{align} 
and its $A \leftrightarrow B$ permutation. All other functions of $x_A, x_B$ which can appear in a tree-level amplitude yield averages which $\sim t^a$ for integer $a$, see for instance the Appendix of~\cite{Holstein:2016cfx}.

\paragraph{General Relativity.}
From the exchange of one and two gravitons, in the Appendix we use the above formalism to evaluate the first PM corrections in GR and find,
\begin{align}
 a_1 (s) &= 4 \left(  2 \sigma^2 - 1 \right) \; , &a_2 (s) &= 3 \left( 5 \sigma^2 - 1 \right)
 \label{eqn:a2_GR}
\end{align}
in agreement with the many existing calculations (see e.g. \cite{Bjerrum-Bohr:2002gqz, Bjerrum-Bohr:2002fji}).
Substituting these into the formula~\eqref{eqn:amp_to_pot} gives the two-body potential in this gauge,
\begin{align}
 v_1 &= \frac{m_A m_B}{E_A E_B} \left( 1 - 2 \sigma^2 \right)  \; ,   \label{eqn:v2_GR}  \\
 v_2 &=  \frac{m_A m_B }{E_A E_B} \left\{ \frac{3 - 15 \sigma^2}{4}  - \frac{\mu}{\sqrt{s}}   ( 1 - 2 \sigma^2 )  \left[   \frac{4 \sigma s}{E_A E_B} + \frac{m_A m_B}{E_A E_B}  \left( \frac{s}{E_A E_B} -1 \right) \frac{ 1 - 2 \sigma^2}{2} \right] \right\}
\end{align}
which are the $V_{\rm GR}^{(1)}$ and $V_{\rm GR}^{(2)}$ appearing in \eqref{eqn:intro_V}.
These potentials capture all orders in $p^2$ (i.e. $v^2/ c^2$) at a fixed order in $\kappa$. 
Expanding in small $p^2$ reproduces the existing PN results \cite{Buonanno:1998gg, Damour:1988mr, Damour:1999cr},
\begin{align}
 v_1  &= -1 -  ( 3 + 2 \nu ) \frac{ p^2 }{ 2 mu^2} + \frac{ 5 - 20 \nu - 8 \nu^2 }{2} \left( \frac{ p^2 }{ 2  \mu^2 } \right)^2  -
 \frac{  ( 2 \nu -1) (-7 + 28 \nu + 8 \nu^2) }{2}  \left( \frac{ p^2 }{2 \mu^2} \right)^3 + ... 
 \nonumber  \\ 
 v_2  &= \frac{ 1 + \nu }{2} + \frac{ 10 + 27 \nu + 3 \nu^2 }{2} \frac{p^2}{2 \mu^2} 
 + \frac{ 
 -27 + 147 \nu + 222 \nu^2 + 15 \nu^3 }{4}  \left( \frac{ p^2 }{2 \mu^2 } \right)^2 + ... 
\end{align}

\subsection{Laplace-Runge-Lenz symmetry and orbital precession}

Now we turn to the question of how to extract from a potential (or, equivalently, an amplitude) the orbital precession. 
Recent approaches proceed via the scattering angle / radial action, using e.g. the boundary-to-bound dictionary \cite{Kalin:2019rwq, Kalin:2019inp} or the amplitude-action relation from eikonal exponential \cite{Kol:2021jjc, Bern:2022jvn}, or the KMOC in-in formalism \cite{Kosower:2018adc}.
Here, we adopt a different approach: we use the breaking of the Laplace-Runge-Lenz symmetry to extract the orbital precession. 
A similar method was used in \cite{Nabet:2014kva} for the Post-Newtonian expansion (see also \cite{Moser1970,Reeb52}), and we shall now develop this in an analogous way for the Post-Minkowskian expansion.

\paragraph{Newtonian theory.}
In the non-relativistic limit, the two-body dynamics is described by the Newtonian Hamiltonian (per reduced mass),
\begin{align}
    H_N = \frac{p^2}{2\mu^2} - \frac{\kappa}{\mu r}
    \label{eqn:HN_def}
\end{align}
The Laplace-Runge-Lenz (LRL) vector is defined by,
\begin{align}
   \bfK = \frac{ \bfp \cross \bfL }{\kappa} - \hat{\bfr}
   \label{eqn:K_def}
\end{align}
where $\bfL = \bfr \cross \bfp$ is the usual angular momentum, and we have chosen a normalisation in which, 
\begin{align}
 K^2  = 1 + \frac{2 L^2 }{\kappa^2}  H_N 
 \label{eqn:K2_def}
\end{align}
Various properties of this vector are reviewed in Appendix~\ref{app:LRL}.
The most important ones are:
\begin{itemize}
    \item[(i)] it is conserved, since $\{ H_N , \bfK \} = 0$, 

    \item[(ii)] it is directed towards a turning point of the orbit, since when $\bfp \cdot \bfr = 0$ we have $ \bfK \propto \hat{\bfr}$. 


\end{itemize}

\paragraph{Symmetry breaking.}
The conservation of $\bfK$ in the Newtonian theory is tied to a so-called ``hidden symmetry'', which has been the subject of numerous studies dating back at least to Pauli's study of the Hydrogen atom~\cite{Pauli:1926qpj, Fock1935,Bander:1965rz} (see \cite{10.1119/1.9745} for an historical account).
From a modern perspective, this symmetry can be understood as a dual conformal transformation (i.e. conformal symmetry in momentum space) \cite{Caron-Huot:2014gia}, and can be written explicitly in terms of a special conformal transformation ($b_\mu$), translation ($a_\mu$) and dilation ($\alpha$), 
\begin{align}
 p_\mu \to p_\mu' = \alpha  \left( \frac{ p_\mu - b_\mu p^2  }{ 1 + 2 b^\mu \bfp_\mu + b^2 p^2} + a_\mu \right)
\end{align}
where $a_\mu$ and $b_\mu$ are spatial vectors directed along $\bfK \cross \bfL$ and satisfying $b/a = \alpha$ and $2 -\alpha = \alpha \sqrt{  1 - 4 a^2 } $, as described in \cite{Caron-Huot:2014gia, Nabet:2014kva}.
This transformation preserves the energy but generally changes the angular momentum. This is one reason why the symmetry is often ``hidden'': it is spontaneously broken once $L^2$ is fixed.

Beyond the Newtonian limit, $\bfK$ is not generically conserved, since for a general potential we do not expect invariance under dual conformal transformations.
The rate of change of the LRL vector provides an order parameter for the symmetry breaking, and is given explicitly by,
\begin{align}
 \dot{\bfK} = \frac{ \hat{\bfr} \cross \bfL}{r^2} \, \left. \frac{\partial H}{\partial  \left( \frac{\kappa}{r} \right) } \right|_{H_N} 
 = \frac{ \hat{\bfr} \cross \bfL}{r^2}  \sum_{n=1} n \,  \tilde{h}_n ( H_N ) \, \left( \frac{\kappa}{r} \right)^{n-1} \; .
 \label{eqn:LRL_from_pot}
\end{align}
In particular, note that the special feature of the LRL gauge~\eqref{eqn:LRL_gauge} is that, since $\tilde{h}_1 = 0$, the time variation of the LRL vector starts at $\mathcal{O} ( \kappa )$. 

Since $\bfK$ is associated with a turning point of the orbit, this precession in $\bfK$ is directly tied to the precession of the orbit. 
Our goal for the remainder of this subsection is therefore to express orbital invariants, such as the periastron-to-periastron precession and period, in terms of this LRL vector. 
To do so, we will first review how $K$ is related to the trajectory $\{ r (t) , p (t) \}$ which satisfies the equations of motion.

\paragraph{Going on-shell.}
The equation of motion,
\begin{align}
 H [ r, p ] = E \equiv  M + \mu \E 
 \label{eqn:H=E}
\end{align}
can be solved for the momentum as a function of $r$ and $\E$.
Conceptually, one may then use,
\begin{align}
 L^2 [r, p] = L^2 
\end{align}
to at least partially fix $r$ in terms of $\E$ and $L$, the two constants of motion.
One way to do this is via the Laplace-Runge-Lenz vector. 
In particular, the on-shell solution for $r (t)$ will obey a relation of the form, 
\begin{align}
\hat{\bfr} \cdot \hat{\bfK} [ r, p ]  = \cos \th (t)
\label{eqn:th_def}
\end{align}
for some angle $\th (t)$. 
But now using the definition~\eqref{eqn:K_def}, which implies
\begin{align}
 \bfr \cdot \bfK  = \frac{L^2}{\mu \kappa} -  r \; 
\end{align}
we see that we can write $r$ in terms of $\th (t)$ via the implicit equation,
\begin{align}
 \frac{\kappa}{\mu r} = \frac{ \kappa^2}{ L^2} \left( 1  + K [r ,p] \cos \th (t) \right) \; . 
 \label{eqn:r_onshell}
\end{align}
We can then insert the on-shell expressions for $p$ and $r$ into the relation \eqref{eqn:K2_def} for $K^2$, and solve perturbatively for $K ( E, L , \th )$. 
Given \eqref{eqn:r_onshell}, this is equivalent to solving for $r (E, L , \th )$. 

Of course, this doesn't completely fix the dynamical $\{ r(t), p(t) \}$, since we have not determined the function $\th (t)$. 
Even in the Newtonian limit, this is determined by a transcendental equation (the Kepler equation). 
However, even without explicit knowledge of $\th (t)$, we may still compute certain orbital properties (often referred to as ``adiabatic invariants''). 
This is because, from \eqref{eqn:r_onshell}, we see that $r$ will undergo periodic motion in $\th$, and in particular will return to the value\footnote{
$r_c$ is the size of a circular orbit with this angular momentum, and it also sets a lower bound on the relative energy $2 \mu \E > - \kappa/r_c$. 
},
\begin{align}
 r_c \equiv L^2/(\kappa \mu ) 
 \end{align} 
at every $\th = (2n+1)\pi$ (assuming that $K$ is a finite function of $\th$, which we confirm below). 
Therefore each $2 \pi$ revolution of $\th$ defines a complete orbit, so we can extract orbital averages via $\int_{-\pi}^{\pi} d \th$ even without knowing the explicit time dependence of $\th (t)$.

\paragraph{Orbital precession.}
Suppose that in a time interval $dt$ the LRL vector changes by $d \bfK$. Generally, this represents both a change in direction and in magnitude. 
The vector,
\begin{align}
 d \boldsymbol{\Theta} = \frac{ \bfK \cross d \bfK }{ K^2 }
\end{align}
represents the infintesimal rotation of the LRL vector (i.e. $| d \boldsymbol{\Theta}|$ is the angle through which the LRL vector has rotated in the plane orthogonal to $d \boldsymbol{\Theta}$).  
The total angular precession of the periastron is thus obtained by integrating $d \boldsymbol{\Theta}$ around one orbit, and then taking its magnitude,
\begin{align}
\Delta \Theta =  \left| \oint dt  \frac{ \bfK \cross \dot{\bfK} }{ K^2 }  \right|  
\label{eqn:Dth_def}
\end{align}

To evaluate this, note that projecting \eqref{eqn:LRL_from_pot} onto the two independent components parallel and perpendicular to $\hat{\bfK}$,
\begin{align}
\hat{\bfK} \cross \dot{\bfK} &= \frac{\bfL}{r^2} \cos \th \; \left. \frac{\partial H}{\partial  \left( \frac{\kappa}{r} \right) } \right|_{H_0} \; , \;\; &\dot{K} &= \frac{L}{r^2} \, \sin \th \; \left. \frac{\partial H}{\partial  \left( \frac{\kappa}{r} \right) } \right|_{H_0}
\end{align}
we see that the precession \eqref{eqn:Dth_def} can be written as,
\begin{align}
\Delta \Theta &=   \oint dt \; \frac{\dot{K} \cos \th}{K \sin \th}    
=  \oint d \th \; \frac{\cos \th}{K} \frac{\partial K}{\partial \cos \th}
\label{eqn:pre_from_LRL}
\end{align}
where we have used that $\dot{K} ( E , L , \th ) = \dot{\th} \partial_{\th} K (E, L , \th )$ for the on-shell LRL vector.

\paragraph{Orbital period.}
The time dependence of $\th (t)$ is related to the angular momentum by\footnote{
This can be proven by writing the $r$ equation of motion, $\dot{r} =  \gamma_r \, p_r$, and then using the identity $p_r = \frac{\kappa}{L}  \, A \sin \th$. 
}, 
\begin{align}
  r^2  \dot{\th}  = L \left. \frac{\partial \mathcal{H}}{\partial \mathcal{H}_N } \right|_{\th}  \; . 
  \label{eqn:th_eom}
\end{align}
This can be written in terms of the LRL vector, since once on-shell
\begin{align}
 \left. \frac{\partial \mathcal{H}_N }{\partial \mathcal{H} } \right|_{\theta}   =   \frac{\kappa^2}{2 L^2} \frac{\partial K^2}{\partial E} \; .
\end{align}
This means that, while we may not have solved \eqref{eqn:th_eom} for the explicit $\th (t)$, we can nonetheless extract the time which elapses during a complete orbit (e.g. from $r= r_c$ to $r=r_c$, or equivalently from periastron to periastron),
\begin{align}
 T = L \oint \frac{d \th }{K} \; \frac{r^2}{\kappa^2} \frac{\partial E}{\partial K} \; .
\label{eqn:per_from_LRL}
\end{align}

Therefore a computation of $K (E, L , \th)$ is sufficient to determine, via the integral relations~\eqref{eqn:pre_from_LRL} and \eqref{eqn:per_from_LRL}, both the orbital precession and period.

\paragraph{PM expansion.}
To compute $K (E, L , \th )$ and in particular to match with a scattering amplitude, we will invoke the Post-Minkowski expansion in powers of $\kappa/r$. 
For the momentum, this is straightforward: we can write the on-shell solution as,
\begin{align}
 \frac{p^2}{2 \mu^2} &= \sum_{n=0} P_n ( \E ) \left(  \frac{\kappa}{\mu r} \right)^n 
 \label{eqn:p_onshell}
\end{align}
and then solve \eqref{eqn:H=E} perturbatively for the $P_n (\E)$ coefficients. 
The first few orders are defined by, 
\begin{align}
 h_0 ( P_0 ) &= E   \nonumber \\ 
 h_0' (P_0 )  P_1 &= - h_1 ( P_0 )     \nonumber \\ 
 h_0' (P_0 )  P_2 
 &= - h_2 (P_0 ) - \frac{ h_1 (P_0) }{ h_0' (P_0)}  \left(  \frac{h_0'' (P_0) }{2 h_0' (P_0) } - \partial_{P_0} \right) h_1 (P_0 ) \label{eqn:Pn_from_amp}
\end{align}
In the amplitude gauge,
\begin{align}
 P_0 (E) = \frac{1}{8 E^2} \left( E - (m_A + m_B )^2 \right) \left( E - (m_A - m_B )^2 \right)    \; ,
 \label{eqn:P0_def}
\end{align}
which satisfies $E^2 = s |_{p = P_0} $ and $v_0' = \sqrt{s} / 2 E_A E_B $. 
As observed in \cite{Bern:2019crd}, in this gauge the $P_n$ are proportional to the scattering amplitude coefficients (compare \eqref{eqn:Pn_from_amp} with the solution of the Lippmann-Schwinger equation \eqref{eqn:amp_to_pot}), and in fact \cite{Kalin:2019rwq} showed that the general result for non-derivative couplings is,
\begin{align}
 P_n ( \E ) = \frac{M}{4 E }  a_n (s = E^2)
\end{align}
at any PM order.
Finally, note that we have defined $M$ and $\mu$ such that $P_0 (0) = 0$ and $P'_0 (0) = 1$.  It is also customary to define $\kappa$ such that $P_1 (0) = 1$ in the Newtonian limit, so that $H$ reduces to $H_N$ at small $p^2$. However, we will find it more convenient to retain $\kappa \equiv G_N m_A m_B$, even in cases where the effective strength of gravity differs from $G_N \equiv 1/(8 \pi M_P^2)$ due to additional massless fields. The main reason is that, as can be seen in e.g. \cite{Julie:2022qux}, beyond leading order an expansion in terms of $\kappa_{\rm eff} = P_1 (0) \kappa$ rather than $\kappa$ leads to PN coefficients that are non-analytic functions of the field theory couplings.   

Once the momentum is put on-shell, \eqref{eqn:K2_def} becomes,
\begin{align}
 K^2 = \frac{\mu r_c }{ \kappa} 2 P_0 + 1  + 2 (P_1 - 1 ) \frac{r_c}{r}  + \sum_{n=1} \left( \frac{\kappa}{\mu r_c} \right)^n  \, 2 P_{n+1} \left( \frac{r_c}{r} \right)^{n+1}
 \label{eqn:LRL_from_amp}
\end{align}
Expanding in small $\kappa/(\mu r_c) = \kappa^2/L^2$ we would find a leading order solution,
\begin{align}
 K ( E, L , \th ) = \frac{L}{\kappa} \sqrt{P_0 (E) } \;\; \Rightarrow \;\; \frac{r_c}{r}  =  \frac{L}{\kappa} \sqrt{P_0 (E) } \cos \th \; .
 \label{eqn:r_inertial}
\end{align}
This corresponds to inertial (unbound) trajectories: the two particles fly past each other in straight lines. 
Computing the corrections to this from their interaction gives the PM expansion,
\begin{align}
 K ( E , L,  \th ) = \frac{L}{\kappa}  \sqrt{P_0} \sum_{n=0} K_n^{\rm PM} \left( E , \th \right) \, \left(  \frac{\kappa^2}{L^2} \right)^n 
\end{align}
For instance, the leading correction is,
\begin{align}
 K_1^{\rm PM}  (E , \th ) = \frac{  (P_1 - 1)^2 \cos^2 \th }{ 2  P_0  }  + P_2  \cos^2 \th \; . 
\end{align}
In this expansion scheme, the integration over $\th$ in \eqref{eqn:pre_from_LRL} is straightforward (the integrand is a polynomial in $\cos \th$). However, each $K_n \sim P_{n+1} \cos^{n+1} \th$ plus a number of lower-order terms involving $P_1 - 1$ and $P_0$, which is a consequence of expanding around the ``wrong'' background (perturbing inertial trajectories rather than a bound orbit).

\paragraph{PN expansion.}
Alternatively, we could note that the relative energy $|\E| < \frac{\kappa}{2\mu  r_c}$ for bound orbits, and therefore expand in both $\kappa/L$ and $\E$ simultaneously. 
This corresponds to the usual PN expansion. 
The leading order solution would then be\footnote{
Note that in deriving \eqref{eqn:K0_PN} from \eqref{eqn:LRL_from_amp} one treats $P_1 - 1$ as $\mathcal{O} ( p^2 )$. This is manifestly the case if $\kappa$ is chosen such that $P_1 (0) = 1$, but otherwise this expansion requires some care. 
},
\begin{align}
 K^2 = 1 + \frac{2 L^2}{\kappa^2} \E \equiv e_N^2  \qquad \Rightarrow \qquad \frac{r_c}{r} =  1 + e_N \cos \th
 \label{eqn:K0_PN}
\end{align}
which is an ellipse with eccenticity $e_N$ (to be treated $\mathcal{O} (1)$) and a semi-latus rectum $r_c$.  
Corrections can then be computed to this Keplerian orbit, and take the form, 
\begin{align}
 K ( E , L,  \th ) =  e_N \sum_{a,b = 0} K_{ab}^{\rm PN} ( e_N, \th ) \,  \left(  \frac{\kappa^2}{L^2} \right)^a  \, \E^b \; .
\end{align}
For instance, the leading corrections are,
\begin{align}
 K_{01}^{\rm PN} &= P_0'' (0) \frac{e_N^2 - 1 }{2 e_N^2} +  P_1' (0) \frac{1 + e_N \cos \th}{e_N^2}
 \;\; , \;\;\;\; 
&K_{10}^{\rm PN} &= P_2 (0)  \frac{ \left( 1 + e_N \cos \th \right)^2 }{ e_N^2}  
\end{align}
The $\th$ integrals are again fairly straightforward in this expansion scheme (\eqref{eqn:per_from_LRL} requires two simple residue integrals, but is otherwise polynomial), and one advantage is that there is now a direct connection with observations (for which one expects $e_N \lesssim 1$, whereas the naive PM expansion above treats $e_N \gg 1$).
However, the original $P_n ( \E)$ coefficients contain a good deal of information about the interactions at arbitrary energies/relative velocities, and in trying to engineer the ``right'' background (an ellipse) we have expanded them unnecessarily.

\paragraph{P1PM expansion.}
Instead of the naive PM or PN expansions above, we will evaluate the LRL vector using the following scheme. 
The free-theory $P_0$ will be treated as small ($\sim \mathcal{O} ( \kappa / r_c)$), but all other $P_n (\E)$ will be kept general. 
This amounts to defining the relativistic eccentricity,  
\begin{align}
 e^2 = 1 + \frac{r_c}{ \kappa } 2 P_0 (E)
\end{align}
and treating it as $\mathcal{O}(1)$, but we will refrain from expanding in small $\E$.
The result is an expansion of the form, 
\begin{align}
 K ( E , L,  \th ) = \sum_n K_n \left( E , e , \th \right) \, \left(  \frac{\kappa^2}{L^2} \right)^n 
 \label{eqn:K_P1PM}
\end{align}
which is subtly distinct from either PM or PN. 
In effect, this can be thought of as \emph{either} a PM expansion in which all of the $P_0$ and $P_1$ terms have been resummed, \emph{or} a PN expansion in which the $\E$ terms have been resummed. 
To recover the PM expansion we simply expand in large $e \sim \mathcal{O} \left( L^2/\kappa^2 \right)$ at fixed $\E$, and to recover the PN expansion we simply expand in small $\E \sim  \mathcal{O} \left( \kappa^2/L^2 \right)$ at fixed $e$. 

The leading order coefficient, $K_0$, is determined by the quadratic equation,
\begin{align}
 K_0^2 = e^2 + (P_1 - 1)  \left( 1 + K_0 \cos \th  \right) \; . 
 \label{eqn:K_1PM}
\end{align}
Introducing an effective eccentricity,
\begin{align}
 \tilde{e}^2 = \frac{ e^2 +  2 ( P_1 - 1 ) }{ (P_1 - 1)^2 } 
\end{align}
the solution can be written as,
\begin{align}
 K_0 =  (P_1 - 1 ) \left(  \sqrt{ \tilde{e}^2 + \cos^2 \th } + \cos \th \right) \; . 
 \label{eqn:K_1PM_sol}
\end{align}
This shows the resummation of all of the $P_1$ terms explicitly: in both the PM or PN expansion, this $K_0$ would have been expanded in powers of $\cos \th$ (i.e. in small $P_1 - 1$)
For lack of a better name, we will refer to the expansion \eqref{eqn:K_P1PM} as the \emph{Post-1PM} (P1PM) expansion of $K$, i.e. the Post-1PM expansion is a systematic expansion about the 1PM theory \eqref{eqn:K_1PM}, which has the exact solution \eqref{eqn:K_1PM_sol}.

It is now straightforward to solve \eqref{eqn:K2_def}, together with the on-shell solutions \eqref{eqn:r_onshell} and \eqref{eqn:p_onshell}, for each $K_n$ coefficient. This gives, 
\begin{align}
 K_1 (E , e^2, \th ) &= \frac{ P_2 ( 1 + K_0 \cos \th )^2 }{  (P_1 - 1) \sqrt{ \tilde{e}^2 + \cos^2 \th } } \nonumber \\ 
 K_2 (E , e^2, \th ) &=  \frac{ P_3 ( 1 + K_0 \cos \th )^3 + 2 P_2 K_1 \cos \th ( 1 + K_0 \cos \th ) - \tfrac{1}{2} K_1^2  }{  (P_1 - 1) \sqrt{ \tilde{e}^2 + \cos^2 \th } } 
\end{align}
and so on. 
This expansion scheme has the advantage that each $K_n$ contains far fewer terms (since there is no longer any proliferation of $P_0$ or $P_1$ terms)\footnote{
Another advantage is that, since $P_1 - 1$ is no longer treated as small, there is no need to redefine $\kappa$ in order to set $P_1 (0) = 1$ and one can keep a fixed expansion parameter \eqref{eqn:kappa_def} set by $M_P$ and independent of any further couplings.
}. The only drawback is that now the integration over $\th$ requires an integrand that $\sim \sqrt{ \cos^2 \th + \tilde{e}^2}$, but in practice these are always tractable\footnote{
A helpful identity is, $\tilde{e}^2 (P_1 - 1)/K_0 = \sqrt{\tilde{e}^2 + \cos^2 \th } - \cos \th$.
}.

\paragraph{P1PM precession.}
When the expansion \eqref{eqn:K_P1PM} above is used with the integral formula \eqref{eqn:pre_from_LRL} for the precession, it produces, 
\begin{align}
 \frac{ \Theta }{2 \pi } = 1 + \sum_{n=0} \Theta_n ( E , e ) \left( \frac{\kappa^2}{L^2} \right)^n 
\end{align}
The first few (from integrating $K_0, K_1, K_2$ and $K_3$) are given by,
\begin{align}
 \Theta_0 (E, e) &= 0 \nonumber \\ 
 \Theta_1 (E , e ) &= 2 \pi P_2    \nonumber \\ 
 \Theta_2 (E , e ) &= 3 \pi \left(  P_2^2 + 2 P_1 P_3 \right)    \nonumber \\ 
 \Theta_3 (E, e) &= 5 \pi \left( P_2^3 + 6 P_1 P_2 P_3 + 3 P_1^2 P_4  +  \frac{3}{5} \left( e^2 - 1 \right) d P_4   \right) \;  \label{eqn:pre_from_amp}  .
\end{align}
Now, these can be straightforwardly matched onto the standard PM expansion, 
\begin{align}
 \frac{ \Theta }{2 \pi} = 1 + \sum_n  \Theta_n^{\rm PM} ( E ) \left(  \frac{\kappa^2}{L^2} \right)^n 
\end{align}
by expanding $e^2 = 1 + 2 P_0 L^2/\kappa^2$. 
This gives $\Theta_0^{\rm PM} = \Theta_0$ and $\Theta_1^{\rm PM} = \Theta_1$, but at 3PM we have,
\begin{align}
  \Theta_2^{\rm PM} &= 3 \pi \left( P_2^2 + 2 P_1 P_3 + 2 P_0 P_4 \right)  \;  , 
  \label{eqn:pre_from_amp_PM}
\end{align}
which agrees with the scattering angle calculation of \cite[(3.13)]{Kalin:2019inp}. Note that when going from P1PM to PM, the expansion in large $e^2$ has the effect of \emph{mixing} different perturbative orders: in particular the naively 4PM coefficient $P_4$ appears in the 3PM precession. 
Our new observation is that, if the LRL symmetry is exact and hence $\Theta_2^{\rm PM} = 0$, then this mixing of orders places a constraint on amplitudes of different loop orders (more on this below).

\paragraph{Higher orders.}
Carrying out our LRL procedure to higher orders is straightforward.
We find that the general structure of each $\Theta_n$ is,
\begin{align}
 \Theta_n (E, e) \sim \pi  \left[ \sum_{ \substack{ i_{\rm Total} = 2n \\ i_a \geq 1 } } P_{i_1} ... P_{i_n} + (e^2 - 1 ) \sum_{ \substack{ i_{\rm Total} = 2n -2 \\ i_a \geq 1 } } P_{i_1} ... P_{i_{n-2}} + (e^2 - 1 )^2 \sum_{ \substack{ i_{\rm Total} = 2n - 4 \\ i_a \geq 1 } } P_{i_1} ... P_{ i_{n-4}}  + ...   \right] 
\end{align}
neglecting numerical factors.
For instance, at 4PM,
\begin{align}
 \Theta_4 \sim \pi \left[   
 P_2 P_2 P_2 P_2 + P_1 P_2 P_2  P_3 + P_1 P_1 P_2 P_4 + P_1 P_1 P_3 P_3  + P_1 P_1 P_1 P_5
 + (e^2 - 1 ) \left(  P_1 P_5 + P_3 P_3 + P_2 P_4 \right) 
 \right] 
\end{align}
Interestingly, the general structure of each $\Theta_n^{\rm PM}$ is therefore simply,
\begin{align}
 \Theta_n^{\rm PM} \sim \pi \sum_{ \substack{ i_{\rm Total} = 2n \\ i_a \geq 0 } }  P_{i_1} ... P_{i_n} \; , 
\end{align}
again up to numerical factors. 

Therefore to build from the $\Theta_n$ a complete answer for $\Theta_3^{\rm PM}$, we require the $P_1 P_5$, $P_3 P_3$ and $P_2 P_4$ terms from $\Theta_4$ and also the $P_6$ term from $\Theta_5$. 
Altogether, this gives,
\begin{align}
\Theta_3^{\rm PM} = 5 \pi \left( P_2^3 + 6 P_1 P_2 P_3 + P_1^2 P_4 + 3 P_0 P_3 P_3 + 6 P_0 P_2 P_4 +  6 P_0 P_1 P_5 +  3  P_0 P_0 P_6 \right) \; . 
\end{align}
Note that since $\bfK$ is closely related to the impact parameter, and the impact parameter is closely related to the scattering angle, the method adopted here for determining the $\Theta_n$ is surely equivalent to existing scattering angle methods \cite{Kalin:2019rwq, Kalin:2019inp}. A relatively simple closed formula for an arbitrary $\Theta_n (E, e)$ in terms of the $P_n$ therefore seems likely, but at present all we have explored in that direction is the leading PM contribution to each $\Theta_n$, which is given for odd/even $n$ by, 
\begin{align}
 \Theta_{2k- 1} (E, e) &= 2 \pi \left( \begin{array}{c} k - \tfrac{1}{2} \\ \tfrac{1}{2} \end{array} \right) (e^2 - 1 )^{k-1} P_{2k}  +  \mathcal{O}  \left(  \left( e^2 - 1 \right)^{k-2} \right)   \nonumber \\ 
 \Theta_{2k} (E, e) &=  3 \pi \left( \begin{array}{c} k + \tfrac{1}{2} \\ \tfrac{3}{2} \end{array} \right) (e^2 - 1 )^{k-1} 2 P_1  P_{2k+1}  +  \mathcal{O}  \left(  \left( e^2 - 1 \right)^{k-2} \right) \; . 
\end{align}

\paragraph{General Relativity.}
Applying the general formula~\eqref{eqn:pre_from_amp} to the General Relativity amplitudes~\eqref{eqn:a2_GR} gives the leading precession from \eqref{eqn:intro_pre}\footnote{
A non-relativistic expansion in small $\E$ reproduces (the $1/L^2$ part of) the familiar PN expansion \cite{Damour:1999cr},
\begin{align}
\theta_{\rm GR}  (\E)  = 3 +  \tfrac{1}{2} \left(  15 - 6 \nu \right) \E + \tfrac{3}{4} (5 - 5 \nu +  4 \nu^2 ) \E^2 + \mathcal{O} \left( \E^3 \right) \; . 
\end{align}
},
\begin{align}
 \theta_{\rm GR}  ( \E ) =  \frac{a_2 (E^2) M}{4 E}  = \frac{ 48 + 15 \E (2 + \nu \E ) (4 + 2 \E  + \nu \E^2 ) }{ 16 (1 + \E \nu ) } 
 \label{eqn:pre_GR}
\end{align}
where $E = M + \mu \E$.

\paragraph{Closed orbits.}
Note that in order for the precession to vanish, $K$ must either be conserved, i.e. $\partial_{\th} K (E, L, \th) = 0$, or it must be the canonical transformation of a conserved vector (i.e. any time-dependence must be pure gauge). 
One way to argue this is to write the precession in a general isotropic gauge as, 
\begin{align}
 \Delta \Theta = \oint  d \th     \sum_{a=1} \, \cos \th  \,  a \,  \tilde{h}_a ( H_N ) \left( \frac{\kappa}{r} \right)^{a-1}  \; \left(  \sum_{b=1} \left( \cos \th \, b \,  \tilde{h}_b ( H_N ) + \tilde{h}'_{b-1} (H_N)  \right) \left( \frac{\kappa}{r} \right)^{b-1}   \right)^{-1} \; .
 \label{eqn:pre_from_pot}
\end{align}
The only way for the precession to vanish without $\bfK$ being trivially conserved is if this integrand is finite and yet integrates to zero around a closed orbit. 
Yet, in a PN expansion, the only cancellation which can take place between the orbital averages $\langle p^{2a} / r^{b} \rangle$ is between $\langle p^2 \rangle$ and $\langle 1/r \rangle$ (i.e. the virial theorem, $\langle p^2 / \mu \rangle = \langle \kappa/r \rangle$).
In terms of $H_N$, we can only cancel $\langle H_N \rangle$ and $\langle \kappa/ r \rangle$. But if we transform to the LRL gauge \eqref{eqn:LRL_gauge}, there are no terms linear in $\kappa/r$ and so no such cancellation is possible. 
So either $\bfK$ is conserved, or one can transform to a gauge in which it is conserved. 

Note that the period generically receives relativistic corrections even when the angular precession vanishes, i.e. even when $\dot{K} \propto \partial H / \partial ( \frac{\kappa}{r} ) = 0$ and the LRL symmetry is exact, $T$ will differ from its Newtonian value unless $\partial H / \partial H_0 = 1$ (which trivially corresponds to the Newtonian theory).

\paragraph{LRL bootstrap.}
The mixing of different orders that takes place when going to the PM expansion could be seen as a nuisance: for instance, in order to reliably determine the 3PM precession would require the 4PM amplitude in $P_4$.
However, in theories with an LRL symmetry (i.e. dual conformal invariance), this curse becomes a blessing. 
Given that each $\Theta_n^{\rm PM} = 0$ on symmetry grounds, we require that,
\begin{align}
 \Theta_1^{\rm PM} &= 0 \; &&\Rightarrow \; &a_2 &= 0  \nonumber \\ 
 \Theta_2^{\rm PM} &= 0 \; &&\Rightarrow  &a_4 &= - \frac{ a_1 a_3 }{ a_0}  \; ,   \label{eqn:LRL_bootstrap} \\
 \Theta_3^{\rm PM} &= 0 \; &&\Rightarrow  &a_6 &= - \frac{2 a_1 a_5 }{a_0} - \frac{a_3 a_3}{a_0} - \frac{ a_1 a_1 a_4 }{3 a_0 a_0}
\nonumber 
\end{align}
and so on, where,
\begin{align}
 a_0 (s) =  4 E P_0 (\E)  = \frac{1}{ 2 \sqrt{s}} \left( s - (m_A - m_B )^2 \right) \left( s - (m_A + m_B )^2 \right) 
 \label{eqn:a0_def}
\end{align}
is fixed by the free theory evolution. 
One consequence of the LRL symmetry is therefore that every even PM order (every odd-loop amplitude) is fixed in terms of the odd orders (even-loop amplitudes), e.g. \eqref{eqn:LRL_bootstrap} fixes the one-loop amplitude, then relates the three-loop amplitude to the two-loop amplitude, then relates the five-loop amplitude to the four-loop and two-loop amplitudes. This pattern continues indefinitely.

\paragraph{Quasi-circular orbits.}
It is instructive to compare the above approach for determining the angular precession with the more traditional approach of solving the radial equation of motion. 
Given a Hamiltonian $\mathcal{H}$ as a function of the dimensionless $p^2$ and $\kappa/(\mu r))$, Hamilton's equations can be written in second order form as,
\begin{align}
\gamma^2 \ddot{r} = \frac{L^2}{r^3} + \frac{\kappa}{r^2} \left(  1 - 2 \dot{r}^2 \frac{\partial}{\partial p^2}  \right) \gamma \frac{\partial H}{ \partial \left( \frac{\kappa}{ \mu r} \right)} \Bigg|_{p^2 = \gamma^2 \dot{r}^2 + L^2 / r^2 }
\label{eqn:r_eom}
\end{align}
where $1/\gamma = 2 \mu \frac{\partial H}{\partial p^2} |_r$  plays the role of a Lorentz dilation factor.
In the non-relativistic limit ($\gamma \to 1$ and $\partial H/ \partial \left( \frac{ \kappa}{\mu r} \right) \to -1$), we recognise the usual $L^2/r^3 - \kappa/r^2$ effective potential of the Newtonian theory. 

The textbook procedure for solving this equation is to introduce the Binet variable $1/u(\varphi ) = \mu r(t)$ with $\dot{\varphi} (t) r^2 (t) = \tilde{L}$ for some fixed $\tilde{L}$, so that \eqref{eqn:r_eom} becomes that of a forced harmonic oscillator,
\begin{align}
 u'' = - F  \; . 
 \label{eqn:u_eom}
\end{align}
For quasi-circular orbits, for which $\dot{r}$ is small and can be neglected, the effective force becomes, 
\begin{align}
 F (u) =  \frac{L^2}{\gamma^2  \tilde{L}^2} u + \frac{1}{\gamma \mu \tilde{L}^2}  \frac{\partial H}{\partial u} \big |_{ p^2 = L^2 u^2}  \; ,
\end{align}
and we can write the solution as,
\begin{align}
 u ( \varphi ) = \tilde{u}_c \left( 1 + \delta \, \cos \varphi + \mathcal{O} ( \delta^2 ) \right)
\end{align}
where the constants $\tilde{u}$ and $\tilde{L}$ are fixed by the conditions $F ( \tilde{u}_c ) = 0$ and $F' (\tilde{u}_c ) = 1$.
In the PN limit, these conditions give,
\begin{align}
\tilde{u}_c &= - \frac{\kappa}{L^2} \left[  v_1 (0)  + \mathcal{O} \left( L^2 \tilde{u}_c^2 \right) \right]  \; &\tilde{L}^2 &= L^2 + \mathcal{O} ( L^2 \tilde{u}_c^2 ) \; .  
\label{eqn:tildeu_PN}
\end{align}

Suppose the Hamiltonian is now perturbed by a potential $v_n \left( \frac{p^2}{2 \mu^2} \right) \left( \frac{\kappa}{\mu r} \right)^n$.
Solving the perturbed \eqref{eqn:u_eom} to leading order in $v_n$, the resulting precession is, 
\begin{align}
\frac{ \Delta \Theta }{2 \pi} = - \left( \frac{ \kappa }{ \mu \tilde{r}_c} \right)^n \left( 1 + \mathcal{O} \left( \frac{p^2}{\mu^2} \right)  \right) \left[  \frac{ n ( n -1 ) }{2} \,  \frac{\mu^2}{p^2}  v_n   + \frac{ 3n + 2 }{2}  v_n'   +     \frac{p^2}{ \mu^2} v_n''   \right]_{p = L/\tilde{r}_c } \; . 
 \label{eqn:dTh_QC}
\end{align}
where $\mu \tilde{r}_c = 1/\tilde{u}_c$ and the $\mathcal{O} (p^2)$ terms we have neglected correspond to the PN corrections in \eqref{eqn:tildeu_PN}.
\eqref{eqn:dTh_QC} shows that the leading term in the PN expansion is,
\begin{align}
\frac{ \Delta \Theta }{2 \pi } = - \frac{\kappa^2}{L^2} \left( \frac{ \kappa }{ \mu \tilde{r}_c} \right)^{n-2}  \frac{n (n-1)}{2} \, v_n (0 ) \; . 
\end{align}
Note that when $n=2$, this captures the $v_2 (0)$ contribution from $P_2 (0) =  \tfrac{1}{4} a_2 |_{p^2=0} \supset - v_2 (0)$ in the above PM result for $\frac{\kappa^2}{L^2} \Theta_1$. Expanding \eqref{eqn:u_eom} to second order would produce the $v_1' (0) v_1 (0)$ term contained in the PM result.

Note that we have included the additional terms in \eqref{eqn:dTh_QC} since if $v_n (0)$ vanishes then the leading term becomes,
\begin{align}
\frac{ \Delta \Theta }{ 2\pi } = - \frac{1}{4} \left( \frac{ \kappa }{ \mu \tilde{r}_c} \right)^n  (n +1 ) (n+4)  \, v_n' (0)  \; .
\label{eqn:dTh_QC_vn'}
\end{align}
This expression will be useful in the next section when we compute the leading effects of the disformal interaction for quasi-circular orbits.

\paragraph{Derivative coupling.}
Finally, note that for derivative couplings, the double expansion in $M_P$ and $M_{\partial}$ would be implemented by replacing the $P_n (\E)$ in \eqref{eqn:p_onshell} with, 
\begin{align}
 P_{n} \left( \E , \frac{1}{M_{\partial}^2 r^2}  \right)  = \sum_{n' = 0}  P_{n, 2n'} \left( \E  \right) \, \left(  \frac{1}{M_{\partial}^2 r^2} \right)^{n'} \; . 
 \label{eqn:Pnn'_def}
\end{align}
where these coefficients are related to the amplitude coefficients \eqref{eqn:ann'_def} by,
\begin{align}
 P_{n,n'} (\E) = \frac{M}{4E} \,  a_{n,n'} (E^2)  \; . 
\end{align}
In practice, the various $P_{n,n'}$ with a common $n+n'$ share the same $r$ dependence and therefore affect the precession in the same way as $P_{n+n'}$ in the above formulae. 
For instance, consider the amplitude $a_{2,2}$. This contributes to the precession in the same way as $P_4$ in the PM expansion \eqref{eqn:pre_from_amp_PM}, 
\begin{align}
\frac{\Delta \Theta}{2\pi}  \supset   \frac{\kappa^2}{L^2} \frac{\mu^2}{ M_{\partial}^2 L^2} \,  3 P_0 (\E) P_{2,2} (\E) \; . 
\label{eqn:Th_P22}
\end{align} 

Interestingly, while the P1PM coefficients \eqref{eqn:pre_from_amp} can be written in an analogous way to \eqref{eqn:V_PM_der},
\begin{align}
\frac{\Theta}{2\pi} = 1 + \sum_{n, n'} \left( \frac{\kappa}{\mu r_c} \right)^n \left( \frac{1}{M_{\partial}^2 r_c^2} \right)^{n'} \Theta_{n, n'} \left(  E, e \right)
\end{align}
with $\mu r_c = L^2/\kappa$, a mixing of different orders takes place when passing to the PM expansion by expanding out $e$. 
For instance, the leading PM precession at $\mathcal{O} \left( \kappa^2/L^2 \right)$ is,
\begin{align}
\frac{\Delta \Theta}{2\pi} =  \frac{\kappa}{\mu r_c} \left[  \theta_{1,0} (\E) +    \frac{\kappa \mu}{ r_c M_{\partial}^2}  \theta_{1,1} (\E)   + \mathcal{O} \left(   \frac{1}{ r_c^2 M_{\partial}^2 } \right)  \right]  + \mathcal{O} \left(  \left( \frac{\kappa}{\mu r_c} \right)^2 \right)
\end{align}
where $\theta_{1,1}$ is given by~\eqref{eqn:Th_P22}.

~\\

That completes the network of relations shown in Figure~\ref{fig:overview}: we now have explicit relations between the two-body potential, scattering amplitude, LRL symmetry and orbital precession.

\section{Scalar-tensor theories}
\label{sec:ST}

Having established a general formalism, we now apply this to the particular scalar-tensor theory described in the Introduction.
Concretely, we consider an EFT of the form~\eqref{eqn:EFT_form} which depends on the metric $g_{\mu\nu}$ and a massless scalar field $\phi$ with canonical kinetic terms,
\begin{align}
 \mathcal{L} = \frac{ \sqrt{-g} }{2} \left( M_P^2 R [ g_{\mu\nu} ] + ( \partial \phi )^2 \right)
\end{align}
and which couple to matter via a common effective metric,
\begin{align}
 \tilde{g}_{A \, \mu \nu} = \tilde{g}_{B \, \mu \nu}= e^{C \left( \frac{\phi}{M_P} \right) } g_{\mu\nu} + D \left( \frac{\phi}{M_P} \right) \frac{ \nabla_\mu \phi \nabla_\nu \phi }{ M_P^2 M_{\partial}^2 } \; .
 \label{eqn:geff_def_2}
\end{align}
We expand around a Minkowski spacetime background, so that $g_{\mu\nu} = \eta_{\mu\nu} + h_{\mu\nu}/M_P$ defines the metric perturbation $h_{\mu\nu}$. 
Note that in the main text we assume the equivalence principle, so both $\tilde{g}_A$ and $\tilde{g}_B$ are the same. 
In Appendix~\ref{app:ST}, we describe the situation with body-dependent couplings $C_A$ and $D_A$ in the effective metric, which is designed to capture possible strong gravity effects which can violate the equivalence principle, such as spontaneous scalarisation \cite{Damour:1992we, Damour:1993hw, Damour:1996ke} or dynamical scalarisation \cite{Barausse:2012da, Shibata:2013pra, Palenzuela:2013hsa} around neutron stars, or time-dependent hair around black holes \cite{Sotiriou:2013qea, Sotiriou:2014pfa}. 

At the 2PM order to which we will work, the only terms in $C$ and $D$ which are relevant are,
\begin{align}
 C \left( \phi \right) &= \sqrt{2} \alpha \phi + 2 \beta \phi^2 + \mathcal{O} \left( \phi^3 \right) \; , \;\; &D ( \phi ) &= \lambda + \mathcal{O} \left(  \phi^2  \right)
\end{align}
The curious normalisation of $\sqrt{2}$ is so that our notation matches that of \cite{Julie:2022qux}, for ease of comparison. 
To keep the main narrative as simple as possible, we will set $\beta = 0$ for this section. 
More general expressions with non-zero $\beta$ can be found in Appendix~\ref{app:ST}. 

The PM expansion of the two-body potential in this theory can then be organised according to the scalar couplings $\alpha$ and $\lambda$ as in \eqref{eqn:intro_V}, 
\begin{align}
 v_1 \left( \frac{p^2}{2 \mu^2} \right)  &=  V_{\rm GR}^{(1)} ( p^2) + \alpha^2 \, V_{\alpha^2}^{(1)} (p^2)  \nonumber \\
 v_{2,0} \left( \frac{p^2}{2 \mu^2} \right) &= V_{\rm GR}^{(2)} ( p^2) + \alpha^2 \, V_{\alpha^2}^{(2)} + \alpha^4 V^{(2)}_{\alpha^4} (p^2) \nonumber \\ 
 v_{2,2} \left( \frac{p^2}{2 \mu^2} \right)  &= \lambda \alpha^2  V^{(2)}_{\lambda \alpha^2} (p^2)
\end{align}
Our goal in this section is to determine each of these functions of $p^2$, and find the resulting orbital precession, which can be analogously written as, 
\begin{align}
\frac{ \Theta  }{ 2\pi} = 1 + \frac{\kappa^2}{L^2}  \left[
  \theta_{\rm GR} (\E) +  \alpha^2  \theta_{\alpha^2} (\E) + \alpha^4 \theta_{\alpha^4} (\E) \right]  + \lambda \alpha^2 \frac{ \Theta_{\lambda \alpha^2} }{2\pi} 
\end{align}
where $\Theta_{\lambda \alpha^2}$ captures the leading order correction from the disformal interaction and will scale with different powers of $L$ in the PM versus PN expansion.

\subsection{Conformal (non-derivative) coupling}
\label{sec:con}

Let us first consider the conformally coupled scalar (i.e. set the disformal coupling $D (\phi) = 0$ in \eqref{eqn:geff_def_2}). 
The Post-Newtonian effects of such a non-derivative coupling have been studied in some detail \cite{Damour:1992we, Yagi:2011xp, Mirshekari:2013vb, Lang:2013fna, Sennett:2016klh, Julie:2017pkb, Bernard:2018hta, Bernard:2018ivi, Bernard:2022noq, Julie:2022qux}.
Here we study the complementary PM expansion.

The relevant Feynman rules for computing the amplitudes up to 2PM are,
\FloatBarrier
\begin{figure}[htbp!]
\centering
\begin{tikzpicture}[baseline=-0.1cm]
			\begin{feynman}
				\vertex (a1) at (0,1) {$\tilde{p}_1^\mu$};
				\vertex (b2) at (1,0);
				\vertex (c1) at (0,-1) {$\tilde{p}_2^\mu$};
				\vertex (b3) at (2.4,0) {$\tilde{\ell}^\mu$};
				
				\diagram*{
                                (a1) -- [fermion] (b2),
                                (c1) -- [fermion] (b2),
				(b2)  -- [scalar] (b3),
				};					
			\end{feynman}
                \end{tikzpicture} $=  \frac{ \sqrt{2} \alpha }{M_P} \left(  \tilde{p}_1 \cdot \tilde{p}_2 + 2 m_A^2 \right)  , $ \hfill
             \begin{tikzpicture}[baseline=-0.1cm]
			\begin{feynman}
				\vertex (a1) at (0,1) {$\tilde{p}_1^\mu$};
				\vertex (b2) at (1,0);
				\vertex (c1) at (0,-1) {$\tilde{p}_2^\mu$};
				\vertex (a3) at (2,1) {$\tilde{\ell}_1^\mu$};
				\vertex (c3) at (2,-1) {$\tilde{\ell}_2^\mu$};
				
				\diagram*{
                                (a1) -- [fermion] (b2),
                                (c1) -- [fermion] (b2),
				(b2)  -- [scalar] (a3),
				(b2)  -- [scalar] (c3),
				};					
			\end{feynman}
                \end{tikzpicture} $= - \frac{2 \alpha^2}{M_P^2} \left(  \tilde{p}_1 \cdot \tilde{p}_2 + 4 m_A^2  \right) , $ \qquad
\end{figure}
\FloatBarrier

\FloatBarrier
\begin{figure}[htbp!]
\centering
\begin{tikzpicture}[baseline=-0.1cm]
			\begin{feynman}
				\vertex (a1) at (0,1) {$\tilde{p}_1^\mu$};
				\vertex (b2) at (1,0);
				\vertex (c1) at (0,-1) {$\tilde{p}_2^\mu$};
				\vertex (b3) at (2.4,0) {$\tilde{\ell}^\mu$};
				
				\diagram*{
                                (a1) -- [fermion] (b2),
                                (c1) -- [fermion] (b2),
				(b2)  -- [graviton] (b3),
				};					
			\end{feynman}
                \end{tikzpicture} $=  \frac{2}{M_P}  \epsilon_{\mu \nu}^* ( \tilde{\ell} ) \tilde{p}_1^\mu \tilde{p}_2^\nu  , $ \hfill
             \begin{tikzpicture}[baseline=-0.1cm]
			\begin{feynman}
				\vertex (a1) at (0,1) {$\tilde{p}_1^\mu$};
				\vertex (b2) at (1,0);
				\vertex (c1) at (0,-1) {$\tilde{p}_2^\mu$};
				\vertex (a3) at (2,1) {$\tilde{\ell}_1^\mu$};
				\vertex (c3) at (2,-1) {$\tilde{\ell}_2^\mu$};
				
				\diagram*{
                                (a1) -- [fermion] (b2),
                                (c1) -- [fermion] (b2),
				(b2)  -- [scalar] (a3),
				(b2)  -- [graviton] (c3),
				};					
			\end{feynman}
                \end{tikzpicture} $= -\frac{ \sqrt{2} \alpha}{M_P^2} \epsilon_{\mu\nu}^* ( \tilde{\ell}_2 )  \tilde{p}_1^\mu \tilde{p}_2^\nu , $ \qquad
\end{figure}
\FloatBarrier

\paragraph{1PM results.}
The cubic $\chi_A \chi_A \phi$ interaction in the centre-of-mass frame of Figure~\ref{fig:FS} is simply,
\begin{align}
 A_{\chi_A \chi_A \to \phi} \left( \tilde{\bfp}_1, \tilde{\bfp}_2 ; \tilde{\bfl} \right) = \frac{ \sqrt{2} \alpha  }{M_P} \left( m_A^2 + \mathcal{O} (t) \right) 
\end{align}
and so the unitarity cut~\eqref{eqn:a1_unit} immediately gives $ a_1 (s) \supset 4 \alpha^2$ from the scalar exchange. 
Of course, at this order the $\mathcal{A}_{\chi_A \chi_B \to \chi_A \chi_B}$ amplitude at tree-level is simply,
\begin{align}
\mathcal{A}_{\chi_A \chi_B \to \chi_A \chi_B} = \frac{2 \alpha^2}{M_P^2}  \frac{m_A^2 m_B^2}{ -t - i \epsilon} 
\end{align}
and the $\text{Disc}_t$ indeed gives this value of $a_1$, since $\text{Disc}_t  ( -t - i \epsilon )^{-1} =  2 \pi i  \delta (t)$. 
When combined with the GR contribution \eqref{eqn:a2_GR}, this gives,
\begin{align}
 a_1 (s) = 4 \left( 2 \sigma^2 - 1 +  \alpha^2 \right) \; .
\end{align}
The corresponding 1PM potential mediated by the scalar exchange then takes the form \eqref{eqn:intro_V}, with,
\begin{align}
 V^{(1)}_{\alpha^2} (p^2) = - \frac{m_A m_B}{E_A E_B}
 \label{eqn:v1_con}
\end{align}
and where $E_{A,B} = \sqrt{p^2 + m_{A,B}^2}$.

\paragraph{2PM results.}
In this scalar-tensor theory, the optical theorem expression~\eqref{eqn:a2_unit} for the 2PM amplitude becomes,
\begin{align}
 \rho_{hh}^{\rm cl} + \rho_{\phi h}^{\rm cl} + \rho_{h \phi}^{\rm cl} + \rho_{\phi \phi}^{\rm cl}  = \frac{m_A^2 m_B^2}{M_P^4} \frac{\pi M}{\sqrt{t} } a_{2} (s)
\end{align}

For the $\rho_{\phi \phi}$ contribution, we first compute,
\begin{align}
 A_{\chi_A \chi_A \to \phi \phi } 
 &=  \frac{ 2 \alpha^2 m_A^2 }{M_P^2} \left(   1 - \frac{1}{ x_A^2 + \frac{t}{4 m_A^2}  }  \right)  \; ,
\end{align}
from the above Feynman rules, and then evaluate the classical part of $\rho$ by picking out all terms of the form~\eqref{eqn:xBdA_id}, 
\begin{align}
 \rho_{\phi \phi}^{\rm cl} &=   - 4 \alpha^4  \frac{ m_A^2 m_B^2 }{ M_P^4 } \left\langle \frac{1}{x_A^2 + \frac{t}{4m_A^2}} \right\rangle   + \left( A \leftrightarrow B \right) \nonumber \\
 &= - 4 \alpha^4  \frac{ m_A^2 m_B^2 }{ M_P^4 } \frac{\pi M}{\sqrt{-t} }
\end{align}

For the $\rho_{\phi h}$ contribution, we first compute,
\begin{align}
 A_{\chi_A \chi_A \to \phi h } 
 &=   \frac{ 2 \sqrt{2} \alpha m_A^2 }{M_P^2} \frac{ \hat{\bfp}_A^i \hat{\bfp}_A^j \epsilon_{ij}^* ( \hat{\bfl} ) }{ x_A^2 + \frac{t}{4m_A^2} }
\end{align}
from the above Feynman rules, and then evaluate the classical part of $\rho$ using the polarisation sum~\eqref{eqn:N_02} and again pick out the terms of the form~\eqref{eqn:xBdA_id}, 
\begin{align}
\rho_{\phi h}^{\rm cl}  &=  \frac{ 4 \alpha^2 m_A^2 m_B^2 }{ M_P^4 } \left\langle    \frac{   1  }{   x_A^2 + \frac{t}{4m_A^2}  } \right \rangle + ( A \leftrightarrow B ) \nonumber \\
 &=  + 4 \alpha^4  \frac{ m_A^2 m_B^2 }{ M_P^4 } \frac{\pi M}{\sqrt{-t} } \; . 
\end{align}
The $\rho_{h \phi}^{\rm cl}$ gives an identical contribution. 

Finally, for the GR contribution, we show in the Appendix that:
\begin{align}
 \rho_{hh}^{\rm cl} = 3 (5 \sigma^2 - 1 ) \frac{m_A^2 m_B^2}{M_P^4}  \frac{\pi M}{\sqrt{t}} \; . 
\end{align}
Putting these all together, we arrive at the 2PM amplitude for a conformally coupled scalar,
\begin{align}
 a_2 (s) &= 3 ( 5 \sigma^2 - 1 ) + 8 \alpha_1^2  - 4 \alpha_1^4 \; .
 \label{eqn:a2_con}
\end{align}
The potential in this gauge can then be constructed from $a_2$ and $a_1$ using \eqref{eqn:amp_to_pot}, and we find that it takes the form~\eqref{eqn:intro_V} with, 
\begin{align}
 V^{(2)}_{\alpha^2} &= \frac{m_A m_B }{E_A E_B} \left\{ -2  + \frac{\mu}{\sqrt{s}}  \left[  \frac{4 \sigma s}{E_A E_B}  +  \frac{m_A m_B}{E_A E_B}  \left( \frac{s}{E_A E_B} -1 \right) ( 1 - 2 \sigma^2) \right] \right\} \nonumber \\
  V^{(2)}_{\alpha^4} &=  \frac{m_A m_B }{E_A E_B} \left\{ 1   - \frac{\mu}{\sqrt{s}} \frac{m_A m_B}{E_A E_B} \left( \frac{s}{2 E_A E_B} - \frac{1}{2} \right)  \right\} \; . 
  \label{eqn:v2_con}
\end{align}

\paragraph{Comparison with PN expansion.}
Expanding each of these potentials at small $p^2$ gives,
\begin{align}
 V^{(1)}_{\alpha^2} &=  -1 + (1 - 2 \nu ) \frac{p^2}{2 \mu^2} + \frac{ -3 + 12 \nu - 8 \nu^2 }{2} \left( \frac{p^2}{2 \mu^2} \right)^2   + ... \nonumber \\ 
  V^{(2)}_{\alpha^2} &=  1 + \nu   +
 \left( -3 + 11 \nu + 3 \nu^2 \right) \frac{p^2}{2 \mu^2 }
+ \frac{  15 - 81 \nu + 74 \nu^2 + 15 \nu^3 }{2}  \left( \frac{p^2}{2 \mu^2}  \right)^2+ ... \label{eqn:V_PN_con} \\ 
  V^{(2)}_{\alpha^4} &=  \frac{1 + \nu}{2}   +
 \left( 1 - 5 \nu + 3 \nu^2 \right) \frac{p^2}{2 \mu^2 }
+ \frac{ -9 + 51 \nu - 74 \nu^2 + 15 \nu^3  }{4}  \left( \frac{p^2}{2 \mu^2}  \right)^2+ ... \nonumber 
\end{align}
These PN coefficients coincide with the relevant part of the 3PN result recently reported in \cite{Julie:2022qux} (see also \cite{Bernard:2018hta, Bernard:2018ivi}), focussing on their $\alpha_A \alpha_B$ terms and performing a canonical transform (as described in Appendix~\ref{app:canon}) to map to the amplitude gauge~\eqref{eqn:amp_gauge}. 

\paragraph{Ultra-relativistic limit.}
The virtue of our PM result is that it captures all orders in velocity, and so can be used to investigate the behaviour at large $p^2$. 
Interestingly, we find in the ultra-relativistic limit $p/\mu \gg 1$, the conformal scalar interaction \emph{vanishes} as,
\begin{align}
 V_{\alpha^2}^{(1)} &\sim -\frac{ \mu M }{p^2}  &V^{(2)}_{\alpha^2} &\sim \frac{4 \mu}{p}    &V^{(4)}_{\alpha^4} &\sim \frac{ \mu M }{p^2} 
\end{align} 
By contrast, the GR potentials grow like $V_{\rm GR}^{(1)} \sim - 8 p^2 / (\mu M)$ and $V_{\rm GR}^{(2)} \sim 80 p^3/ (\mu M^2 )$. So although the scalar gives a comparable fifth force in the non-relativistic limit when $\alpha \sim \mathcal{O} (1)$, it is always subdominant at high energies, for any value of $\alpha$ (at least at this PM order).

\paragraph{Orbital precession.}
From the amplitude $a_2$, we can immediately write down the leading periastron-to-periastron precession angle using \eqref{eqn:pre_from_amp},
\begin{align}
 \Theta_2^{\rm PM} 
 &=  \theta_{\text{GR}} (\E) +  \frac{ 2 \alpha^2 -  \alpha^4  }{1 + \nu \E} \; ,
 \label{eqn:pre_con_PM}
\end{align}
where the GR contribution is given in \eqref{eqn:pre_GR}. 
Expanding at non-relativistic again generates the entire PN series at $\mathcal{O} ( \kappa/ r )$, 
\begin{align}
\Theta_2^{\rm PM}  &= ( 1 + \alpha^2 ) (3-\alpha^2)  +   \left( \frac{15}{4} + \nu \left( - \frac{3}{2}  - 2 \alpha^2 +  \alpha^4\right)  \right) \mathcal{E} + \mathcal{O} \left( \E^2 \right) \; .
\end{align}
As a consistency check, note that the leading PN coefficient in this expansion matches the known PPN result \cite{Will:2004nx}.

\paragraph{Hidden symmetry.}
One intriguing feature of the PN result is that it has a zero at $\alpha^2 = 3$. 
For this special value, the leading relativistic correction to the orbital precession vanishes. 
Ordinarily, the LRL symmetry is broken at 2PN by the relativistic effects of the metric, but for $\alpha^2 = 3$ the scalar precisely cancels these effects and the symmetry is preserved. 
In terms of the scattering amplitude, this enhanced symmetry is tied to a soft limit for the amplitude,
\begin{align}
\lim_{p^2 \to 0} a_2 = 0 \; ,
\end{align}
which is easily verified by setting $\alpha^2 = 3$ in \eqref{eqn:a2_con}. 

However, this is where our post-Minkowskian result complements the PN expansion in a particularly useful way: since the metric and scalar exchange introduce different functional dependence on the velocity, the LRL symmetry \emph{must} be broken at higher PN orders for this conformally coupled scalar-tensor theory: no tuning of $\alpha$ could even cancel the $2 \sigma^2 -1$ dependence of GR.
But it is natural to ask: could additional couplings between $\phi$ and the compact object be tuned to restore the LRL symmetry to even higher PN orders? 
In fact, by repeating the calculation with general body-dependent $C_A$ and $D_A$ in the Appendix, we find that this is not possible: when GR is modified by a single light scalar field, the LRL symmetry is inevitably broken at $2$PN and orbits always precess at this order. 
The reason is ultimately simple: from an amplitude perspective, to fully restore the symmetry at 1PM would require $a_2 (s)= 0$ at any $s$, but this cannot be achieved in a scalar-tensor theory since the tensor polarisation sum can introduce $s$ dependence in the $t$-channel discontinuity (e.g. $s^2/t$ at tree-level) while the scalar cannot. 

To achieve closed orbits beyond 1PN, one must add additional spinning fields to GR. 
Building on the closed orbit solutions which exist for maximally supersymmetric Yang-Mills \cite{Caron-Huot:2014gia, Alvarez-Jimenez:2018lff}, it was recently shown that black hole scattering in $\mathcal{N}=8$ supergravity satisfies $a_2 = 0$ for all $s$ and therefore classical orbits do not precess at 1PM \cite{Caron-Huot:2018ape}.
We explore this further in Section~\ref{sec:integrable}.

\subsection{Disformal (derivative) coupling}

Now we include the effects of a disformal coupling between matter and the dark scalar $\phi$. 
The effects of this interaction on the orbits of binary systems was recently studied using the PN expansion in a series of works \cite{Brax:2018bow, Kuntz:2019zef, Davis:2019ltc, Brax:2020vgg, Brax:2021qqo}.
The derivative nature of the interaction complicates the PN expansion since each interaction introduces at least two powers of the relative velocity $v$, and is further suppressed by $1-e^2$ for circular orbits.

Here, we consider the problem from the PM/amplitude perspective.
In terms of relativistic scattering amplitudes, the disformal interaction is simply the addition of a further quartic vertex,
\FloatBarrier
\begin{figure}[htbp!]
\centering
             \begin{tikzpicture}[baseline=-0.1cm]
			\begin{feynman}
				\vertex (a1) at (0,1) {$\tilde{p}_1^\mu$};
				\vertex (b2) at (1,0);
				\vertex (c1) at (0,-1) {$\tilde{p}_2^\mu$};
				\vertex (a3) at (2,1) {$\tilde{\ell}_1^\mu$};
				\vertex (c3) at (2,-1) {$\tilde{\ell}_2^\mu$};
				
				\diagram*{
                                (a1) -- [fermion] (b2),
                                (c1) -- [fermion] (b2),
				(b2)  -- [scalar] (a3),
				(b2)  -- [scalar] (c3),
				};					
			\end{feynman}
                \end{tikzpicture} $= \frac{ \lambda }{  M_P^2 M_{\partial}^2 } \left( 
                \tilde{\ell}_1 \cdot \tilde{p}_1 \, \tilde{\ell}_2 \cdot \tilde{p}_2 
                +
                \tilde{\ell}_1 \cdot \tilde{p}_2 \, \tilde{\ell}_2 \cdot \tilde{p}_1
                -
                2 \tilde{\ell}_1 \cdot \tilde{\ell}_2 \left(   p_1 \cdot p_2 + m^2  \right) \right) , $ \qquad
\end{figure}
\FloatBarrier
\noindent which arises due to the term $\frac{\lambda}{2 M_P^2 M_{\partial}^2 } \nabla_\mu \phi \nabla_\nu \phi \, T^{\mu\nu}$ in the linearised action. 
In the centre-of-mass frame shown in Figure~\ref{fig:FS}, this vertex contributes simply,
\begin{align}
 A_{\chi_A \chi_A \to \phi \phi} \supset  \frac{ \lambda m_A^2 t }{ 2 M_P^2 M_{\partial}^2 }  x_A^2 + \mathcal{O} ( t^2 ) 
\end{align}
Combined with the conformally coupled $\chi_B \chi_B \to \phi \phi$ of the previous section, we can then evaluate the $\mathcal{O} ( \lambda )$ part of $\rho_{\phi \phi}$ which contributes to the classical limit,
\begin{align}
 \rho_{\phi \phi} &\supset  \left(  \frac{ \lambda m_A^2 t }{ 2 M_P^2 M_{\partial}^2 } \right) \left( - \frac{2 \alpha^2 m_B^2}{M_P^2} \right)   \left\langle \frac{ x_A^2}{ x_B^2 - \frac{t}{4m_B^2}} \right\rangle  + ( A \leftrightarrow B ) \nonumber \\
 &=  - \frac{ \alpha^2 \lambda m_A^2 m_B^2 t}{ M_P^4 M_{\partial}^2 } \frac{\pi M}{\sqrt{t}} \frac{  1 - y^2 }{2} \; .
\end{align}
Note that at $\mathcal{O} ( \lambda^2 )$, the unitarity integral $\rho^{\rm cl}_{\phi \phi}$ vanishes because the disformal interaction does not introduce any poles in $x_A$ (no terms of the form~\eqref{eqn:xBdA_id}) and therefore cannot produce any $\sqrt{t}$-type discontinuity\footnote{
This is analogous to the fact that it is only triangle diagrams which contribute to the classical potential in GR \cite{Bjerrum-Bohr:2018xdl}.
}. 
We therefore find that the 2PM amplitude for a conformally- and disformally-coupled scalar is given by,
\begin{align}
 a_{2,2} (s) &= \lambda \alpha^2  ( \sigma^2 - 1 ) \; .
\end{align}
The corresponding 2PM disformal potential in \eqref{eqn:intro_V} is therefore,
\begin{align}
V^{(2)}_{\lambda \alpha^2} (p^2) = \frac{ m_A m_B }{E_A E_B} \; \frac{ 1 - \sigma^2 }{4} \; . 
\label{eqn:v2_dis}
\end{align}
Note that in the PN expansion,
\begin{align}
V^{(2)}_{\lambda \alpha^2} (p^2) = - \frac{1}{2} \frac{p^2}{ 2 \mu^2} + \frac{ 1 - 4 \nu }{2}  \left( \frac{p^2}{2 \mu^2} \right)^2 + ... 
\end{align}
Since this potential therefore $\sim p^2/r^4$, it is formally 4PN order with the usual counting. Similarly, a naive PM counting might label this effect as 4PM given the $r$ dependence, but since $M_{\partial} \ll M_P$ in most phenomenologically relevant theories we prefer to describe this potential as a $( M_{\partial} r )^{-2}$ correction to the 2PM potential. 

Interestingly, unlike its conformal counterpart, this potential does not vanish in the ultra-relativistic limit,
\begin{align}
 V^{(2)}_{\lambda \alpha^2} (p^2) \sim - \frac{p^2}{\mu M} 
\end{align}
though it does remain subdominant to GR's $p^3$ scaling.

\paragraph{Precession.}
Finally, we come to the orbital precession induced by the disformal coupling. 
The leading effect depends on which expansion is used,
\begin{align}
 \frac{\Theta_{\lambda \alpha^2}}{2\pi} = \begin{cases}
 \frac{\kappa^2}{L^2} \frac{\mu^2}{ L^2 M_{\partial}^2 } \, \theta_{\lambda \alpha^2} ( \E ) \;\; &\text{ in PM expansion},  \\ 
  \left( \frac{\kappa}{\mu \bar{r}_c} \right)^{2} \frac{1}{ \bar{r}_c^2 M_{\partial}^2 } \bar{\theta}_{\lambda \alpha^2} \;\; &\text{in PN expansion} , 
 \end{cases}
\end{align}
where $\bar{r}_c = r_c ( 1 + \alpha^2 )$ is the radius of circular orbits in this scalar-tensor theory at leading PN order.

For the PM expansion, the $\theta_{\lambda \alpha^2}$ coefficient is given by \eqref{eqn:Th_P22}, 
\begin{align}
 \theta_{\lambda \alpha^2} 
 &=  
 \frac{ P_0 (\E)  a_{2,2} (E^2) M }{ 4 E}  
 =
  - \frac{ \E^2  (2 + \nu \E )^2 (4 + 2 \E  + \nu \E^2  )^2 }{ 128 (1 + \nu \E )^3 }
 \label{eqn:pre_dis_PM}
\end{align}
In particular, due to the derivative nature of the interaction, this effect vanishes as $\E^2 \to 0$ (i.e. $p^2 \to 0$). 
Unlike the conformal coupling, a straightforward expansion of \eqref{eqn:pre_dis_PM} at small $p^2$ does \emph{not} fully capture the leading PN correction to the orbit, since it will be of the same order as the $\theta_{2,1} \sim \E^1$ and $\theta_{3,1} \sim \E^0$ terms, which we have not computed\footnote{
These depend on the disformal coupling $\lambda$ via the $a_{3,2}$ and $a_{4,2}$ coefficients, i.e. two- and three-loop Feynman diagrams containing a single disformal, two conformal vertices, and GR vertices. 
}. 
Instead, we can infer the leading PN correction directly from the potential $V^{(2)}_{\lambda \alpha^2}$ using \eqref{eqn:dTh_QC_vn'} for quasi-circular orbits, 
\begin{align}
 \bar{\theta}_{\lambda \alpha^2} 
 &=  
- 10 v_{2,2}' (0)
 =
 5
 \label{eqn:pre_dis_PN}
\end{align}
which reproduces the previous result of \cite{Brax:2018bow}.
To make the comparison more explicit, recall that with our definitions we have,
\begin{align}
\frac{ \Delta \Theta }{2 \pi}  = \frac{ G_N M }{ \bar{r}_c}  \left[
\frac{ \theta_{\rm GR} +  \alpha^2  \theta_{\alpha^2} + \alpha^4 \theta_{\alpha^4}  }{ 1 + \alpha^2 }
 + \frac{ \lambda \alpha^2 M }{ 8 \pi  M_P^2 M_{\partial}^2 \bar{r}_c^3}  \bar{\theta}_{\lambda \alpha^2} 
\right] \; . 
\end{align}
The non-relativistic limits $\{ \theta_{\rm GR}, \, \theta_{\alpha^2} , \, \theta_{\alpha^4} , \, \bar{\theta}_{\lambda \alpha^2} \} \to \{ 3, \, 2, \, -1, \, 5 \}$ therefore exactly match \cite[(4.38)]{Brax:2018bow}, since $(\beta^2/M^4)_{\rm there} = (\alpha^2 / 4 M_P^2 M_{\partial}^2)_{\rm here}$. 

The virtue of the PM amplitude calculation presented here is two-fold: firstly, it captures the full $\E$ dependence of the precession at a fixed $L$, and secondly, it is computationally simpler than integrating out degrees of freedom from a reduced Lagrangian. 
The price to be paid is that, unlike for non-derivative couplings where the amplitudes directly determine adiabatic invariants of the orbit, for derivative couplings we are still reliant on constructing a two-body potential in order to reliably determine the PN expansion of quantities like the orbital precession, and so to make contact with observations the relatively simple scattering amplitudes must be transformed back into potentials.
Nonetheless, \eqref{eqn:pre_con_PM} and \eqref{eqn:pre_dis_PM} represent an important step forward in understanding the behaviour of scalar-tensor theories, since they shed light on the high-energy behaviour of classical scattering. They also provide a concrete connection between the EFT coefficients $\alpha$ and $\lambda$ and the emergence (or breaking) of the Newtonian LRL symmetry in the fully relativistic theory.

\section{Supersymmetric (integrable) examples}
\label{sec:integrable}

Finally, we briefly comment on the possibility of extending the scalar-tensor theory considered above with additional fields in such a way that the LRL symmetry is restored and the orbital precession vanishes.

\subsection{$\mathcal{N}=4$ super Yang-Mills}

As a warm-up, consider $\mathcal{N}=4$ supersymmetric Yang-Mills.
Two-body scattering in the bosonic sector of the theory has been considered in \cite{Sakata:2017pue, Alvarez-Jimenez:2018lff}, and once all auxilliary fields are integrated out there is a remarkably simply EFT for a vector field $A_\mu$ and a complex scalar $\phi$,
\begin{align}
 \mathcal{L} = - \frac{1}{4} F^{\mu\nu} F_{\mu\nu} - \frac{1}{2} ( D_\mu \phi )^* ( D^\mu \phi ) - \frac{1}{2} \left( m - \frac{\kappa}{r}  \right) \phi^* \phi
\end{align}
where $D_\mu$ is the covariant derivative $\partial_\mu - i A_\mu$ and $r$ is the relative separation between two bodies that couple minimally to $A_\mu$.
The two-body Hamiltonian for this scalar-vector theory was found in \cite{Alvarez-Jimenez:2018lff}, and can be written as,
\begin{align}
    H [r,p] = \sqrt{ p^2 + \left( \mu - \frac{\kappa}{r} \right)^2  } -  \frac{\kappa}{r} \; . 
     \label{eqn:H_SYM}
\end{align}

\paragraph{Precession.}
Solving $H [r, p] = E$ for the $P_n$ coefficients of \eqref{eqn:p_onshell} gives,
\begin{align}
2 P_0 &= \frac{ E^2 }{\mu^2} - 1   \; ,  &P_1 &= \frac{ E}{\mu} + 1  \; ,  & P_{n > 1} &= 0 \; . 
\end{align}
Although this is not written in the amplitude gauge, our expressions for $\Theta$ and $T$ are valid in any gauge.
In particular, since $P_{n>1} = 0$ in the gauge~\eqref{eqn:H_SYM}, it immediately implies that all $\Theta_n = 0$. 
Classical orbits are therefore closed, at any order in perturbation theory.

\paragraph{Hidden symmetry.}
There must therefore be a conserved LRL vector in this theory\footnote{
The observation that a scalar-induced precession could exactly cancel a vector-induced precession and produce closed orbits (and hence restore the LRL symmetry) was also made in \cite{doi:10.2991/jnmp.1996.3.3-4.15}. 
}.
\cite{Alvarez-Jimenez:2018lff} identified this vector for the Hamiltonian~\eqref{eqn:H_SYM}, and found a modified version of the Newtonian \eqref{eqn:K_def}.
Here, we have argued that there is always a gauge in which the usual Newtonian LRL vector is conserved.
To see this explicitly, consider transforming \eqref{eqn:H_SYM} into the LRL gauge, in which $\tilde{h}_1 = 0$. 
We find that the particular structure of \eqref{eqn:H_SYM} is such that all $\tilde{h}_{n > 1} = 0$ also vanish, and so in this gauge the Hamtilonian becomes a function of $H_N$ only,
\begin{align}
 H = H_N + \frac{1}{2} H_N^2 + \frac{1}{4} H_N^3 + \frac{1}{8} H_N^8 + ... 
\end{align}
Since the LRL vector $\bfK$ defined in \eqref{eqn:K_def} commutes with $H_N$, we see that in this gauge $\bfK$ is also conserved in the fully relativistic theory. 
As described in \cite{Caron-Huot:2014gia}, this is a consequence of the dual conformal invariance of $\mathcal{N}=4$ super-Yang Mills, and also related to the so-called ``no-triangle property'' (certain triangle Feynman diagrams vanish) in perturbation theory.

\subsection{$\mathcal{N}=8$ supergravity}

Now a brief comment about $\mathcal{N} = 8 $ supergravity. 
The addition of supergravity's graviphoton enables the cancellation of precession in bound orbits at any PN order (unlike the scalar-tensor theory above, which could only cancel the precession at leading PN order).

The details of the scattering problem are given in \cite{Caron-Huot:2018ape}. In short, half-BPS (extremal) black hole states in this theory are characterised by a mass and a number of electric charges (we set the magnetic charges to zero). 
The simplest non-trivial scattering problem is to consider a pair of such black holes with masses $m_A$ and $m_B$ and with their electric charge vectors misaligned by an angle $\varphi$.
Then the 1PM (tree-level) amplitude is given by \cite{Caron-Huot:2018ape},
\begin{align}
 a_1 (s) =  4 ( \sigma - \cos \varphi )^2  
 \label{eqn:a1_SUGRA}
\end{align}
where note that the force vanishes in the static limit $\sigma \to 1$ when the two charge vectors are aligned. 
The 2PM (one-loop) scattering amplitude was also found by \cite{Caron-Huot:2018ape}, and the classical part vanishes identically
\begin{align}
 a_2 (s) = 0 \; 
\end{align}
for any value of $\varphi$. 
The classical orbits of these extremal black holes are therefore closed ellipses at this 2PM order\footnote{
Note, however, that the resulting closed orbit is \emph{not} simply the Newtonian one, since the period still experience time dilation due to the non-zero $a_1$. 
}.

We reproduce these results here in order to ask the following question.
The vanishing of $a_2$ at one-loop implies no precession and a conserved LRL vector at 1PM, but what happens at higher PM orders? 
From \eqref{eqn:pre_from_amp_PM}, we see that to have no precession at 4PM in this theory we must have,
\begin{align}
 a_4 = - \frac{a_1 a_3}{ a_0}
 \label{eqn:LRL_boot_a4}
\end{align}
where $a_0$ is given in \eqref{eqn:a0_def} and $a_1$ is given in \eqref{eqn:a1_SUGRA}. 
The 3PM (two-loop) amplitude for extremal black hole scattering was recently calculated in \cite{Parra-Martinez:2020dzs}, and we reproduce the classical part of it here for convenience:
\begin{align}
a_3 (s)  = 16 ( \sigma - \cos \phi )^4 \left[ \frac{ (m_1^2 + m_2^2 + 2 \sigma m_1 m_2 ) ( \sigma - \cos \phi )^2  }{ 2 m_1 m_2 ( \sigma^2 - 1 )^2 } - 2 \frac{\text{arcsinh} \left( \sqrt{\frac{\sigma-1}{2} } \right) }{ \sqrt{\sigma^2 - 1}}   \right]  \;  .
\end{align}
So while the three-loop amplitude $a_4$ is not currently known, if the LRL (dual conformal) symmetry remains unbroken at that order then \eqref{eqn:LRL_boot_a4} gives a unique prediction for its classical part (the coefficient of $\sqrt{t}$). 
And this pattern continues to higher orders: an exact LRL symmetry in supergravity would mean that every even PM order is effectively fixed in terms of lower orders. 
It would be interesting to use this fact to explore the universality of ultra-relativistic scattering \cite{Bern:2020gjj}.

As a final comment, we note that it is not yet known whether orbits in supergravity really do remain closed at higher PM orders. 
It seems likely, given the connection between precession and dual conformal invariance. 
And there is not yet any evidence to the contrary: in fact, it was recently shown in \cite{deNeeling:2023egt} that, at least up to 5PN for general masses (and to all PN orders in the test mass limit), the two-body Hamiltonian can be mapped to a Kepler-like problem with conserved LRL vector by a suitable canonical transformation.  
Nonetheless, neither of these constitute a proof, and it would be interesting to explicitly compute the 4PM amplitude $a_4$ and compare with the prediction above to verify whether the precession indeed vanishes at higher orders.

\section{Discussion}
\label{sec:disc}

These results pave the way for a more accurate modelling of gravitational waveforms in scalar-tensor theories. 
Since such models are ubiquitous in high-energy model building, this will improve our ability to search for new fundamental physics with current and future gravitational wave data. 
Particularly for the disformal coupling, for which there is still a relatively large region of viable parameter space, a comparison with current and future gravitational wave measurements could probe this coupling in a qualitatively new range of scales.  \\

Furthermore, on the theory side, the emergence of a new symmetry in these systems raises a number of interesting questions. For instance: what other phenomenological consequences does it have; whether it is radiatively stable; whether it could be further enhanced beyond 2PN by a suitable tuning of addition interactions or fields (without going all the way to supergravity); and so on.  
Completing the full 2PM calculation also sheds light on the validity of perturbative expansions involving the disformal coupling, and in particular the range of relative velocities for which the disformal coupling may dominate over the conformal coupling (and yet remain under perturbative control).  \\

We should stress that we are not proposing the scalar-tensor theory~\eqref{eqn:geff_def} with $\alpha^2 \approx 3$ as a realistic model of our Universe. For one thing, fifth force constraints from Solar System probes such as Cassini \cite{Bertotti:2003rm} effectively constrain $|\alpha | \lesssim 10^{-5}$. 
Rather, we believe this scalar-tensor theory provides an interesting proof of principle. 
Supergravity enjoys a sufficiently large degree of symmetry that many difficult problems become tractable, but the price to be paid is that it introduces many new degrees of freedom beyond GR. 
One might wonder: if we make the most minimal modification and add just a single scalar degree of freedom to GR, how much of this symmetry could we recover? 
The tuning of \eqref{eqn:geff_def} we have described for which the precession vanishes at 1PN provides one answer to this question: it corresponds to a simple scalar-tensor theory which, at least at leading PN orbit, shares the closed orbits and LRL symmetry (dual conformal invariance) of $\mathcal{N}=8$ supergravity.  \\

There are a number of promising future directions, including

\begin{itemize}

\item[(i)] to find a Lagrangian formulation of the LRL symmetry which is restored in this scalar-tensor theory at $\alpha^2 = 3$. 

\item[(ii)] to use the $t$-channel dispersion relation to derive sum rules for the precession and period in terms of new UV physics/states (along the lines of \cite{Brax:2017xho, Banks:2020gpu}),

\item[(iii)] to include spin for the compact objects, along the lines of the PN analysis in \cite{Brax:2021qqo} which identified the leading disformal coupling to the binary's spin and orbital angular momentum,

\item[(iv)] to go to higher PM orders: one of the main takeaways from the above should be that adding a scalar is relatively straightforward in this amplitude framework, and since 3PM and 4PM GR amplitudes are already known it should be possible to include a scalar to that order as well,

\item[(v)] to include the dissipative effects from radiation, including the emitted power and gravitational wave phase evolution, to make contact with waveform templates. For instance, this could be done in the PM framework by adapting the EFT approach of \cite{Kalin:2022hph} to include non-minimal coupling to a scalar field. 

This is beyond the scope of the current investigation and is left for the future. \\

\end{itemize}

\paragraph{Acknowledgements.}
S.M. is supported by a UKRI Stephen Hawking Fellowship (EP/T017481/1), and thanks Diederik Roest for useful discussions at an early stage in the project. 
This work has been partially supported by STFC consolidated grant ST/T000694/1.

\appendix
\section{Canonical transformation details}
\label{app:canon}

In this Appendix we detail the canonical transformations which move between the different gauges described in Section~\ref{sec:overview}. 
Concretely, a canonical transformation is a redefinition of the form,
\begin{align}
	\bfr &\to \bfr + \{ X , \bfr \} + \frac{1}{2!} \{ X , \{ X , \bfr \} \} + \frac{1}{3!} \{ X , \{ X , \{ X , \bfr \} \} \} + ... \\
	\bfp &\to \bfp + \{ X , \bfp \} + \frac{1}{2!} \{ X , \{ X , \bfp \} \} +\frac{1}{3!}\{ X,  \{ X , \{ X , \bfp \} \}  \} +  ... 
\end{align}
which preserves the Poisson bracket $\{ \bfr_i, \bfp_j \} = \delta_{ij}$. 
These are the classical analogue of unitary transformations in quantum mechanics, $\mathcal{O} \to e^{X} \mathcal{O} e^{-X}$. 
The time evolution of these new variables is described by the Hamiltonian,
\begin{align}
	H \to H + \{ X , H \} + \frac{1}{2!} \{ X , \{ X , H \} \} + \frac{1}{3!} \{ X,  \{ X , \{ X , H \} \}  + ... 
	\label{eqn:H_can_trans}
\end{align}

$X$ can be written in the same PM/PN form as $H$,
\begin{align}
 X [r, p ] = \left( \sum_{n=0}  \epsilon_n \left( \frac{p^2}{2 \mu^2} , \frac{p_r^2}{2 \mu^2} \right) \left( \frac{\kappa}{\mu r} \right)^n \right) \frac{ \bfp \cdot \bfr }{\mu^2}  =  \left( \sum_{ \substack{ n=0 \\ a,b = 0 } }  \epsilon_n^{(a,b)} \left( \frac{p^2}{2 \mu^2} \right)^a \left( \frac{p_r^2}{2 \mu^2} \right)^b \left( \frac{\kappa}{\mu r} \right)^n \right) \frac{ \bfp \cdot \bfr }{\mu^2} 
 \label{eqn:X_def}
\end{align}
where $\bfp \cdot \bfr$ is the generator of the identity transformation (i.e. $\bfr \to e^{\epsilon_0^{(0,0)}} \bfr$, $\bfp \to e^{\epsilon_0^{(0,0)}} \bfp$ ). 
Once $\mu$ and $\kappa$ are fixed, we henceforth set $\epsilon_0^{(0,0)} = 0$. 
The canonical transformation~\eqref{eqn:H_can_trans} of the Hamiltonian \eqref{eqn:H_PM} with this $X$ corresponds to a redefinition of the $H_n^{(a,b)}$ coefficients. 

\paragraph{1PN.}
At leading Post-Newtonian order, the most general canonical transformation depends on just two parameters, $\{ \epsilon_0^{10} , \epsilon_1^{00} \}$. 
It implements the following redefinition of the 1PN Hamiltonian\footnote{
As a useful consistency check, note that the sum of the $n$-indices in each term must match, and the sum of all $a, b$ and $n$ indices in each term must match.
},
\begin{align}
	H_{0}^{20} &\to H_{0}^{20} + 2 H_{0}^{10} \epsilon_{0}^{10}  \nonumber \\ 
	H_{1}^{10} &\to H_{1}^{10} + H_{1}^{00} \epsilon_{0}^{10} + 2 H_{0}^{10} \epsilon_{1}^{00}  \nonumber \\
	H_{1}^{01} &\to H_{1}^{01} + 2 H_{1}^{00} \epsilon_{0}^{10} - 2 H_{0}^{10} \epsilon_{1}^{00} \nonumber \\ 
	H_{2}^{00} &\to H_{2}^{00} + H_{1}^{00} \epsilon_{1}^{00} 
	\label{eqn:1PN_can_tran}
\end{align}
Various expressions for the Hamiltonian are given in the literature (for both GR and ST theories), and since they use a variety of different co-ordinate systems one can find a variety of different values quoted for the $H_{n}^{ab}$ coefficients. The physically relevant (gauge-invariant) quantities at each order are those which are invariant under \eqref{eqn:1PN_can_tran}. 
There are only two such combinations at 1PN, which we can take for instance to be,
\begin{align}
	\tilde{b}_{0}^{2} &= 2 H_{1}^{10} + 2 H_{1}^{01} - 3 H_{0}^{20}   \; , \;\; 
	&\tilde{b}_{2}^{0} &= H_{2}^{00} - H_{1}^{10} - \tfrac{1}{2} H_{1}^{01} + H_{0}^{20} \; ,  
\end{align}
or any combination thereof. 
These correspond to the (PN expansion of the) potential coefficients in the LRL gauge \eqref{eqn:LRL_gauge}.
The coefficients in the amplitude gauge are given by,
\begin{align}
	v_0^{2} &= \frac{3\nu-1}{2} \; ,  &v_{1}^1 &=  H_1^{10} + H_1^{01} - \frac{3}{2} H_0^{20}  + \frac{ 9 \nu - 3 }{4}   \; , \;\; &v_{2}^0 &=   H_2^{00} + \frac{1}{2}  H_1^{01}  - \frac{1}{2}H_0^{20} + \frac{3 \nu - 1 }{4} \; . 
\end{align}

Finally, note that within a general isotropic gauge \eqref{eqn:isotropic_gauge}, we have the residual gauge freedom $\epsilon_0^{01} = \epsilon^{10}_0 \equiv \epsilon^1_0$ which does not regenerate any $p_r$ dependence, and therefore implements the redefinition,
\begin{align}
  h_0^2 &\to h_0^2 + 2 \epsilon_0^1   \; , &h_1^1 &\to h_1^1 + 3 \epsilon_0^1  \; ,  &h_2^0 &\to h_2^0 + \epsilon_0^1 \; . 
\end{align}
This can be used to set $h_0^2$ to any desired value, e.g. the amplitude gauge value $v_2^0$.

\paragraph{3PN.}
At third PN order, the general canonical transformation depends on four parameters, $\{  \epsilon_0^{20} , \epsilon_1^{10} , \epsilon_1^{01} , \epsilon_2^{00}  \}$, and implements the following redefinition of the Hamiltonian, 
\begin{align} 
	H_{0}^{30} &\to H_0^{30} + 4 H_{0}^{20} \epsilon_{0}^{10} + 4 H_{0}^{10} \epsilon_{0}^{10} \epsilon_0^{10} + 2 H_{0}^{10} \epsilon_{20} \nonumber \\ 
	H_{1}^{20} &\to  
			H_{1}^{20} + 4 H_{0}^{20} \epsilon_{1}^{00} + 3 H_{1}^{10} \epsilon_{0}^{10} + 7 H_{0}^{10} \epsilon_{1}^{00} \epsilon_{0}^{10} + \frac{3}{2} H_{1}^{00} \epsilon_{0}^{10} \epsilon_0^{10}  + 2 H_{0}^{10} \epsilon_{1}^{10} + H_{1}^{00} \epsilon_{0}^{20} \nonumber \\ 
		H_{1}^{11} &\to H_{1}^{11}- 4 H_{0}^{20} \epsilon_{1}^{00} + 2 H_{1}^{10} \epsilon_{0}^{10} - H_{1}^{01} \epsilon_{0}^{10} - 4 H_{0}^{10} \epsilon_{1}^{00} \epsilon_{0}^{10} - 
			2 H_{0}^{10} \epsilon_{1}^{10} + 6 H_{0}^{10} \epsilon_{1}^{01} + 4 H_{1}^{00} \epsilon_{0}^{20}  \nonumber \\ 
	H_{1}^{02} &\to H_{1}^{02} + 6 H_{1}^{01} \epsilon_{0}^{10} - \frac{3}{2}  H_{0}^{10} \epsilon_{1}^{00} \epsilon_{0}^{10} + 6 H_{1}^{00} \epsilon_{0}^{10} \epsilon_0^{10}  - 
			6 H_{0}^{10} \epsilon_{1}^{01}  \nonumber \\ 
	H_{2}^{10} &\to H_{2}^{10} + 3 H_{1}^{10} \epsilon_{1}^{00} + 3 H_{0}^{10} \epsilon_{1}^{00} \epsilon_1^{00} + 2 H_{0}^{10} \epsilon_{2}^{00} + 2 H_{2}^{00} \epsilon_{0}^{10} 
	+ \frac{5}{2}  H_{1}^{00} \epsilon_{1}^{00} \epsilon_{0}^{10}  + H_{1}^{00} \epsilon_{1}^{10} \nonumber \\ 
	H_{2}^{01} &\to H_{2}^{01} - 2 H_{1}^{10} \epsilon_{1}^{00} + H_{1}^{01} \epsilon_{1}^{00} - 3 H_{0}^{10} \epsilon_{1}^{00} \epsilon_1^{00} - 4 H_{0}^{10} \epsilon_{2}^{00} + 
			4 H_{2}^{00} \epsilon_{0}^{10} + 2 H_{1}^{00} \epsilon_{1}^{00} \epsilon_{0}^{10} + 2 H_{1}^{00} \epsilon_{1}^{10} + 3 H_{1}^{00} \epsilon_{1}^{01} \nonumber \\ 
	H_{3}^{00} &\to H_{3}^{00} + 2 H_{2}^{00} \epsilon_{1}^{00} + H_{1}^{00} \epsilon_{1}^{00} \epsilon_1^{00} + H_{1}^{00} \epsilon_{2}^{00}  \label{eqn:2PN_can_tran}
\end{align}
There are now three independent combinations that are invariant under this transformation, which we can take to be,
\begin{align}
	\tilde{b}_{0}^{3} &=  5 ( H_1^{10} + H_1^{01} - 2 H_0^{20} )^2 + 2 ( H_1^{20} + H_1^{11}  + H_1^{02} ) - 5 H_0^{30}  \; , \;\; \nonumber \\
	\tilde{b}_{2}^{1} &=  -5 H_1^{10} H_1^{10} - 7 H_1^{10} H_1^{01} - \frac{19}{8}  H_1^{01}  H_1^{01}  + H_2^{10} + \frac{1}{2} H_2^{01}  + 
  4 H_2^{00} ( H_1^{10} + H_1^{01} - 2 H_0^{20} ) \nonumber \\
  &\;\; + 16 H_1^{10} H_0^{20} + 10 H_1^{01} H_0^{20} - 12 H_0^{20} H_0^{20} - 
  2 H_1^{20}  - H_1^{11} - \frac{3}{4} H_1^{02}  + 3 H_0^{30} \; ,    \nonumber \\
	\tilde{b}_3^0 &= H_{3}^{00} - H_{2}^{00} ( H_{1}^{10} - 2 H_{0}^{20} ) + 
  H_{1}^{10} H_{1}^{10} + \frac{1}{4} H_{1}^{10} H_{1}^{01} + \frac{1}{16} H_{1}^{01} H_{1}^{01} - \frac{1}{2} H_{0}^{20} (6 H_{1}^{10} + H_{1}^{01}) + 
     2 H_{0}^{20} H_{0}^{20} \nonumber \\ 
     &\;\; + \frac{1}{8} (-8 H_{2}^{10} - 2 H_{2}^{01} + 8 H_{1}^{20} + 2 H_{1}^{11} + H_{1}^{02} - 8 H_{0}^{30} )
\end{align}
or any combination thereof. 
These correspond to the (PN expansion of the) potential coefficients in the LRL gauge \eqref{eqn:LRL_gauge}.
The coefficients in the amplitude gauge are given by,
\begin{align}
v_{0}^3 &= \frac{1}{2} (1  - 5\nu + \nu^2 )   \nonumber \\
v_1^{2} &= \frac{5}{32} (7 + -34 \nu + 31 \nu^2 ) + \frac{1}{2} \tilde{b}_0^{3} - \frac{5}{8} \tilde{b}_0^2 \left(  \tilde{b}_0^2 + 3\nu -1 \right)  \nonumber \\ 
v_2^1 &= \frac{1}{4} ( \nu -1 ) ( 11 \nu -3 ) + \tilde{b}_0^3 + \tilde{b}_2^1 - \tilde{b}_0^2 \tilde{b}_0^2 -2 \tilde{b}_0^2 \tilde{b}_2^0 + (3 \nu - 1 ) \left( \tilde{b}_0^2 + \tilde{b}_2^0 \right)   \nonumber \\ 
v_3^0 &=  \frac{1}{32} (5 + -22 \nu + 13 \nu^2 ) + \frac{1}{2} \tilde{b}_0^3 + \tilde{b}_2^1 + \tilde{b}_3^0 - \frac{3}{8} \tilde{b}_0^2 \tilde{b}_0^2 - \frac{3}{2} \tilde{b}_0^2 \tilde{b}_2^0 + \frac{3}{8}  (3 \nu -1 ) \left( \tilde{b}_0^2 + 2 \tilde{b}_2^0  \right)
\end{align}

Finally, within a general isotropic gauge there is a residual gauge freedom,
\begin{align}
h_0^{3} &\to h_0^{3} + 2 \epsilon_0^2 + 4 \epsilon_1^{0} ( h_0^{2}  + \epsilon_1^0 )   \nonumber \\ 
h_1^{2} &\to h_1^2 + 5 \epsilon_0^2 + 5 \epsilon_1^0 \left( h_1^1   + \frac{3}{2} \epsilon_1^0 \right)   \nonumber \\
h_2^1 &\to h_2^1 + 4 \epsilon_0^2 + 
    4 \epsilon_1^0  ( h_2^0  + h_1^1 - h_0^2  + \epsilon_1^0 )   \nonumber \\
h_3^0 &\to h_3^0  + \epsilon_0^2 +  
    \frac{1}{2} \epsilon_1^0  (6 h_2^0  - 2 h_0^2 + h_1^0 )
\end{align}
which can be used to set e.g. $h_0^3$ to any desired value.

\paragraph{Scalar-Tensor Theory.}
Finally, in order to compare our results with the recent 3PN calculation of \cite{Julie:2022qux}, we write the $H_n^{(a,b)}$ coefficients of their Hamiltonian and transform them explicitly into the amplitude gauge.
At 0PN, the dynamics is governed by a single quantity: the effective Newton's constant,
\begin{align}
    G = 1 + \alpha^2 \; . 
\end{align}
At 1PN, the dynamics is governed by just two functions of $\alpha$ and $\beta$, 
\begin{align}
    \gamma = \frac{- 2 \alpha^2}{1 + \alpha^2} \; , \;\; \bar{\beta} = \frac{\beta \alpha^2}{  ( 1 + \alpha^2 )^2}
\end{align}
in addition to $G$. Explicitly, the Hamiltonian coefficients are,
\begin{align}
 H_0^{10} &= 1  \; , &H_1^{00} &= -G   \\ 
H_0^{20} &= \frac{3 \nu - 1}{2}  \; , & H_1^{10} &= -G ( 3 + 2 \gamma + \nu ) \; , &H_1^{01} &=  -G \nu \;  ,  &H_2^{00} &= -G^2 \frac{1+\beta}{2}  \nonumber 
\end{align}
At 2PN, the relevant coefficients which overlap with our 2PM expansion are,
\begin{align}
H_0^{30} &= \frac{1}{2} (1 - 5 \nu + 5 \nu^2)  \; ,  \nonumber \\
 H_1^{20} &= - \frac{G}{2} (-5 - 4 \gamma (1 - 4 \nu) + 22 \nu + 3 \nu^2) \; ,   \nonumber \\
H_1^{11} &= -G \nu (\nu-1)   \;  ,  \nonumber \\ 
H_1^{02} &= - \frac{3G }{2} \nu^2 \; ,  \nonumber \\
H_2^{10} &= \frac{G^2}{4} (22  + 28 \gamma + 9 \gamma^2 + 58 \nu + 36 \gamma \nu )   \; ,  \nonumber \\ 
 H_2^{01} &= \frac{G^2}{4} ( - \gamma^2 - 4 \gamma (1 + 6 \nu) - 4 (1 + 8 \nu - 3 \bar{\beta} \nu))
 \label{eqn:PN_coefs_ST}
\end{align}
Transforming these to the amplitude gauge gives the result \eqref{eqn:V_PN_con} found in the main text.

Note that \cite{Julie:2022qux} also provides the full 3PN Hamiltonian for general body-dependent conformal couplings, and these can also be mapped to the amplitude gauge using the above formulae. 
In particular, we will show below that just two,
\begin{align}
 v_1^0 &= - 1 - \alpha_A \alpha_B \; , 
 &v_2^0 &= ( 1 + \alpha_ A \alpha_B )^2 \, \frac{1 + \nu + \bar{\beta}_+ - m_- \bar{\beta}_- }{2} 
 \label{eqn:PN_coefs_ST_2}
\end{align}
are enough to completely fix the full 2PM scalar-tensor potential.

\section{Scattering amplitude details}
\label{app:amp}

Here we provide a longer, more pedagogical, account of the Lippmann-Schwinger equation and Feinberg-Sucher trick.
As an illustrative example, we also review the calculation for Maxwell electromagnetism and General Relativity. 
These theories have of course been studied extensively and all of the above ingredient are well-known and have been verified through various methods---this makes it a useful consistency check of our general relations. 
The amplitudes given there also form part of the more general scalar-tensor calculations performed in the final section.

\subsection{Lippmann-Schwinger equation}

\paragraph{Two-particle scattering.}
The idea is to momentarily promote the Hamiltonian to a quantum mechanical operator\footnote{
The precise operator ordering will not matter, since different choices will produce the same classical limit at the end of the calculation.
},
$\hat{H} = \hat{H}_0  +  \hat{V}$. 
The eigenstates of the free Hamiltonian $\hat{H}_0$ are simply momentum eigenstates: if $| \bfp_1 ... \bfp_n \rangle$ represents a state of $n$ particles with the momenta indicated, then $\hat{H}_0 | \bfp_1 ... \bfp_n \rangle = \sum_{a=1}^n E_a | \bfp_1 ... \bfp_n \rangle$ with an energy $E_a ( p^2_a ) = \sqrt{ p_a^2 + m_a^2}$ for each particle. 
As a result of the interactions in $\hat{V}$, an initial two-particle state $| \bfp_1 \bfp_2 \rangle$ at $t = -\infty$ will evolve into a new state $\hat{S} | \bfp_1 \bfp_2 \rangle = | \bfp_1 \bfp_2 \rangle + i \hat{T} | \bfp_1 \bfp_2 \rangle$ at $t= + \infty$. 
Providing the underlying field theory interactions are invariant under spacetime translations, the $\hat{S}$ and $\hat{T}$ operators conserve both spatial momentum and energy. Their matrix elements will therefore contain a Dirac $\delta$ function, and it is convenient to factor this out explicitly and define the scattering amplitude as,
\begin{align}
    \langle \bfp_3 \bfp_4 | \hat{T} | \bfp_1 \bfp_2 \rangle = (2 \pi )^{d+1} \delta^{d} \left( \bfp_1 + \bfp_2 - \bfp_3 - \bfp_4 \right) \delta \left( E_1 + E_2 - E_3 - E_4 \right) \, A ( \bfp_1 , \bfp_2 ; \bfp_3 , \bfp_4 ) \; .
    \label{eqn:A_def}
\end{align}
On the other hand, since $\hat{V}$ and $\hat{H}_0$ do not generally commute, a matrix element of $\hat{V}$ will conserve spatial momentum but not necessarily energy, so we write,
\begin{align}
    \langle \bfp_3 \bfp_4 | \hat{V} | \bfp_1 \bfp_2 \rangle = (2 \pi )^{d} \delta^{d} \left( \bfp_1 + \bfp_2 - \bfp_3 - \bfp_4 \right) \; \tilde{V} ( \bfp_1 , \bfp_2 ; \bfp_3 , \bfp_4 ) \; ,
    \label{eqn:V_def}
\end{align}
without the energy-conserving delta function. 
The goal is then to relate the matrix element $A ( \bfp_1, \bfp_2 ; \bfp_3 \bfp_4 )$, which can be computed via the usual Feynman diagram machinery, to the matrix element $\tilde{V} (\bfp_1, \bfp_2 ; \bfp_3 , \bfp_4 )$, which is the quantum mechanical version of the potential~\eqref{eqn:V_PM}.

\paragraph{Time evolution.}
To connect these matrix elements of $\hat{T}$ and $\hat{V}$, we use the fact that $\hat{S}$ can be written as,
\begin{align}
	\hat{S} = \hat{U} ( + \infty, - \infty )
\end{align}
where $\hat{U} (t_+, t_-)$ is the time evolution operator from time $t_-$ to time $t_+$.
In the interaction picture, the operators $\hat{V}$ and $\hat{U}$ are given an additional time dependence from the free part of the Hamiltonian, 
\begin{align}
	\hat{U} (t_+, t_- ) &= e^{i \hat{H}_0 t_+} e^{- i \hat{H} (t_+ - t_- )} e^{- i \hat{H}_0 t_-} \; , \;\; 
	&\hat{V} (t) &= e^{i \hat{H}_0 t} \hat{V} e^{-i \hat{H}_0 t}
\end{align}
and therefore satisfy the differential equation,
$i \tfrac{d}{dt} \hat{U} (t, t_- ) = \hat{V} (t) \hat{U} (t, t_- )$, 
which has an immediate integral solution,
\begin{align}
	\hat{U} (t_+, t_- ) = 1  - i \int_{t_-}^{t_+} d t \, \hat{V} (t)  \, \hat{U} (t, t_-) \; ,
	\label{eqn:U_from_V}
\end{align}
since $\hat{U} = 1$ is the trivial evolution of the free theory in this picture.  

To evaluate the two-particle matrix element of \eqref{eqn:U_from_V}, it is useful to define an object which is closely related to the amplitude by writing $\hat{U} ( t_+ , -\infty ) = 1 + i \hat{T} (t_+)$ at a \emph{finite} $t_+$. 
The analogue of \eqref{eqn:A_def} is then,
\begin{align}
	\langle \bfp_3 \bfp_4 |  \hat{T} (t_+ ) | \bfp_1 \bfp_2 \rangle = (2 \pi )^d \delta ( \bfp_1 + \bfp_2 - \bfp_3 - \bfp_4 ) \, \int_{-\infty}^{t_+} dt \;  e^{-i t ( E_1 + E_2 - E_3 - E_4 ) } \; \tilde{A} ( \bfp_1 , \bfp_2 ; \bfp_3 , \bfp_4 ) \; . 
	\label{eqn:tA_def}
\end{align}
The time-dependence here is fixed by the Ward identity of time translations, together with the boundary condition that $\hat{U} ( - \infty, - \infty) = 1$.
Note in particular that energy is not conserved at a finite $t_+$, which is essentially a consequence of the time-energy uncertainty relation.
However, taking $t_+ \to + \infty$ removes the finite cut-off in time, restoring manifest time-translation invariance and producing an energy-conserving $\delta$-function. 
In that limit, comparing with \eqref{eqn:A_def} shows that $\tilde{A}$ is simply the usual amplitude $A$ but \emph{without energy conservation}\footnote{
This $\tilde{A}$ object will be useful in the intermediate steps below, but ultimately we would like to impose energy conservation and get back to $A$ for any final result. 
This is particularly since, unlike $A$, this $\tilde{A}$ is \emph{not} uniquely defined by a Lagrangian, since it is sensitive to total time derivatives. These no longer vanish in the action due to the presence of a temporal boundary, and instead produce terms in $\tilde{A}$ proportional to $(E_1 + E_2 - E_3 - E_4)$. 
}.

\paragraph{Lippmann-Schwinger equation.}
To evaluate the two-particle matrix element of \eqref{eqn:U_from_V} with $t_- \to - \infty$, we insert a complete set of states between $\hat{V}$ and $\hat{U}$ and then use the integral identity, 
\begin{align}
 \int_{-\infty}^{t} d t' e^{- i E t'} = \frac{i e^{- i E t}}{ E + i \epsilon}
 \label{eqn:exp_int}
\end{align}
where the $+ i \epsilon$ (with $\epsilon > 0$) arises from the contour prescription which ensures that the time integral converges at $t' \to - \infty$. 
Upon factoring out momentum and an overall time integral, this produces from \eqref{eqn:U_from_V} the Lippmann-Schwinger equation,
\begin{align}
    \tilde{A} ( \bfp_1, \bfp_2 ; \bfp_3, \bfp_4 ) 
    = 
    - \tilde{V} ( \bfp_1, \bfp_2 ; \bfp_3, \bfp_4 )  
    + 
    \sum_N \frac{ \tilde{A} ( \bfp_1, \bfp_2 ; N ) \tilde{V} ( N ; \bfp_3, \bfp_4)     }{ E_1 + E_2 - E_N + i \epsilon } 
    \label{eqn:LS}
\end{align}
where $| N \rangle$ is a complete set of $\hat{H}_0$ eigenstates (where $\hat{H}_0 | N \rangle = E_N | N \rangle$) and the sum implicitly includes integrals over all continuous quantum numbers, subject to the constraint that the total momentum of $| N \rangle$ is $\bfp_1 + \bfp_2 = \bfp_3 + \bfp_4$.

\paragraph{Conservative LS equation.}
So far, we have not made any approximations: \eqref{eqn:LS} is exact, and describes the full non-perturbative scattering amplitude in quantum field theory. 
However, it also involves an infinite number of amplitudes, for all $2 \to N$ processes. 
So to proceed, we make our first assumption: we focus on the ``conservative'' part of the problem, which is the effect of the $N =2$-particle states only (this is often called the ``elastic" contribution in particle physics).
Then \eqref{eqn:LS} involves only the two-particle matrix elements \eqref{eqn:tA_def} and \eqref{eqn:V_def}, and simplifies into\footnote{
Note that while the full theory will contain both the massive particles we are scattering and various massless force-carriers (e.g. graviton/photon), for the elastic contribution to \eqref{eqn:LS} we focus on the states containing two massive particles only. This is because any other number of massive particles would represent an annihilation/creation of the compact objects in the binary (not a classical effect), and any number of additional force-carriers would correspond to radiation/dissipative effects.  
},
\begin{align}
    \tilde{A} ( \bfp_1, \bfp_2 ; \bfp_3, \bfp_4 ) 
    = 
    - \tilde{V} ( \bfp_1, \bfp_2 ; \bfp_3, \bfp_4 )  + \int \frac{d^d \bfk}{ (2\pi )^d} \frac{ \tilde{A} ( \bfp_1, \bfp_2 ; \bfk , \bfk' ) \tilde{V} (  \bfk , \bfk' ; \bfp_3, \bfp_4 )   }{ E_A ( p_1^2) + E_B (p_2^2) - E_A ( k^2 ) - E_B ( k'^2 ) + i \epsilon } \; ,
	\label{eqn:cons_LS}
\end{align}
where $\bfk' = \bfp_1 + \bfp_2 - \bfk$ is fixed by momentum-conservation. 

The amplitude is therefore equal to the potential in momentum space, up to an integral correction which is often called the ``Born subtraction''. 
As an aside, the physical reason for this subtraction is that, quantum mechanically, the interaction potential changes not just the dynamical evolution but also what we identify as a particle eigenstate. 
Concretely, suppose that we introduce another ``two-particle state'' which is an eigenstate of the full Hamiltonian, $\hat{H} | \Psi_{\bfp_1 \bfp_2} \rangle = (E_1 + E_2) | \Psi_{\bfp_1 \bfp_2} \rangle$, and coincides with the free theory eigenstate $| \bfp_1 \bfp_2 \rangle$ in the far past (where the interactions can be neglected by the adiabatic hypothesis).  
It is straightforward to verify that,
\begin{align}
    | \Psi_{\bfp_1 \bfp_2} \rangle = | \bfp_1 \bfp_2 \rangle  +   \frac{1}{  E_1 + E_2 - \hat{H}_0 + i \epsilon }  \hat{V} | \Psi_{\bfp_1 \bfp_2} \rangle
\end{align}
is both an eigenstate of $\hat{H}$ and satisfies the desired boundary condition (where again the $+i \epsilon$ is needed to ensure convergence of the far past limit). 
Consequently, the Lippmann-Schwinger equation shows that $\tilde{A} ( \bfp_1 \bfp_2 ; \bfp_3 \bfp_4)$ is proportional to $\langle \bfp_3 \bfp_4 | \hat{V} | \Psi_{\bfp_1 \bfp_2} \rangle$, and the Born subtraction in \eqref{eqn:cons_LS} is precisely accounting for this different in-state compared to \eqref{eqn:V_def}.

\paragraph{Two-particle kinematics.}
To make contact with the reduced two-body problem, we will now specialise to the centre-of-mass frame where the on-shell 4-momenta of each particle are given by \eqref{eqn:p_CoM}, as shown in Figure~\ref{fig:FS}. 
Providing the underlying interactions are invariant under spatial rotations, the functions $\tilde{A}$ and $\tilde{V}$ can depend on at most three independent scalars built from the momenta $\bfp$ and $\bfp'$. 
We chose $p^2 = | \bfp |^2$, $p'^2 = | \bfp' |^2$ and $q^2 = | \bfp - \bfp'|^2$.
Furthermore, the amplitude $A$ is defined on the support of an energy-conserving $\delta$-function, which enforces that $p'^2 = p^2$ since $m_1 = m_3$ and $m_2 = m_4$.  
So in the centre-of-mass frame, we can write,
\begin{align}
	A ( \bfp_1 , \bfp_2 ; \bfp_3, \bfp_4 ) &= A ( p^2, q^2 ) \; , \;\; 
	&\tilde{A} ( \bfp_1 , \bfp_2 ; \bfp_3, \bfp_4 ) &= \tilde{A} ( p^2, p'^2 , q^2 ) \; , \;\; 
	&\tilde{V} ( \bfp_1 , \bfp_2 ; \bfp_3, \bfp_4 ) &= \tilde{V} ( p^2, p'^2, q^2 )
\end{align}

Usually, when computing a relativistic scattering amplitude, one appeals to Lorentz-invariance when fixing the kinematics---if the amplitude is the same in all inertial frames, we might as well pick one in which the algebra is easy (e.g. the centre-of-mass frame).
However, note that here we are using states with the normalisation\footnote{
Crucially, this is what allows us to resolve the identity as $1 = \int \frac{d^d \bfk d^d \bfk'}{  (2 \pi)^{d}  (2 \pi)^{d} } | \bfk \bfk' \rangle \langle \bfk \bfk' | + ...$ in \eqref{eqn:cons_LS}.
},
\begin{align}
	\langle \bfp' | \bfp \rangle = (2 \pi)^d \delta^d \left( \bfp - \bfp' \right)  \; ,
\end{align}
which is \emph{not} Lorentz-invariant. 
Had we instead used states with a Lorentz-invariant normalisation in \eqref{eqn:A_def} to define an amplitude $\mathcal{A}$, we would have,
\begin{align}
	\mathcal{A} ( \bfp_1, \bfp_2 ; \bfp_3 , \bfp_4 ) 
	=
	\left( \prod_{i=1}^4 \sqrt{2 E_i} \right) \, A ( \bfp_1, \bfp_2 ; \bfp_3, \bfp_4 ) \; .
\end{align} 
Since the left-hand-side is a Lorentz-scalar, it is a function only of the two Mandelstam invariants $s = ( p_1^\mu + p_2^\mu )^2$ and $t = (p_1^\mu - p_3^\mu)^2$. 
Comparing with the right-hand-side, we see that $A ( \bfp_1, \bfp_2 ; \bfp_3 , \bfp_4)$ does depend on the inertial frame (through the energy of each particle), and if we focus on the centre-of-mass frame then we have,
\begin{align}
	A ( p^2, q^2 ) = \frac{ \mathcal{A} (s,t) }{ 4 E_A (p^2) E_B ( p^2) }
\end{align}
where,
\begin{align}
	s &= ( E_A (p^2) + E_B (p^2) )^2  \; , \;\; & t &= - q^2 \; . 
\end{align}

\paragraph{Classical LS equation.}
We are now in a position to extract the classical potential \eqref{eqn:V_PM} from the scattering amplitude.
The conservative Lippmann-Schwinger equation \eqref{eqn:cons_LS} captures both the classical and quantum parts of the potential. 
Since we are only interested in the classical part, our goal is to now take a suitable classical limit.

To do this, we perform a perturbative (Post-Minkowski) expansion of both the amplitude and the potential in powers of the coupling $\kappa$. 
Since we wish to extract a classical potential of the form \eqref{eqn:V_PM}, which $\sim r^{-n}$ at order $\kappa^n$, we simply \emph{define} the classical part of each PM order as the part which $\sim q^{n-d}$ at small $q$, where $d$ is the number of spatial dimensions\footnote{
This definition is a bit subtle, since this is \emph{not}, in general, the leading contribution in small $q$ limit. The ``super-classical'' terms which grow faster than $q^{n-d}$ contain IR divergences and are guaranteed to cancel out by perturbative unitarity.
}.
Explicitly, the Fourier transform of the classical potential~\eqref{eqn:V_PM} can be evaluated using the identity~\eqref{eqn:FT_id}, and gives,
\begin{align}
	V (p^2, q^2) = \frac{ 4 \pi \kappa }{q^2} v_1 (p^2) +  \frac{2 \pi^2 \kappa^2}{q} \sum_{n=0}^{\infty} \kappa^{2n} v_{2+2n} (p^2) \frac{ (- q^2)^{n} }{ (2n)! } 
	- 2 \pi 
	\kappa^3 \log ( q^2 ) \sum_{n=0}^{\infty} \kappa^{2n} v_{3+2n} (p^2) \frac{ (-q^2)^{n} }{ (2n+1)! } \; .
	\label{eqn:V_PM_2}
\end{align}
In the quantum theory, these terms must be contained within $\tilde{V} ( p^2, p'^2, q^2)$. 
In particular, note that since its Fourier transform can be written as $\langle \bfp_3 \bfp_4 |  \hat{V} ( \hat{\bfp} , \hat{\bfr} ) | \bfp_1 \bfp_2 \rangle$, and any re-ordering of the $\hat{\bfp}$ and $\hat{\bfr}$ operators will not affect the classical part, we can simply choose to order all $\hat{\bfp}$'s to the right of all $\hat{\bfr}$'s. This amounts to setting\footnote{
Note that \eqref{eqn:op_order} can also be viewed as judiciously adding boundary terms to the action to cancel any $p^2 - p'^2$ dependence.
},
\begin{align}
	\left[ \tilde{V} ( p^2, p'^2, q^2 ) \right]^{\rm cl} &= V ( p^2, q^2 ) \;  ,
	\label{eqn:op_order}
\end{align} 
where all $p'^2$ have been replaced with $p^2$.
The same operator ordering can be used to replace $ \tilde{A} ( p^2, p'^2 , q^2 )$ with $A ( p^2, q^2)$, and in order to match the desired PM expansion in $(\kappa/r)^n$ we can similarly define the classical part of this amplitude as the following non-analytic terms,
\begin{align}
	\left[ \mathcal{A} (s, t) \right]^{\rm cl} =  \frac{ 4 \pi \kappa }{-t} a_1 (s) +  \frac{2 \pi^2 \kappa^2}{\sqrt{-t}} \sum_{n=0}^{\infty} \kappa^{2n} a_{2+2n} (s) \frac{ t^n}{ (2n)!} 
	- 
	2 \pi \kappa^3 \log (- t) \sum_{n=0}^{\infty} \kappa^{2n} a_{3+2n} (s) \frac{ t^n }{ (2n +1 )! }
	\label{eqn:A_PM}
\end{align}
which we have written in terms of small $t$ rather than small $q^2$ for later convenience. 

The Lippmann-Schwinger equation~\eqref{eqn:cons_LS} can then be used to fix each $v_n (p^2)$ in terms of the $a_n (s)$ in these classical PM expansions, as given in the main text.
In particular, the classical part of the Born subtraction integral can be evaluated explicitly using the integral identity~\eqref{eqn:Born_id}.


Note that an alternative operator ordering would be to place all of the $\hat{p}$ operators to the \emph{left} of all $\hat{r}$ operators, which produces the equivalent relation,
\begin{align}
    v_{2,0} \left( \frac{p^2}{2 \mu^2} \right) &= - \frac{ m_A m_B}{4 E_A E_B } \left[ a_{2,0} ( s )  +   \frac{ \mu v_1 (s) }{ 2 \sqrt{s} }  \hat{D}'_{p^2} \; a_{1} \left( s \right)  \right] 
\end{align}
with $\hat{D}'_{p^2} = \tfrac{ E_A E_B - s }{4 E_A E_B }  + E_A E_B \partial_{p^2} $. 
In some cases this identity is the more useful, since e.g. the effect of the conformally coupled scalar of Section~\ref{sec:con} is to simply shift the amplitude $a_2$ by a constant, which does not affect the $\partial_{p^2}$ in this expression.

\paragraph{Sign convention.}
Finally, let us remark on our sign convention for the amplitude.
The overall phase of $\mathcal{A}$ is fixed by \eqref{eqn:A_def} and the relation $\hat{S} = 1 + i \hat{T}$. 
This is the choice which corresponds to a positive $\tfrac{1}{2i} \text{Disc}_s \mathcal{A} = \text{Im} \, \mathcal{A}$ for physical $2 \to 2$ kinematics, thanks to unitarity. 
For instance, with our convention exchanging a field of mass $M$ in the $s$-channel gives an amplitude,
\begin{align}
 \mathcal{A} (s,t) = \frac{Z^2}{M^2 - s - i \epsilon}
\end{align}
where unitarity of the exchanged field requires that $Z^2 > 0$. 
Furthermore, positivity of this discontinuity translates into the positivity of any even $s$-derivative for which the dispersion relation converges,  $\partial_s^{2n} \mathcal{A} > 0$ for $n \geq 1$, thanks to analyticity and Froissart bound. 
However, it is also common to see the definition $\hat{S} = 1 - i \hat{T}'$, particularly in the context of non-relativistic calculations, which corresponds to an amplitude $\mathcal{A}' = - \mathcal{A}$. This removes the relative minus sign in \eqref{eqn:cons_LS}, and in particular at 1PM the potential is simply equal to (the Fourier transform of) the amplitude $A'$. 
It is worth keeping in mind that amplitudes with this alternative convention would have \emph{negative} discontinuities via the optical theorem.

\subsection{Feinberg Sucher trick}

\paragraph{Dispersion relation.}
The $t$-channel discontinuity \eqref{eqn:Disc_def} in fact contains all of the information needed to construct the classical potential. 
It is particularly useful because perturbative unitarity allows us to compute this object using only tree-level Feynman diagrams, and so it is widely used in amplitude-based approaches to the two-body problem to simplify perturbative calculations.
However, it also plays a fundamental role in the fully non-perturbative theory.
Within the analytic $S$-matrix programme, the scattering amplitude can be expressed via contour integration as\footnote{
In the event of divergences at $|t| \to \infty$, a circular integration contour at infinity implements the necessary subtractions. 
}, 
\begin{align}
\mathcal{A} (s, t) =  \int_{-\infty}^{+\infty} \frac{d t'}{2 \pi i} \frac{ \text{Disc}_t \mathcal{A} (s, t')}{t'-t - i \epsilon} \; .
\label{eqn:disp}
\end{align}
This is the so-called ``$t$-channel dispersion relation''. 
The discontinuity therefore encodes all of the information in the scattering amplitude. 

Going to complex values of the momenta is not just a calculational convenience, but is actually forced upon us in the following sense.
In the previous section, we implicitly matched the Fourier transform,
\begin{align}
 A (p^2, r) = \frac{1}{4 E_A E_B} \int \frac{d^d \bfq}{ (2 \pi)^d} e^{-i \bfq \cdot \bfr} \mathcal{A} (s, t = -q^2 ) 
 \label{eqn:A_FT}
\end{align}
onto the classical potential. 
But to perform this Fourier transform, we need to evaluate $\mathcal{A} (s,t)$ for all real $\bfq = \bfp_1 - \bfp_3$, even those which are unphysical (i.e. $s$-channel kinematics requires $(m_1+ m_2)^2 -s < t < 0$, but we need to integrate over all $t = -q^2 < 0$).

It was pointed out by Feinberg and Sucher \cite{Feinberg:1988yw} that the dispersion relation~\eqref{eqn:disp} can be used to make sense of this Fourier transform. 
Plugging \eqref{eqn:disp} into \eqref{eqn:A_FT} allows us to write the long-distance (classical) part of the potential as\footnote{
More precisely, the integration region in \eqref{eqn:disp} contains both positive and negative values, which correspond to $t$-channel and $u$-channel scattering respectively. 
The $u$-channel cut begins from $-t \geq s - (m_1 - m_2 )^2 \geq 4 m_1 m_2$, while the $t$-channel cut begins from $t \geq ( \mu_1 + \mu_2 )^2$, where $\mu$ is the mass of the exchanged field. So when the mass of the exchanged field is much less than the mass of the scattering particles, $(\mu_1 + \mu_2 )^2 \ll 4 m_1 m_2$, it is the $t$-channel cut which dominates the long-distance behaviour of the potential. 
},
\begin{align}
A (p^2 , r)  = \frac{1}{4 E_A E_B} \int_{0}^{\infty} \frac{d t}{2 \pi i} \, e^{- \sqrt{t} r} \text{Disc}_t \, \mathcal{A} (s,t) + \text{small-distance corrections} \; ,
\label{eqn:FT_Disc}
\end{align} 
in $d=3$ dimensions.

\paragraph{Post-Minkowski expansion.}
The classical terms in the amplitude \eqref{eqn:A_PM} all have one thing in common: they all contribute to singularities in the complex $t$-plane at fixed $s$. 
This guarantees that the $\text{Disc}_t \mathcal{A}^{\rm cl}$ has an analogous PM expansion,
\begin{align}
	\text{Disc}_t \mathcal{A}^{\rm cl} (s,t)  = \begin{cases}
	4 \pi^2 i \left[  2 \kappa  a_1 (s)  \delta (t)  +  \frac{\kappa^2}{\sqrt{t} } \sum_{n=0}^{\infty} \kappa^{n} a_{2+n} (s) \frac{ ( \sqrt{t} ) ^{n} }{ n !} 
 \right]
	&\text{if } t > 0\\ 
	0 & \text{otherwise.}
	\end{cases}
\end{align}
as quote in the main text.
Note that the $\log (-t)$ series has naturally combined with the $1/\sqrt{-t}$ series. 
The Fourier transform \eqref{eqn:FT_Disc} can then be immediately performed,
\begin{align}
4 E_A E_B A^{\rm cl} (p^2, r) = \sum_n a_n (s) \left( \frac{\kappa}{r} \right)^n \; .
\end{align}
and analogously for the potential.

\paragraph{General 2PM computation.}
The general structure of every 2PM calculation is then the following. 
First one computes the tree-level Feynman diagrams for the process $\chi_A \chi_A \to X_1 X_2$ with the kinematics shown in Figure~\ref{fig:FS}. Supposing $X_1$ and $X_2$ have spins $S_1$ and $S_2$, this gives a result of the form,
\begin{align}
 \mathcal{A}_{\chi_A \chi_A \to X_1 X_2}
 = \left[ \sum_{a}^{1 + \text{min} (S_1, S_2) } F_A^{(a)} ( t ,  x_A ) \;  \mathcal{O}_a^{ I ; J} [\hat{\bfp}_A ] \right] \epsilon^*_{ I  } ( \hat{\bfl} ) \epsilon^*_{ J } ( \hat{\bfl} ) 
\end{align}
where $I = \{ i_1, ... , i_{S_1} \}$ and $J = \{ j_1, ... , j_{S_2} \}$ are a list of spatial indices, $\epsilon_I$ and $\epsilon_J$ are the polarisation tensors for $X_1$ and $X_2$, and the $\{ \mathcal{O}^{I ; J}_a [ \hat{\bfp}_A ] \}$ are a complete basis of tensors built from $\hat{\bfp}_A^i$ and $\delta^{ij}$. 
The dynamical information is encoded in the functions $F_A^{(a)}$.
$\tilde{p}_1 \leftrightarrow \tilde{p}_2$ permutation invariance and locality dictate that the amplitude can have at most simple poles in $p_1 \cdot \ell_1$ and $p_2 \cdot \ell_1$ with equal residue, and so the $F_A^{(a)}$ take the form,
\begin{align}
 F_A^{(a)} ( t , x_A )  &=  \frac{ m_A^2 Z_A^{(a)}  }{ m_A^2 -  (p_1^\mu - \ell_1^\mu )^2  } + \frac{ m_A^2 Z_A^{(a)}  }{ m_A^2 -  (p_2^\mu - \ell_1^\mu )^2  } + R_A^{(a)} ( t , x_A )   
\end{align}
where $R_A^{(a)} (t, x_A)$ is an analytic function of $x_A$ and $Z_A^{(a)}$ does not depend on the kinematics. 
The only part of $\rho_{X_1 X_2}$ which can contribute to the classical $a_2$ coefficient is then,
\begin{align}
 \rho_{X_1 X_2} ( s, t) \supset \sum_{a,b} \,  Z_A^{(a)} \left\langle \frac{ C_B^{(a)} (t, x_B) N_{a,b} (x_A, x_B, y) }{ x_A^2 + \frac{t}{4m_A^2} }   \right\rangle  + \left( A \leftrightarrow B \right) 
 \label{eqn:rho_ans}
\end{align}
where the $N^{(a,b)}$ are a fixed basis of spin structures,
\begin{align}
N_{a,b} (x_A, x_B, y ) = \mathcal{O}_a^{I; I'} [ \hat{\bfp}_A ] P_{I ; I'} [ \hat{\bfl} ] P_{J ; J'} [ \hat{\bfl} ] \mathcal{O}_b^{J ; J'} [ \hat{\bfp}_B ] \; .
\end{align}
and we have defined the sum over polarisations,
\begin{align}
 P_{I ; I'} [ \hat{\bfl} ] = \sum_h \epsilon_I^{(h) *} ( \hat{\bfl} ) \epsilon_{I'}^{(h)} ( \hat{\bfl} ) \; . 
\end{align}
These can be computed ahead of time, and applied universally to any EFT that contains fields of spin $S_1$ and $S_2$. 
In particular, each $N_{a,b} (x_A, x_B, y)$ is simply a polynomial of order at most $S_1 + S_2$ in each variable. 

Since this may seem somewhat abstract, let us list a few examples,
\begin{itemize}

\item $S_1 = S_2 = 0$. Then $\mathcal{O} [ \hat{\bfp} ] = 1$ and $N (x_A, x_B, y) = 1$, so $\rho_{X_1 X_2}$ is simply the angular average over $C_B (t, x_A)$. 

\item $S_1 = 0, S_2 = 1$. Then $\mathcal{O} [ \hat{\bfp} ] = \bfp^{j}$ is the only available tensor structure, and the corresponding polarisation sum is,
\begin{align}
 N ( x_A, x_B, y ) = y - x_A x_B \; . 
\end{align}
 
\item $S_1= 0, S_2 = 2$. Then $\mathcal{O} [ \hat{\bfp} ] = \bfp^{j_1} \bfp^{j_2}$ is the only available tensor structure (since the polarisation tensor is traceless), and the corresponding polarisation sum is,
\begin{align}
 N ( x_A , x_B , y ) = ( y - x_A x_B )^2 - \frac{1}{2} (1 - x_A^2) ( 1 - x_B^2 ) \; . 
 \label{eqn:N_02}
\end{align}

\item $S_1 = S_2 = 1$. Then there are two possible tensor structures: $\mathcal{O}_1 [ \hat{\bfp} ] = \bfp^{i} \bfp^{j}$ and $\mathcal{O}_0 [ \hat{\bfp} ] = \delta^{ij}$. The corresponding polarisation sums are,
\begin{align}
&N_{1,1}  = ( y - x_A x_B )^2 \; , 
&N_{1,0}  &= 1 - x_A^2  \;, 
&N_{0,1} &= 1 - x_B^2 \; ,
&N_{0,0} &= 2 \; . 
\end{align}

\end{itemize}

To summarise: from a tree-level computation of the scattering amplitude involving a massless spin $S_1$ and $S_2$ field, one should read off the residues $Z_A^{(a)}$ and the analytic remainders $R_A^{(a)}$ with respect to a convenient basis of tensor structures $\mathcal{O}_a [ \hat{\bfp} ]$. The classical part of the unitarity integral $\rho_{X_1 X_2}$ is then found from \eqref{eqn:rho_ans}, using the polarisation sum $N_{a,b}$ listed above and the identity~\eqref{eqn:xBdA_id}.  \\

In principle, the Feinberg-Sucher trick could also be used to find the 3PM amplitude from the exchange of 3 gravitons. 
However, this would require carrying out angular integrals over 5-point kinematics, and also evaluating polarisation sums over products of six polarisation tensors.
Developing these would be worthwhile, since the resulting machinery would be applicable to any field theory. 
In the particular case of GR, the 3PM calculation was completed in \cite{Bern:2019nnu}, and the current state-of-the-art is the 4PM (i.e. three-loop) amplitude recently computed in \cite{Bern:2021dqo} (see also \cite{Bern:2021yeh, Bern:2022jvn}). 
We will not require results beyond 2PM at the order at which we are working here.

\paragraph{Independent structures.}
As an interesting aside, note that at 2PM there are only $1 + \min (J_1, J_2)$ independent functions of the velocity which can possibly arise from the unitarity integral in $\rho_{X_1 X_2}$. 
Consequently, if one knows this many PN orders, one can completely resum the whole $v^2/c^2$ series.
This is particularly useful for the scalar, since for instance the entire velocity dependence of $\rho_{\phi \phi}$ can be determined from simply the first PN correction in which the scalar appears. 
In GR, the number of PN corrections needed would naively be $3$. 
However, the double copy structure effectively reduces this number to just $2$,
i.e. once the $2PN$ and $3PN$ potentials have been determined, in principle this is enough to uniquely fix the whole $2PM$ potential. 

It would be interesting to estimate how many independent structures there are at 3PM. 
To do this counting carefully would require analysing the angular average over 3-particle kinematics.
Naively, it would seem that $\rho_{\phi \phi \phi}$ may also have only a single independent structure. In that case, the $\mathcal{O} ( \alpha_1^6 )$ of the full 3PM potential could be determined directly from the existing 3PN calculations of \cite{Julie:2022qux}.

\subsection{Electromagnetism}

Scattering in this theory is captured by the simple Feynman rules,

\FloatBarrier
\begin{figure}[htbp!]
\centering
\begin{tikzpicture}[baseline=-0.1cm]
			\begin{feynman}
				\vertex (a1) at (0,1) {$\tilde{p}_1^\mu$};
				\vertex (b2) at (1,0);
				\vertex (c1) at (0,-1) {$\tilde{p}_2^\mu$};
				\vertex (b3) at (2.4,0) {$\tilde{\ell}^\mu$};
				
				\diagram*{
                                (a1) -- [fermion] (b2),
                                (c1) -- [fermion] (b2),
				(b2)  -- [photon] (b3),
				};					
			\end{feynman}
                \end{tikzpicture} $=  q_A \,  \tilde{\bfp}_A  \cdot  \epsilon^* ( \tilde{\bfl} )  , $ \hfill
             \begin{tikzpicture}[baseline=-0.1cm]
			\begin{feynman}
				\vertex (a1) at (0,1) {$\tilde{p}_1^\mu$};
				\vertex (b2) at (1,0);
				\vertex (c1) at (0,-1) {$\tilde{p}_2^\mu$};
				\vertex (a3) at (2,1) {$\tilde{\ell}_1^\mu$};
				\vertex (c3) at (2,-1) {$\tilde{\ell}_2^\mu$};
				
				\diagram*{
                                (a1) -- [fermion] (b2),
                                (c1) -- [fermion] (b2),
				(b2)  -- [photon] (a3),
				(b2)  -- [photon] (c3),
				};					
			\end{feynman}
                \end{tikzpicture} $= 2 q_A^2 \, \epsilon^* (\bfl_1) \cdot  \epsilon^{*} ( \bfl_2) , $ \qquad
\end{figure}
\noindent where we have used the centre-of-mass kinematics shown in Figure~\ref{fig:FS}.

\paragraph{1PM amplitudes.}
The tree-level amplitude is,
\begin{align}	
 \mathcal{A}_{\chi_A \bar{\chi}_A \to \gamma} =	 q_A \, \bfp_A^i \epsilon_i ( \bfl )
\end{align}
where we have used the centre-of-mass kinematics for $\chi_A$ shown in Figure~\ref{fig:FS}. 
The unitarity integral then gives,
\begin{align}
 \rho_{\gamma} = q_A q_B 
\end{align}
The expansion parameter for this problem is then taken to be $\kappa = q_A q_B$.

\paragraph{2PM amplitudes.}
Pair production (the analytic continuation of Compton scattering) has three tree-level Feynman diagrams,
\FloatBarrier
\begin{figure}[htbp!]
\centering
$ \mathcal{A}_{\chi_A \bar{\chi}_A \to \gamma_1 \gamma_2} =$
 \begin{tikzpicture}[baseline=-0.6cm]
			\begin{feynman}
				\vertex (a1);
				\vertex [below=0.6cm of a1] (b1);
				\vertex [below=0.6cm of b1] (c1);
				\vertex [left=0.6cm of a1] (a2);
				\vertex [below=0.6cm of a2] (b2);
				\vertex [below=0.6cm of b2] (c2);
				\vertex [left=0.6cm of a2] (a3);
				\vertex [below=0.6cm of a3] (b3);
				\vertex [below=0.6cm of b3] (c3);
				\vertex [left=0.6cm of a3] (a4);
				\vertex [below=0.6cm of a4] (b4);
				\vertex [below=0.6cm of b4] (c4);
				
				\diagram*{
                                (a4) -- [fermion] (b3),
                                (b3) -- [fermion] (c4),
				(b3)  -- [photon] (a2),
				(b3)  -- [photon] (c2),
				};					
			\end{feynman}
                \end{tikzpicture}
                $+$
                \begin{tikzpicture}[baseline=-0.6cm]
			\begin{feynman}
				\vertex (a1);
				\vertex [below=0.6cm of a1] (b1);
				\vertex [below=0.6cm of b1] (c1);
				\vertex [left=0.6cm of a1] (a2);
				\vertex [below=0.6cm of a2] (b2);
				\vertex [below=0.6cm of b2] (c2);
				\vertex [left=0.6cm of a2] (a3);
				\vertex [below=0.6cm of a3] (b3);
				\vertex [below=0.6cm of b3] (c3);
				\vertex [left=0.6cm of a3] (a4);
				\vertex [below=0.6cm of a4] (b4);
				\vertex [below=0.6cm of b4] (c4);
				
				\diagram*{
                                (a4) -- [fermion] (a3),
                                (c3) -- [fermion] (c4),
                                (a3) -- [fermion] (c3),
				(a3)  -- [photon] (a2),
				(c3)  -- [photon] (c2),
				};				
			\end{feynman}
                \end{tikzpicture}
                $+$
                \begin{tikzpicture}[baseline=-0.6cm]
			\begin{feynman}
				\vertex (a1);
				\vertex [below=0.6cm of a1] (b1);
				\vertex [below=0.6cm of b1] (c1);
				\vertex [left=0.6cm of a1] (a2);
				\vertex [below=0.6cm of a2] (b2);
				\vertex [below=0.6cm of b2] (c2);
				\vertex [left=0.6cm of a2] (a3);
				\vertex [below=0.6cm of a3] (b3);
				\vertex [below=0.6cm of b3] (c3);
				\vertex [left=0.6cm of a3] (a4);
				\vertex [below=0.6cm of a4] (b4);
				\vertex [below=0.6cm of b4] (c4);
				
				\diagram*{
                                (a4) -- [fermion] (a3),
                                (c3) -- [fermion] (c4),
                                (a3) -- [fermion] (c3),
				(a3)  -- [photon] (c2),
				(c3)  -- [photon] (a2),
				};				
			\end{feynman}
                \end{tikzpicture}   \\[14pt]
                 ${\color{white} \mathcal{A}_{\chi_A \bar{\chi}_A \to \gamma_1 \gamma_2}} = 2 q_A^2 \left( \epsilon_1 \cdot \epsilon_2 + \frac{2 \tilde{p}_1 \cdot \epsilon_1 \; \tilde{p}_2 \cdot \epsilon_2 }{ ( p_1^\mu - \ell_1^\mu )^2 - m_A^2 } +  \frac{2 \tilde{p}_1 \cdot \epsilon_2 \; \tilde{p}_2 \cdot \epsilon_1 }{ ( p_2^\mu - \ell_1^\mu )^2 - m_A^2 } \right) $
\end{figure}
\FloatBarrier
This can be written as,
\begin{align}
	 A_{\chi_A \chi_A \to \gamma_1 \gamma_2} ( \tilde{\bfp}_1 , \tilde{\bfp}_2 ; \tilde{\bfl}_1 , \tilde{\bfl}_2 ) 
	 &= 2 q_A^2 \left( \epsilon_1 \cdot \epsilon_2 + \frac{2 \tilde{p}_1 \cdot \epsilon_1 \; \tilde{p}_2 \cdot \epsilon_2 }{ ( p_1^\mu - \ell_1^\mu )^2 - m_A^2 } +  \frac{2 \tilde{p}_1 \cdot \epsilon_2 \; \tilde{p}_2 \cdot \epsilon_1 }{ ( p_2^\mu - \ell_1^\mu )^2 - m_A^2 } \right)  \nonumber \\ 
	 &= 2 q_A^2  \epsilon_{1}^i  \epsilon_{2}^{j} \mathcal{O}_{ij} ( \tilde{\bfp}_A, \tilde{\bfl} )  
\end{align}
where in the centre-of-mass frame shown in Figure~\ref{fig:FS}, 
\begin{align}
	\mathcal{O}^{ij} ( \tilde{\bfp}_A, \tilde{\bfl} ) = \delta^{ij} + \frac{ \hat{\bfp}_A^i \hat{\bfp}_A^j }{ ( \hat{\bfp}_A \cdot \hat{\bfl} )^2 - \frac{t}{\tilde{p}_A^2} }
\end{align}

\paragraph{One-loop amplitudes.}
The unitarity cut is then,
\begin{align}
\rho_2 (s,t)  = 4 e^4 \left\langle  \mathcal{O}_{ij} ( \bfp_A, \bfl ) P^{i i'} ( \hat{\bfl} )  P^{j j'} ( \hat{\bfl} ) \mathcal{O}^*_{ i' j'} ( \bfp_B, \bfl )  \right\rangle_{\hat{\bfl}}  
\label{eqn:EM_OPPO}
\end{align}
where the polarisation sum,
\begin{align}
	P^{i j} ( \hat{\bfl} ) = \sum_{h = \pm} \epsilon_h^i ( \hat{\bfl} ) \epsilon_h^{* \, j} ( \hat{\bfl} ) = \delta_{ij} - \hat{\bfl}^i \hat{\bfl}^j \; . 
\end{align}
We discuss this polarisation sum and the angular average in more detail in Appendix~\ref{app:int_id}. 
In short, the only angular average which contributes to the classical part is, 
\begin{align}
	\left\langle \frac{ ( \tilde{\bfp}_B \cdot \hat{\bfl} )^{2n} }{ ( \tilde{\bfp}_A \cdot \hat{\bfl} )^2 - t }  \right\rangle_{\hat{\bfl}}^{\rm cl}  
	= \frac{ \pi m_A}{\sqrt{-t}} 
	\label{eqn:xBdA_id}
\end{align}
and its $A \leftrightarrow B$ permutation, and so in practice we can quickly pick out the two terms from \eqref{eqn:EM_OPPO} which contribute to the classical amplitude and find,
\begin{align}
	\frac{a_2 (s)}{\sqrt{-t}} = \rho_2^{\rm cl} (s, t) = \frac{ \pi (m_A + m_B )}{\sqrt{-t}}  \; .
\end{align}



\subsection{General Relativity}

The scattering amplitudes for spinless particles in General Relativity have been computed via a variety of methods, and the majority make use of unitarity and the KLT double copy relation in some way. 
Crossing \'{a} la Feinberg and Sucher, as described in the previous section, was applied to General Relativity in \cite{Holstein:2016fxh}, and it is this calculation which we briefly review here. 

In particular, while the relevant Feynman rules would be,
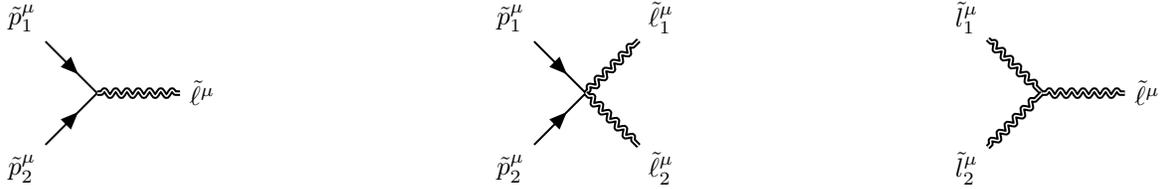
\FloatBarrier
\begin{figure}[htbp!]
\centering
\begin{tikzpicture}[baseline=-0.1cm]
			\begin{feynman}
				\vertex (a1) at (0,1) {$\tilde{p}_1^\mu$};
				\vertex (b2) at (1,0);
				\vertex (c1) at (0,-1) {$\tilde{p}_2^\mu$};
				\vertex (b3) at (2.4,0) {$\tilde{\ell}^\mu$};
				
				\diagram*{
                                (a1) -- [fermion] (b2),
                                (c1) -- [fermion] (b2),
				(b2)  -- [graviton] (b3),
				};					
			\end{feynman}
                \end{tikzpicture}  \hfill
             \begin{tikzpicture}[baseline=-0.1cm]
			\begin{feynman}
				\vertex (a1) at (0,1) {$\tilde{p}_1^\mu$};
				\vertex (b2) at (1,0);
				\vertex (c1) at (0,-1) {$\tilde{p}_2^\mu$};
				\vertex (a3) at (2,1) {$\tilde{\ell}_1^\mu$};
				\vertex (c3) at (2,-1) {$\tilde{\ell}_2^\mu$};
				
				\diagram*{
                                (a1) -- [fermion] (b2),
                                (c1) -- [fermion] (b2),
				(b2)  -- [graviton] (a3),
				(b2)  -- [graviton] (c3),
				};					
			\end{feynman}
                \end{tikzpicture}  \hfill
                \begin{tikzpicture}[baseline=-0.1cm]
			\begin{feynman}
				\vertex (a1) at (0,1) {$\tilde{l}_1^\mu$};
				\vertex (b2) at (1,0);
				\vertex (c1) at (0,-1) {$\tilde{l}_2^\mu$};
				\vertex (b3) at (2.4,0) {$\tilde{\ell}^\mu$};
				
				\diagram*{
                                (a1) -- [graviton] (b2),
                                (c1) -- [graviton] (b2),
				(b2)  -- [graviton] (b3),
				};					
			\end{feynman}
                \end{tikzpicture} 
\end{figure}
\FloatBarrier
\noindent although we will not require their specific values, since the KLT double copy relation allows us to construct the GR amplitude directly from the electromagnetism amplitudes of the previous subsection. 

\paragraph{1PM Amplitude.}
From the action for fluctuations,
\begin{align}
 A_{\chi_A \chi_A \to h} ( \bfp_1 , \bfp_2 ; \bfl ) = \bfp_A^i \bfp_A^j \epsilon_{i j} ( \bfl )
\end{align}
Note that the double copy structure is manifest,
\begin{align}
 \mathcal{O}^{ij} ( \bfp_A ) = \mathcal{O}^i (\bfp_A)  \mathcal{O}^j ( \bfp_A ) \; . 
\end{align}
since polarisation tensor is traceless. 

The unitarity integral then gives,
\begin{align}
\mathcal{A}_{\rm cl}^{(1)} (s,t) = - \frac{ 8 \pi G_N }{t}  \left( (s - m_1^2 - m_2^2)^2 - 2 m_1^2 m_2^2   \right)
\end{align}

\paragraph{2PM Amplitude.}
As above, the tree-level $2\to 2$ amplitude takes the form,
\begin{align}
	A_{\chi_A \chi_A \to h_1 h_2} ( \tilde{\bfp}_1 , \tilde{\bfp}_2 ; \tilde{\ell}_1 , \tilde{\ell}_2 ) 
	=
	\kappa^2 \mathcal{O}_{ij, ab} ( \tilde{\bfp}_A, \tilde{\ell} )  \epsilon_1^{ij} \epsilon_2^{ab} 
\end{align}
Rather than compute the various Feynman diagrams explicitly (which requires the rather lengthy three-graviton vertex), we simply use the KLT double copy relation to express it as,
\begin{align}
	\mathcal{O}_{ij, ab} ( \tilde{\bfp}_A, \tilde{\ell} ) = \mathcal{O}_{ia} ( \tilde{\bfp}_A, \tilde{\ell} ) \mathcal{O}_{jb} ( \tilde{\bfp}_A, \tilde{\ell} )  
\end{align}
in terms of the electromagnetism answer of the previous subsection. 

The unitarity cut is then,
\begin{align}
	\rho_2 (s,t)  = \left\langle  \mathcal{O}_{ij, ab} ( \bfp_A, \bfl ) P^{i j, i' j'} ( \hat{\bfl} )  P^{ab, a' b'} ( \hat{\bfl} ) \mathcal{O}^*_{ i' j' ,a' b'} ( \bfp_B, \bfl )  \right\rangle_{\hat{\bfl}}  
	\label{eqn:EM_OPPO}
\end{align}
where the polarisation sum is now\footnote{
Note that partially fixing to a covariant gauge (e.g. harmonic/de Donder) would give $
 \mathcal{P}_{\mu\nu} = \frac{1}{2} ( \eta \eta + \eta \eta - \eta \eta )$ and retain manifest Lorentz invariance, however it would also require the addition of ghost fields beyond 1PM (see e.g. \cite{Bjerrum-Bohr:2002gqz}).
},
\begin{align}
	P^{i j ; a b} ( \hat{\bfl} ) = \sum_{h = \pm} \epsilon_h^{ij} ( \hat{\bfl} ) \epsilon_h^{* \, ab} ( \hat{\bfl} ) = \frac{1}{2} \left( P_{ia} P_{jb} + P_{ib} P_{ja} - P_{ij} P_{ab}  \right) \; . 
\end{align}
Once again we defer any discussion of this polarisation sum to the Appendix, and simply note that again there are very few (only four) terms of the form~\eqref{eqn:xBdA_id} which contribute to the classical part\footnote{
Note that picking out these classical terms from the angular average corresponds to evaluating the two triangle Feynman diagrams,
\begin{align}
a_2 =  c_{>} \mathcal{I}_{>} +  c_{<} \mathcal{I}_{<}
\end{align}
in the notation of \cite{Bjerrum-Bohr:2018xdl}. 
}, 
\begin{align}
\frac{ N (x_A, x_B, y ) }{d_A d_B} \supset \frac{ (16 y^2 -4 ) x_A^2  + 2 x_A^4 }{d_A d_B} + ( A \leftrightarrow B ) 
\end{align} 
The result is, 
\begin{align}
 \rho_{hh}  \sqrt{t} &= \pi m_A 3 ( 5 y^2 - 1)  + ( A \leftrightarrow B )  \nonumber \\
 \Rightarrow \quad	\frac{ \kappa^2 }{\sqrt{-t}} a_2 (s) &= \frac{\kappa^2 \pi}{\sqrt{-t}} m_A m_B (m_A + m_B)   ( 1 - 5 \sigma^2 ) \; .
\end{align}

\paragraph{3PM Amplitude.}
In principle, the Feinberg-Sucher trick could also be used to find the 3PM amplitude from $\mathcal{A}_{2 \to 3}$. 
However, this would require carrying out angular integrals over 5-point kinematics, and also evaluating polarisation sums over products of six polarisation tensors.
Developing these would be worthwhile, since the resulting machinery would be applicable to any field theory. 
However, for the specific case of General Relativity, we can instead use modern double copy techniques to find the amplitude. 
This was done in \cite{Bern:2019nnu} (see also \cite{Cheung:2020gyp, Kalin:2020fhe}), who found at 3PM, 
\begin{align}
\mathcal{A}_{\rm cl}^{(3)} (s,t) &=  3 - 54 \sigma^2 + \nu \left( -6  + 206  \sigma  + 108  \sigma^2 + 4 \sigma^3 \right)  \nonumber \\
&-48 \nu (3 + 12 \sigma^2 - 4 \sigma^4) \frac{ \text{arcsinh} \left( \sqrt{  \frac{ \sigma - 1 }{2 }} \right) }{ \sqrt{ \sigma^2 - 1} } - 
 18 \nu \frac{ \gamma (1 - 2 \sigma^2) (1 - 5 \sigma^2) }{ (1 + \gamma ) (1 + \sigma ) }
\end{align}
For completeness, note that the current state-of-the-art is the 4PM (i.e. three-loop) amplitude computed in \cite{Bern:2021dqo} (see also \cite{Bern:2021yeh, Bern:2022jvn}).

\subsection{Scalar-tensor theory}
\label{app:ST}

Here we collect some details about the scalar-tensor theory discussed in the main text. 

\paragraph{Effective action.}
Expanding the EFT action around a Minkowski background, 
\begin{align}
g_{\mu \nu} = \eta_{\mu \nu} + h_{\mu \nu} \; , \;\;\; \phi = 0 + \varphi
\end{align}
the effective action which governs the perturbations is,
\begin{align}
 S_A = - \beta^2 \varphi^2 \left(  ( \partial \chi )^2 + 4 M^2 \chi^2 \right) 
 - \frac{1}{2} \beta \varphi h^\mu_\mu \chi^2
 + \beta \phi T^\mu_\mu
 + \frac{1}{2} h_{\mu \nu} T^{\mu \nu} + \mathcal{O} ( h^2 \chi^2 )
\end{align}
where the $\mathcal{O} ( h^2 \chi^2 )$ terms are the same as in GR, and the matter stress-energy tensor is,
\begin{align}
 T^{\mu \nu} = \partial^\mu \chi \partial^\nu \chi - \frac{1}{2} \eta^{\mu \nu} \left( ( \partial \chi )^2 + m^2 \chi^2 \right) \; . 
\end{align}
Once the fields are canonically normalised ($\varphi \to \varphi/M_P$ and $h_{\mu \nu} \to h_{\mu \nu}/M_P$), this gives the Feynman rules we need to calculate scattering amplitudes\footnote{
In all diagrams, $\chi$ is depicted by a solid line, $\varphi$ by a dashed line, and $h_{\mu \nu}$ as a wiggly line. 
}.

\paragraph{2PM diagrams.}
The simplest $2 \to 2$ amplitude is $\chi \chi \to \phi \phi$, which is given by,

\FloatBarrier
\begin{figure}[h!]
\begin{flushleft}
$ \mathcal{A}_{\chi \chi \to \varphi \varphi} ( p_1, p_3, k_1, k_2 ) = $
 		\begin{tikzpicture}[baseline=-0.6cm]
			\begin{feynman}
				\vertex (a1);
				\vertex [below=0.5cm of a1] (b1);
				\vertex [below=0.5cm of b1] (c1);
				\vertex [left=0.5cm of a1] (a2);
				\vertex [below=0.5cm of a2] (b2);
				\vertex [below=0.5cm of b2] (c2);
				\vertex [left=0.5cm of a2] (a3);
				\vertex [below=0.5cm of a3] (b3);
				\vertex [below=0.5cm of b3] (c3);
				\vertex [left=0.5cm of a3] (a4);
				\vertex [below=0.5cm of a4] (b4);
				\vertex [below=0.5cm of b4] (c4);
				
				\diagram*{
				(a3)  -- [scalar] (b2),
				(b2) -- [scalar] (a1),
				(c3) -- [plain] (b2),
				(b2) -- [plain] (c1),
				};				
			\end{feynman}
                \end{tikzpicture} $+$ 
 		\begin{tikzpicture}[baseline=-0.6cm]
			\begin{feynman}
				\vertex (a1);
				\vertex [below=0.5cm of a1] (b1);
				\vertex [below=0.5cm of b1] (c1);
				\vertex [left=0.5cm of a1] (a2);
				\vertex [below=0.5cm of a2] (b2);
				\vertex [below=0.5cm of b2] (c2);
				\vertex [left=0.5cm of a2] (a3);
				\vertex [below=0.5cm of a3] (b3);
				\vertex [below=0.5cm of b3] (c3);
				\vertex [left=0.5cm of a3] (a4);
				\vertex [below=0.5cm of a4] (b4);
				\vertex [below=0.5cm of b4] (c4);
				
				\diagram*{
                                (a4) -- [scalar] (b3),
                                (c4) -- [plain] (b3),
				(b3)  -- [plain] (b2),
				(b2) -- [scalar] (a1),
				(b2) -- [plain] (c1),
				};				
			\end{feynman}
                \end{tikzpicture} $+$
 		\begin{tikzpicture}[baseline=-0.6cm]
			\begin{feynman}
				\vertex (a1);
				\vertex [below=0.5cm of a1] (b1);
				\vertex [below=0.5cm of b1] (c1);
				\vertex [left=0.5cm of a1] (a2);
				\vertex [below=0.5cm of a2] (b2);
				\vertex [below=0.5cm of b2] (c2);
				\vertex [left=0.5cm of a2] (a3);
				\vertex [below=0.5cm of a3] (b3);
				\vertex [below=0.5cm of b3] (c3);
				\vertex [left=0.5cm of a3] (a4);
				\vertex [below=0.5cm of a4] (b4);
				\vertex [below=0.5cm of b4] (c4);
				
				\diagram*{
                                (a1) -- [scalar] (b3),
                                (c4) -- [plain] (b3),
				(b3)  -- [plain] (b2),
				(b2) -- [scalar] (a4),
				(b2) -- [plain] (c1),
				};				
			\end{feynman}
                \end{tikzpicture}  \\[10pt]
\mbox{$= \frac{\beta^2}{M_P^2} \left[  
4 p_1 \cdot p_3 - 16 m^2  
+ \frac{ \left( p_1 \cdot (p_1 + k_1) + 2 m^2 \right) \left( p_3 \cdot (p_3 + k_2) + 2 m^2 \right) }{ p_1 \cdot k_1 }  
+ \frac{ \left( p_1 \cdot (p_1 + k_2) + 2 m^2 \right) \left( p_3 \cdot (p_3 + k_1) + 2 m^2 \right) }{ p_3 \cdot k_1 }   
\right]$}
\end{flushleft}
\end{figure}
\FloatBarrier

Next consider the $\chi \chi \to \varphi h$ amplitude,
\FloatBarrier
\begin{figure}[h!]
\begin{flushleft}
$ \mathcal{A}_{\chi \chi \to \varphi h_{\mu \nu}} ( p_1, p_2, k_1, k_2 ) = $
 		\begin{tikzpicture}[baseline=-0.6cm]
			\begin{feynman}
				\vertex (a1);
				\vertex [below=0.5cm of a1] (b1);
				\vertex [below=0.5cm of b1] (c1);
				\vertex [left=0.5cm of a1] (a2);
				\vertex [below=0.5cm of a2] (b2);
				\vertex [below=0.5cm of b2] (c2);
				\vertex [left=0.5cm of a2] (a3);
				\vertex [below=0.5cm of a3] (b3);
				\vertex [below=0.5cm of b3] (c3);
				\vertex [left=0.5cm of a3] (a4);
				\vertex [below=0.5cm of a4] (b4);
				\vertex [below=0.5cm of b4] (c4);
				
				\diagram*{
				(a3)  -- [scalar] (b2),
				(b2) -- [photon] (a1),
				(c3) -- [plain] (b2),
				(b2) -- [plain] (c1),
				};				
			\end{feynman}
                \end{tikzpicture} $+$ 
 		\begin{tikzpicture}[baseline=-0.6cm]
			\begin{feynman}
				\vertex (a1);
				\vertex [below=0.5cm of a1] (b1);
				\vertex [below=0.5cm of b1] (c1);
				\vertex [left=0.5cm of a1] (a2);
				\vertex [below=0.5cm of a2] (b2);
				\vertex [below=0.5cm of b2] (c2);
				\vertex [left=0.5cm of a2] (a3);
				\vertex [below=0.5cm of a3] (b3);
				\vertex [below=0.5cm of b3] (c3);
				\vertex [left=0.5cm of a3] (a4);
				\vertex [below=0.5cm of a4] (b4);
				\vertex [below=0.5cm of b4] (c4);
				
				\diagram*{
                                (a4) -- [scalar] (b3),
                                (c4) -- [plain] (b3),
				(b3)  -- [plain] (b2),
				(b2) -- [photon] (a1),
				(b2) -- [plain] (c1),
				};				
			\end{feynman}
                \end{tikzpicture} $+$
 		\begin{tikzpicture}[baseline=-0.6cm]
			\begin{feynman}
				\vertex (a1);
				\vertex [below=0.5cm of a1] (b1);
				\vertex [below=0.5cm of b1] (c1);
				\vertex [left=0.5cm of a1] (a2);
				\vertex [below=0.5cm of a2] (b2);
				\vertex [below=0.5cm of b2] (c2);
				\vertex [left=0.5cm of a2] (a3);
				\vertex [below=0.5cm of a3] (b3);
				\vertex [below=0.5cm of b3] (c3);
				\vertex [left=0.5cm of a3] (a4);
				\vertex [below=0.5cm of a4] (b4);
				\vertex [below=0.5cm of b4] (c4);
				
				\diagram*{
                                (a1) -- [photon] (b3),
                                (c4) -- [plain] (b3),
				(b3)  -- [plain] (b2),
				(b2) -- [scalar] (a4),
				(b2) -- [plain] (c1),
				};				
			\end{feynman}
                \end{tikzpicture}  \\[10pt]
\mbox{$= \frac{\beta}{M_P^2} \left[  
- M^2 g_{\mu \nu} 
+ 
\frac{ ( p_1 \cdot k_1 - M^2 ) ( p_2^{(\mu} (p_2 + k_2)^{\nu )} + \eta^{\mu \nu} p_2 \cdot k_2 ) }{2 p_1 \cdot k_1} 
+
\frac{ ( p_2 \cdot k_1 - M^2 ) ( p_1^{(\mu} (p_1 + k_2)^{\nu )} +  \eta^{\mu \nu} p_1 \cdot k_2 ) }{2 p_2 \cdot k_1}
\right]$}
\end{flushleft}
\end{figure}
\FloatBarrier
When we project $h_{\mu\nu}$ onto the helicity eigenstate, with a polarisation tensor $\epsilon_{\mu\nu} (k_2) = \epsilon_\mu (k_2) \epsilon_\nu (k_2)$ which is both traceless and orthogonal to $k_2$, we have more simply,
 \begin{align}
\epsilon_{\mu\nu} (k_2) \mathcal{A}_{\chi \chi \to \varphi h_{\mu \nu}} ( p_1, p_2, k_1, k_2 )
= 
\frac{\beta}{M_P^2} \left[  
\frac{ ( p_1 \cdot k_1 - M^2 )  \left( p_2 \cdot \epsilon (k_2) \right)^2  }{ p_1 \cdot k_1} 
+
\frac{ ( p_2 \cdot k_1 - M^2 ) \left(  p_1 \cdot \epsilon (k_2) \right)^2  }{ p_2 \cdot k_1}
\right]
 \end{align}

\paragraph{Body-dependent couplings.}
Finally, consider general body-dependent couplings,
\begin{align}
 \tilde{g}_{A \, \mu \nu} = e^{2 C_A \left( \frac{\phi}{M_P} \right) } g_{\mu\nu} + D_A \left( \frac{\phi}{M_P} \right) \frac{ \nabla_\mu \phi \nabla_\nu \phi }{ M_P^2 M_{\partial}^2 } \; , 
 \label{eqn:geff_def_2}
\end{align}
\begin{align}
 C_A \left( \phi \right) &= \sqrt{2} \alpha_{A} \phi + 2 \beta_A \phi^2 + \mathcal{O} \left( \phi^3 \right) \; , \;\; &D_A ( \phi ) &= \lambda_A + \mathcal{O} \left(  \phi^2  \right)
\end{align}
Rather than compute the scattering amplitudes as in the main text, here we use a slightly different argument.
From our previous discussion of the unitarity cuts that give rise to $a_2$ and therefore $v_2$, we can conclude that the most general potential from \emph{any} scalar-tensor theory with non-derivative couplings takes the following 2PM form in the amplitude gauge,
\begin{align}
V (p^2, r ) = \frac{\kappa}{r} V_{\rm ST}^{(1)} (p^2) + \frac{\kappa^2}{r^2} V_{\rm ST}^{(2)} (p^2) 
\end{align}
where,
\begin{align}
V_{\rm ST}^{(1)}   &= \frac{m_Am_B}{E_A E_B} \left(  2 - 2\sigma^2 - \tilde{G} \right) \nonumber \\
V_{\rm ST}^{(2)}  &= - \frac{m_A m_B}{4 E_A E_B} \left( 3 (5 \sigma^2 - 1) + 4 \tilde{\alpha} \right)
- \frac{\mu}{2 \sqrt{s} }  V_{\rm ST}^{(1)} \left[ V_{\rm ST}^{(1)} \left( \frac{s}{E_A E_B} - 1 \right)  + \frac{ 8 \sigma \, s }{E_A E_B} \right]
\end{align}
are completely fixed by just two constant coefficients $\tilde{\alpha}$ and $\tilde{G}$. 
From the PN expansion, 
\begin{align}
V_{\rm ST}^{(1)} (p^2) &= - \tilde{G} + \mathcal{O} (p^2) \; ,
&V_{\rm ST}^{(2)} (p^2) &= - \tilde{\alpha}  - 3  - \frac{\tilde{G}}{2} ( \tilde{G} -8 ) + \frac{ \tilde{G}^2 \nu }{2} + \mathcal{O} (p^2)
\end{align}
we can therefore \emph{match} the coefficient $a_2$ onto existing 1PN results from \cite{Julie:2022qux} and reproduced in \eqref{eqn:PN_coefs_ST_2}, and find that it is,
\begin{align}
\tilde{G} &= 1 + \alpha_A \alpha_B \; , \;\; &\tilde{\alpha} &= \alpha_A \alpha_B  (2 - \alpha_A \alpha_B ) - \frac{1}{2 M}  \left(   m_B \beta_A \alpha_B^2 \beta_A + m_A \alpha_A^2 \beta_B   \right) \; .
\label{eqn:tildea_2PM}
\end{align}
The first term with no $\beta_{A,B}$ dependence is precisely the $\alpha$-dependent part of $a_2$ found in the main text. 
The second $\beta_{A,B}$-dependent term also coincides with a simple $2 \to 2$ scattering amplitude, and in that language the mass-dependence is straightforward to understand: only the $\alpha_A^2$ (or $\alpha_B^2$) exchange diagrams in $\mathcal{A}_{\chi_A \chi_B \to \phi \phi}$ can have the required poles to generate a classical $1/\sqrt{t}$ dependence, and the angular average \eqref{eqn:xBdA_id} then gives the factor of $m_A$ (or $m_B$) appearing in \eqref{eqn:tildea_2PM}. 

It is then straightforward to compute the leading PM precession in this more general theory, and we also find that no tuning of the EFT couplings can set $\Theta_1^{\rm PM} = 0$ beyond 1PN.

\section{Laplace-Runge-Lenz details}
\label{app:LRL}

Let's first understand how the LRL vector works in Newtonian mechanics (i.e. at 0PN). 

\paragraph{Free particles.}
Suppose that the two particles move inertially without interacting, so that their Hamiltonian is simply\footnote{
We will not write the time dependence of $\bfr$ and $\bfp$ explicitly, but will denote functions of these variables as $f [t]$ to indicate that there is an implicit (rather than explicit) time-dependence.
},
\begin{align}
 H [t] = \frac{p^2}{2 \mu}
\end{align}
where $\mu$ is the reduced mass of the effective one-body system. 
The angular momentum is,
\begin{align}
 L^{ij} [t] = \bfr^i \bfp^j - \bfr^j \bfp^i 
 \label{eqn:L_def}
\end{align}
Invariance under time translations and rotations implies that,
\begin{align}
 H[t] = E \;\; , \;\; L^{ij} [t] = \epsilon^{ijk} \bfL_k
 \label{eqn:HL_fixed}
\end{align}
where $E$ and $\bfL$ are constants. 

But note that fixing $E$ and $\bfL$ does not completely determine the dynamics. 
In fact, since the Hamiltonian does not depend on $\bfr$ we have that $\bfp$ itself is a conserved quantity. 
Of course, the magnitude of $p$ is fixed by $E$ and the condition $\bfp \cdot \bfL = 0$ partially fixes its direction, but there is still one undetermined angular component inside of $\bfp$ which must be specified. 
Altogether, these five constants ($E, \bfL$ and the remaining angle of $\bfp$) fully determine the trajectory $\{ \bfr, \bfp \}$ of the particle, up to the freedom to shift the time co-ordinate. 

Of course, instead of $\bfp$ we could have specified the value of another vector (constructed out of $\bfp$) in order to similarly determine the trajectory. 
One choice is the vector,
\begin{align}
 \bfK^i [t] = \frac{1}{2 \mu} \bfp_j L^{ij} [t] \; , 
 \label{eqn:K_def_inertial}
\end{align}
which is trivially conserved since $\bfp$ and $L^{ij}$ are separately conserved.  
As before, note that $\bfK [t]$ only contains one component which is independent from $H[t]$ and $L^{ij} [t]$, since we have the relations, 
\begin{align}
 L^{ij} [t] \bfK_j [t]  = 0  \; ,\;\; 
 \bfK [t] \cdot \bfK [t] = \frac{1}{2\mu} H[t] L_{ij} [t] L^{ij} [t]    \; .
 \label{eqn:K_indep}
\end{align}
Physically, if we introduce the \emph{impact parameter} $\bfb$ as the minimum value of $\bfr$, such that,
\begin{align}
\left( \bfp \cdot \bfr \right) |_{\bfr = \bfb} = 0
\label{eqn:b_def}
\end{align}
then we find that,
\begin{align}
 \bfK [t] |_{\bfr = \bfb} = H[t] \bfb \; .
\end{align}
Since $\bfK[t]$ is a conserved quantity, it must be $= H[t] \bfb$ at all times. 
This vector (the rescaled impact parameter) is the \emph{Laplace-Runge-Lenz vector}. 

Once the energy and angular momentum are fixed, $\bfb$ is almost fixed by the conditions \eqref{eqn:K_indep}, 
\begin{align}
 \bfb \cdot \bfL = 0 \; , \;\;  b^2 = \frac{L^2}{2 \mu E}
\end{align}
but has one undetermined angular component. 
Altogether, these five constants ($E, \bfL$ and the remaining angular component of $\bfb$) fully determine the trajectory $\{ \bfr, \bfp \}$ of the particle, up to the freedom to shift the time co-ordinate\footnote{
If so desired, one could introduce a sixth conserved constant which fixes this residual time translation freedom. For instance, the vector $\bfr_0 = \bfr - t \bfp/\mu$ is also conserved for inertial motion, but since $\bfr_0 \cdot \bfL = 0$ and $\bfr_0 \cdot \bfb = b^2$, two of its components are fixed by the other conserved quantities and therefore it provides just a single additional constraint (which amounts to fixing the origin for $t$).   
}.

\paragraph{Kepler problem.}
Suppose that the two particles now interact with a $1/r^2$ force law.
The Hamiltonian is,
\begin{align}
 H [t] = \frac{p^2}{2 \mu} + \frac{\kappa}{r} 
\end{align}
where again $\mu$ is the reduced mass of the effective one-body system and $\kappa$ is a small coupling constant\footnote{
For instance, $\kappa = q_1 q_2$ for two point charges interacting electromagnetically or $\kappa = G m_1 m_2 $ for two point masses interacting gravitationally.
}. 
The angular momentum is again given by \eqref{eqn:L_def}, and invariance under time translations and rotations implies that both $H[t]$ and $L^{ij} [t]$ take constant values \eqref{eqn:HL_fixed} for classical trajectories.  

As in the non-interacting case, fixing $E$ and $\bfL$ does not uniquely determine the particle trajectories. 
However, now with an interaction present, $\bfp$ is no longer a conserved quantity, and nor is \eqref{eqn:K_def_inertial}.   
One advantage of the Laplace-Runge-Lenz vector is that it is easily modified to account for the interaction:
\begin{align}
 \bfK^i [t] = \frac{1}{2\mu} \bfp_j L^{ij} [t] + \frac{ \kappa }{2 r} \bfr^i 
 \label{eqn:K_def_Kepler}
\end{align}
This $\bfK [t]$ vector again obeys the three key properties:
\begin{itemize}
 
 \item[(i)] It is conserved, 
 \begin{align}
  \frac{d}{dt} \bfK^i [t] = \frac{1}{2 \mu} \left( - \frac{\kappa}{r^3} \bfr_j \right) L^{ij} [t] + \frac{ \kappa }{ 2 r } \left( \frac{ \bfp^i }{\mu }  - \frac{\bfp \cdot \bfr}{\mu r} \bfr^i  \right)  = 0   
 \end{align}

 \item[(ii)] It contains only one independent component, since it is orthogonal to the angular momentum and its magnitude is fixed by $H[t]$ and $L^{ij}[t]$,
 \begin{align}
  L^{ij} [t] \bfK_j [t] = 0 \; , \;\; \bfK [t] \cdot \bfK [t] = \frac{1}{2 \mu} H[t] L_{ij} [t] L^{ij} [t] + \frac{\kappa^2}{4}
  \label{eqn:K_indep_Kepler}
 \end{align}

 \item[(iii)] It is proportional to the impact parameter, since using \eqref{eqn:b_def}
 \begin{align}
   \bfK^i [t] |_{\bfr = \bfb} = \left( H[t] - \frac{\kappa}{2 b} \right) \bfb^i \; .
   \label{eqn:K_to_b_Kepler}
 \end{align}
  Since $\bfK[t]$ is conserved, it must be equal to the right-hand-side at all times. Note that since $\bfK[t]$ is constrained by \eqref{eqn:K_indep_Kepler}, the impact parameter is constrained by,
  \begin{align}
   \bfb \cdot \bfL = 0 \; , \;\; b = \frac{\kappa}{2 E} \left( 1 - \sqrt{ 1 + \frac{2 E L^2}{\mu \kappa^2} } \right)
  \end{align}
  and contains only one undetermined component.

\end{itemize}

So altogether, specifying the five independent components in $\{ E, \bfL, \bfK \}$ is enough to determine the $\{ \bfr, \bfp \}$ trajectory of the system (again up to shifts in the time co-ordinate). 
We can see this very explicitly by using \eqref{eqn:K_def_Kepler} to write,
\begin{align}
  \bfK \cdot \bfr = \frac{L^2}{2\mu} - \frac{\kappa}{2} r \; .
  \label{eqn:Kr}
\end{align}
So if we introduce the polar angle $\bfr \cdot \bfb = r b \cos \theta$, \eqref{eqn:Kr} and \eqref{eqn:K_to_b_Kepler} fixes $r$ to be,
\begin{align}
 \frac{1}{r} =  \frac{\mu \kappa}{L^2} \left(  1 - \left( 1 - \frac{2 E b}{\kappa} \right) \cos \theta   \right)
\end{align}
which is the usual conic describing the motion of a Keplerian orbit. 
In fact, this tells us that the eccentricity of the orbit is,
\begin{align}
 e =  1 - \frac{2 E b}{\kappa} = \sqrt{ 1 + \frac{2 E L^2}{\mu \kappa^2} } \; , 
\end{align}
so some authors identify the LRL vector with the eccentricity rather than the impact parameter.

\paragraph{General spherical potentials.}
Suppose that the two particles instead interact via the more general Hamiltonian,
\begin{align}
 H [t] = \frac{p^2}{2 \mu} + U (r) \; . 
\end{align}
As before, the angular momentum is given by \eqref{eqn:L_def}, and invariance under time translations and rotations implies that both $H[t]$ and $L^{ij} [t]$ take constant values \eqref{eqn:HL_fixed} for classical trajectories.  
However, this still leaves one undetermined degree of freedom in $\{ \bfr, \bfp \}$. 
We should note immediately that we do not expect to have a conserved quantity like in the previous examples, since that would amount to every non-relativistic potential $U(r)$ being exactly integrable/solvable, and we know this is not the case. 

However, we can nonetheless note that in order to uniquely specify the trajectory of the particles, we need to provide one more datum: for instance the impact parameter, $\bfb$. 
This constant vector can always be given as an initial condition, and since its size and direction are constrained by,
\begin{align}
 \bfb \cdot \bfL = 0 \;, \;\; U(b) + \frac{L^2}{2 \mu b^2} = E \; ,
\end{align}
it again contains the single undetermined component required to solve for $\{ \bfr, \bfp \}$. 

The question is whether there is some analogue of $\bfK [ t]$, built out of $\bfr$ and $\bfp$, which is equal to $\bfb$ throughout the motion. 
It is straightforward to show that,
\begin{align}
 \bfK^i [t] = \frac{1}{2\mu} \bfp_j L^{ij} [t] - \frac{ r U' (r) }{2} \bfr^i + \int d t  ( r U(r) )'' \frac{\bfp \cdot \bfr}{2 \mu} \bfr^i  
 \label{eqn:K_def_U}
\end{align}
is one such vector. It is conserved, thanks to the equations of motion, and proportional to $\bfb$, since using \eqref{eqn:b_def}, 
\begin{align}
 \bfK [t] |_{\bfr = \bfb} = \left( H[t] - U(b) - \frac{b U'(b)}{2}   \right) \bfb \; . 
\end{align}
However, the crucial difference is that this $\bfK$ is now \emph{non-local in time}: only in the particular case of $U(r) = 1/r$ does the $(r U(r) )''$ term vanish in \eqref{eqn:K_def_U}, resulting in a local quantity that depends only on $\bfr$ and $\bfp$ at a single time.  
Otherwise the vector $\bfK$ does not correspond to any local symmetry of the system and does not reduce the number of degrees of freedom.

\section{Useful integral identities}
\label{app:int_id}

\paragraph{Fourier transforms.}
The $d$-dimensional Fourier transform of a radial function coincides with the following Hankel transform,
\begin{align}
\int d^d \bfr \, e^{ i \bfq \cdot \bfr} f(r)  = \int_0^{\infty} d r \, q  ( \frac{ 2 \pi r }{q} )^{d/2}  f(r) J_{\frac{d}{2} -1 } ( q r )
\end{align}
where $J_n$ is the usual Bessel function of the first kind. 
In dimensional regularisation, the Fourier transform of $r^{-n}$ is therefore given by,
\begin{align}
  \int d^d \bfr \, e^{i \bfq \cdot \bfr} r^{-n} = q^{n-d} \; \frac{ \pi^{d/2} \Gamma \left( \frac{d-n}{2} \right) }{ 2^{n-d} \Gamma \left( \frac{n}{2} \right) }
  \label{eqn:FT_id}
\end{align}
For instance, setting $n=1$ and $d=3$ gives the well-known result,
\begin{align}
      \int d^3 \bfr \, \frac{ e^{i \bfq \cdot \bfr} }{ r } = \frac{4 \pi}{q^{2}} \; . 
\end{align}
For every even $n$, we may take $d \to 3$ without encountering a divergence. For instance, when $n=2$, 
\begin{align}
      \int d^3 \bfr \, \frac{ e^{i \bfq \cdot \bfr} }{ r^2 } = \frac{2 \pi^2}{q} \; .
\end{align}
For every odd $n > 1$, the $d \to 3$ limit is singular. However, one virtue of dim reg is that the singular terms are analytic in $q^2$, and thus correspond to ultra-local $\delta^d ( \bfr )$ and its derivatives in position space. 
For instance for $n=3$,   
\begin{align}
   \lim_{d \to 3} \int d^d \bfr e^{i \bfq \cdot \bfr} r^{-3} &= - 2 \pi \log ( q^2 ) + \text{const} \; ,   \nonumber \\
   \lim_{d \to 3} \int \frac{d^d \bfq}{ (2 \pi )^d}  e^{-i \bfq \cdot \bfr} \log ( q^2 )  &= \frac{-1}{2 \pi} r^{-3} + \mathcal{O} \left( \delta^3 ( \bfr ) \right) \; . 
\end{align}

The Feinberg-Sucher trick requires the related identity,
\begin{align}
 \int \frac{d^d \bfq}{ (2 \pi )^d } \frac{ e^{-i \bfq \cdot \bfr} }{ q^2 - t + i \epsilon } 
 = \frac{1}{2\pi} \, \left( \frac{ \sqrt{t} }{ 2 \pi r } \right)^{\frac{d}{2} -1 } \, K_{\frac{d}{2} - 1 } \left( \sqrt{t} r \right)
\end{align}
where $K_n$ is the modified Bessel function of the second kind. 
In $d=3$ dimensions, this gives,
\begin{align}
 \int \frac{d^3 \bfq}{ (2 \pi )^3 } \frac{ e^{-i \bfq \cdot \bfr} }{ q^2 - t + i \epsilon } 
 = \frac{ e^{- \sqrt{t} r} }{4\pi r } \; . 
\end{align}

\paragraph{Non-relativistic loop integration.}
When performing the Born subtractions in the Lippmann-Schwinger equation, we make use of the general identity,
\begin{align}
 \int \frac{d^d \bfl}{( 2 \pi )^d} \frac{1}{ | \bfl |^{n_1} | \bfl - \bfq |^{n_2} } = \frac{1}{ (4 \pi )^{d/2} q^{n_1 + n_2 - d} } \frac{  \Gamma \left( \frac{ d - n_1 }{2} \right) \Gamma \left( \frac{ d - n_2 }{2} \right) \Gamma \left( \frac{n_1 + n_2 - d}{2} \right) }{ \Gamma \left( \frac{n_1}{2} \right) \Gamma \left( \frac{ n_2 }{2} \right) \Gamma \left( d - \frac{ n_1 + n_2 }{2} \right)  } \; . 
 \label{eqn:Born_id}
\end{align}

\begin{align}
	\frac{1}{ E_A (p^2) + E_B (p^2) - E_A( k^2) - E_B (k^2) } = \frac{ 2 \xi \sqrt{s} }{ p^2 - k^2 }  + \frac{3 \xi - 1 }{ 2 \xi s } + \mathcal{O} ( p^2 - k^2 )  
\end{align}

\paragraph{Angular integrals.}
Feinberg and Sucher evaluated the simple seed integrals
\begin{align}
	\langle \frac{1}{d_A} \rangle \;\; , \;\; \langle \frac{1}{d_A d_B} \rangle \; .
\end{align}
From these, general averages over $x_A^a x_B^b / d_A d_B$ can be computed: see e.g. the Appendix B of \cite{Holstein:2016fxh} for a list of the first few\footnote{
There are a few typos in \cite{Holstein:2016fxh} which only become apparent at finite $y-1$, for instance in $K_{22}, J_{02}^A, J_{20}^B, ...$. 
}.

Since we are focussed on the classical component, we are interested in the angular averages which give the non-analytic $1/\sqrt{t}$ at small $t$. 
The only averages which do this are\footnote{
To prove \eqref{eqn:ang_int_id}, integrate over the polar and azimuthal angles which $\bfl$ makes with $\bfp_A$ and $\bfp_B$, using the addition formula $x_B = y x_A + \sqrt{1-y^2} \sqrt{1-x_A^2} \cos \phi$. 
},
\begin{align}
	\left\langle \frac{ x_B^{2n} }{d_A} \right\rangle 
	=
	\frac{m_A}{\sqrt{t}} \frac{  \sqrt{\pi} \Gamma \left( \frac{1}{2} + n \right) }{ \Gamma ( 1 + n ) } ( 1 - y^2 )^n + \mathcal{O} (t^0 ) 
	\label{eqn:ang_int_id}
\end{align} 
and its $A \leftrightarrow B$ permutation (where we have set $d=3$).

\bibliographystyle{JHEP}
\bibliography{refs}

\end{document}